\documentclass[]{aa}
\usepackage{rotating, graphicx}
\usepackage{txfonts}
\usepackage{url}
\usepackage{pdflscape} 
\usepackage{isotope}
\usepackage{color}
\usepackage{morefloats}
\usepackage{setspace}
\usepackage{multirow}
\usepackage[flushleft]{threeparttable}
\usepackage{booktabs,caption}
\usepackage{subfigure}
\usepackage{multirow}
\usepackage{verbatim}
\DeclareCaptionLabelFormat{continued}{#1~#2 (continued)}
\usepackage[breaklinks,colorlinks,citecolor=black,linkcolor=black,urlcolor=blue]{hyperref}
\begin{document} 
\newcommand{\HZ}{HZ\,44}
\newcommand{\HD}{HD\,127493}
\newcommand{\CD}{CD--31$^{\circ}\,$4800}
\newcommand{\RE}{RE\,0503$-$289}
\newcommand{\lsIV}{LS\,IV$-14^\circ116$}
\newcommand{\UVO}{[CW83]\,0825$+$15}
\newcommand{\rev}[1]{\textbf{\textcolor{red}{#1}}}

\title{Heavy metals in intermediate He-rich hot subdwarfs:\\ The chemical
composition of HZ\,44 and HD\,127493}

\author{M. Dorsch\inst{1}
\and M. Latour\inst{2}
\and U. Heber\inst{1}
}

\institute{
Dr. Remeis-Sternwarte \& ECAP, Astronomical Institute, University of Erlangen-N\"urnberg, Sternwartstr. 7, 96049, Bamberg, Germany; 
\email{matti.dorsch@fau.de}
\and Institut f\"ur Astrophysik, Georg-August-Universit\"at, 
Friedrich-Hund-Platz 1, 37077, G\"ottingen, Germany
}
\date{Received ; accepted}
\abstract
{Hot subluminous stars can be spectroscopically classified as subdwarf B (sdB) and O (sdO) stars. While the latter are predominantly hydrogen deficient, the former are mostly helium deficient. The atmospheres of most sdOs are almost devoid of hydrogen, whereas a small group of hot subdwarf stars of mixed H/He composition exists, showing extreme metal abundance anomalies. Whether such intermediate helium-rich (iHe) subdwarf stars provide an evolutionary link between the dominant classes is an open question.  
}
{The presence of strong Ge, Sn, and Pb lines in the UV spectrum of \HZ\ suggests a
strong enrichment of heavy elements in this iHe-sdO star and calls for a detailed quantitative spectral analysis focusing on trans-iron elements.  
}
{Non-LTE model atmospheres and synthetic spectra calculated with \texttt{TLUSTY}/\texttt{SYNSPEC} are combined with high-quality optical, UV and FUV spectra of \HZ\ and its hotter sibling \HD\ to determine their atmospheric parameters and metal abundance patterns. }
{By collecting atomic data from literature we succeeded to determine abundances of 29 metals in \HZ, including the trans-iron elements Ga, Ge, As, Se, Zr, Sn, and Pb and provide upper limits for 10 other metals. This makes it the best described hot subdwarf in terms of chemical composition. For \HD\ the abundance of 15 metals, including Ga, Ge, and Pb and upper limits for another 16 metals were derived. Heavy elements turn out to be overabundant by one to four orders of magnitude with respect to the Sun. Zr and Pb are among the most enriched elements.
}
{
The C, N, and O abundance for both stars can be explained by nucleosynthesis of hydrogen burning in the CNO cycle along with their helium enrichment.
On the other hand, the heavy-element anomalies are unlikely to be caused by nucleosynthesis. Instead diffusion processes are evoked with radiative levitation overcoming gravitational settlement of the heavy elements.
}

\keywords{stars: abundances, stars: atmospheres, stars: individual (\object{HZ\,44}), stars: individual (\object{HD\,127493}),  stars: evolution, stars: subdwarfs}

\titlerunning{The chemical composition of HZ\,44 and HD\,127493}
\maketitle
%
%
\section{Introduction}
\label{Introduction}
\begin{figure}
\centering
\includegraphics[width=0.98\columnwidth]{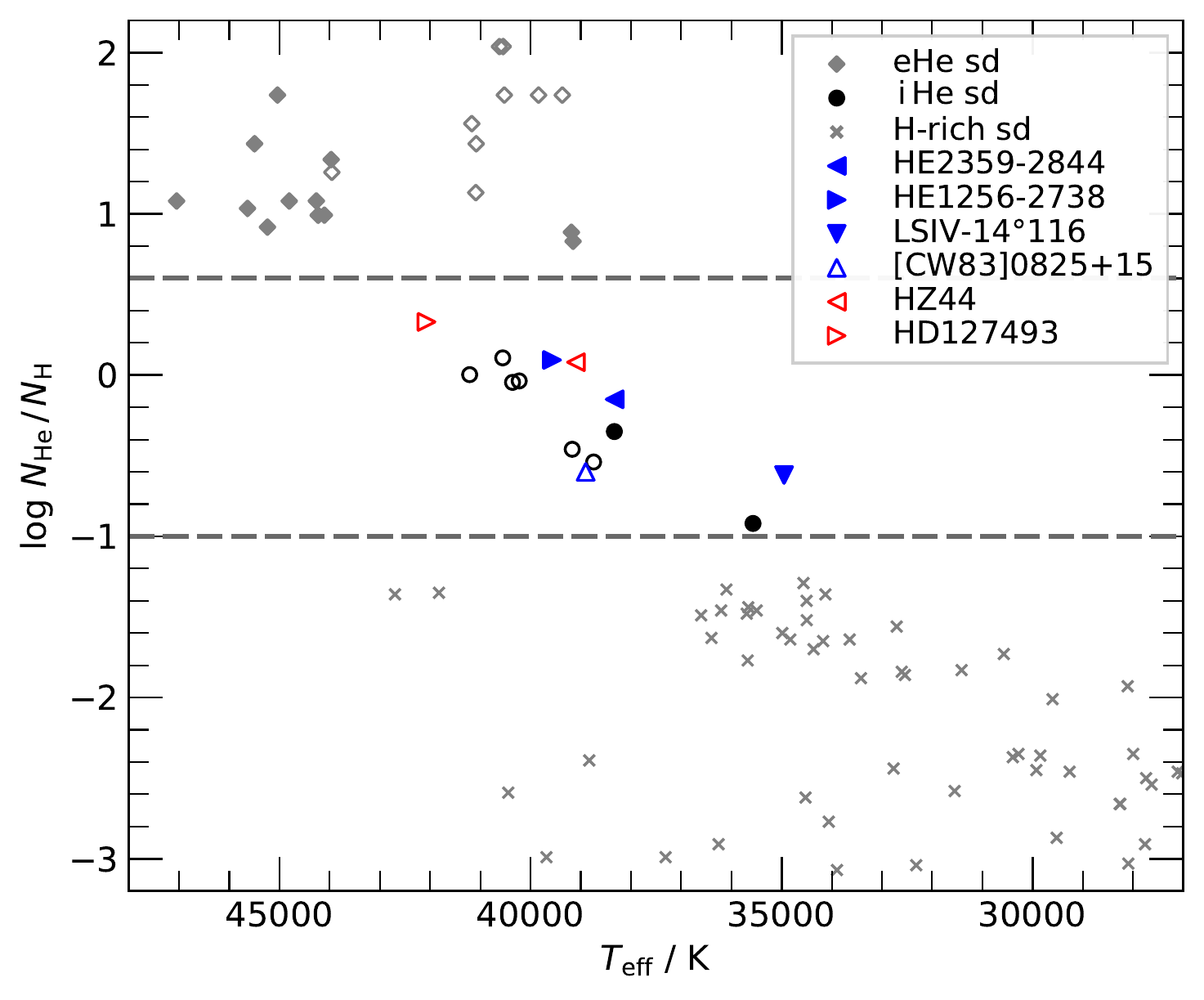}
\vspace{-10pt}
\captionof{figure}{Distribution of helium abundance versus $T_\mathrm{eff}$ for the subdwarf population from the SPY sample \citep{Lisker2005,stroeer07,hirsch09}. 
The dashed lines mark the range of iHe subdwarfs. 
The two heavy-metal subdwarfs from SPY \citep{naslim13} and the prototypical heavy-metal subdwarfs \lsIV\ \citep{naslim11} and \UVO\ \citep{Jeffery2017} are marked by blue triangles, \HD\ and \HZ\ in red.
Open symbols represent C-poor, N-rich stars, filled symbols C-rich stars.
}
\label{fig:tefflogy}
\end{figure}

Hot subdwarf stars of spectral type O and B (sdO and sdB) represent late stages of the evolution of low-mass stars. 
They are characterized by high effective temperatures, ranging from $T_{\mathrm{eff}}=20\,000$\,K to more than 45\,000\,K while their surface gravities are typically between $\log g = 5.0$ and 6.5 \citep{heber16}.
The vast majority of sdB stars are helium-deficient with helium abundances that can reach as low as $\log n(\mathrm{He})/n(\mathrm{H})=-4$ \citep{Lisker2005}. 
Most of these stars are evolving through the core helium-burning phase on the horizontal branch but their remaining hydrogen envelope is too thin to sustain hydrogen shell burning \citep{dor93}. 
This is why they are said to populate the extreme part of the horizontal branch (EHB) in the Hertzsprung-Russell diagram. 
Since the mass of sdB stars is largely dominated by that of the He-burning core (the hydrogen envelope contributes less than 2\%), the mass distribution of these stars is strongly peaked around the mass required for the He-flash ($\sim$0.47 $M_\odot$; \citealt{dor93, han02, Fontaine2012}). 
Unlike normal horizontal branch objects, the sdB stars, due to their lack of H-shell burning, evolve directly to the white dwarf (WD) cooling sequence without an excursion to the asymptotic giant branch (AGB) \citep{dor93}. 
\smallskip\\
The formation and evolutionary history of the sdO stars is not understood very well. Because most sdO stars are hotter and somewhat more luminous than the sdB stars, they can not be associated to the EHB. Whether their evolution is linked to the EHB or not remains an open question. It has been suggested that the He-deficient sdO stars are the descendants of the sdB stars, because they share the peculiar chemical  composition \citep{heber16}. 
However, the majority of sdO stars have atmospheres dominated by helium with hydrogen being a trace element only. The formation of these He-sdO stars is unlikely to be linked to the EHB and two rivaling scenarios have been invoked to explain the hydrogen deficiency, either via internal mixing \citep[]{lanz04,miller08} or via a merger of two helium white dwarfs \citep{zhang12a}.   
A very small number of hot subluminous stars have atmospheres of mixed H/He composition. Because their metal content is very different from that of the extremely He-rich subdwarfs, \cite{naslim13} suggested to distinguish intermediate H/He composition subdwarfs (iHe-sds) with $n(\mathrm{He})/n(\mathrm{H}) < 4$ as a class separate from the extremely He-rich (eHe) hot subdwarfs. 
The SPY project has provided the largest homogeneous sample of hot subdwarfs from high resolution spectroscopy. By quantitative spectral analyses 85 H-rich hot subdwarf stars have been identified, as well as 23 eHe and 10 iHe hot subdwarfs  \citep{Lisker2005,stroeer07,hirsch09}. The latter appear as a transition stage between the cooler H-rich sdB stars and the hotter eHe subdwarf stars (see Fig. \ref{fig:tefflogy}).
As to the carbon and nitrogen abundances a dichotomy exists, both for the eHe and the iHe sds. \cite{stroeer07} classified the line spectra of helium-rich hot subdwarfs in three classes: N-, C-, and C$\&$N-strong. \citet{hirsch09} showed that, indeed, the N strong-lined stars are enriched in nitrogen with respect to the Sun, as are the C strong-lined enriched in carbon and the C\&N strong-lined in both elements.  
This dichotomy is most obvious for the eHe hot subdwarf stars, the N-strong ones being mostly cooler than the C- or C$\&$N- strong ones. For the iHe hot subdwarfs such a separation is less pronounced (see Fig. \ref{fig:tefflogy}).

\cite{naslim11} have discovered trans-iron elements, in particular zirconium and lead, to be strongly overabundant in the iHe-sdB \lsIV.
Since then, three additional intermediate He-sdBs (indicated with blue triangles in Fig.~\ref{fig:tefflogy}), with effective temperature between 35\,000\,K and 40\,000\,K, have been found to be extremely enriched in heavy elements \citep{naslim13,Jeffery2017}.
The origin of the extreme enrichment observed in iHe hot subdwarfs is not yet understood. 
Radiatively driven diffusion 
has been proposed, but is poorly constrained with only four stars ([CW83] 0825+15, \lsIV, and the SPY objects HE\,2359--2844 and HE\,1256--2738) studied so far. Therefore, we decided to extend the sample to higher temperatures by studying \HZ\ (39\,000\,K) and \HD\ (42\,000\,K) for which 
excellent high-resolution spectroscopy is available both for the optical and the ultraviolet spectral range. This makes them excellent targets to perform a comprehensive quantitative abundance analysis and focus on trans-iron elements.
\smallskip\\
\HZ\ and \HD\ were among the first sdOs to be identified in the 1950s.
\HZ\ was discovered in the first survey for faint blue stars in the halo by \cite{Humason1947}. 
The first spectral analysis of the helium line spectrum of \HZ\ was published in the pioneering paper of \cite{Munch1958}.
From a curve of growth analysis \cite{Peterson1970} derived metal abundances for the first time, but we know of no contemporary study. 
\HZ\ is now a spectrophotometric standard star \citep{Massey1988,Oke1990,Landolt2007}, used for the calibration of the HST \citep{Bohlin1990,Bohlin1996,Bohlin2001}, as well as that of \textit{Gaia} \citep{Marinoni2016}, and therefore has frequently been observed.
High resolution spectra are available from the far-UV to the red in the FUSE, IUE, and HIRES@Keck data archives. 
\smallskip\\
\HD\ has been used as secondary spectrophotometric standard star \citep{SpencerJones1985, Kilkenny1998, Bessell1999}. 
Therefore, very accurate photometry is available but spectroscopic observations are not as extensive as for \HZ. Starting with the curve of growth analyses of \citet{Peterson1970} and \citet{Tomley1970} abundances of C, N, O, Ne, Mg, and Si were derived. The first NLTE model atmospheres were calculated by \citet{Kudritzki1976}, who revised the atmospheric parameters. Abundances of carbon \citep{Gruschinske1980} and C, N, O and Si \citep{Simon1980} were derived from equivalent widths of ultraviolet lines. A NLTE analyis of optical spectra allowed \citet{bauer95} to determine the abundances of  C, N, O, Ne, Mg, Al, and Si.
The most recent NLTE analysis by \citet{hirsch09} revised the atmospheric parameters and determined C and N abundances from optical spectra. For completeness we give a comparison of our results with those of previous analyses in the Appendix. Hence, our knowledge of the chemical composition of both stars is rather limited.
\footnote{The abundance analysis performed in this paper is based on, revises, and extends results for \HD\ from \cite{Dorsch2018}.}
\smallskip\\
The paper is organized as follows. In Sect.~\ref{observation} we provide a description of the available spectra followed by a presentation of the atmospheric parameters that we derived for our stars in Sect.~\ref{s:atmos_params}. The spectroscopic masses obtained from the spectral energy distributions and the \textit{Gaia} parallaxes are presented in Sect.~\ref{angular}.
The atomic data used for our abundance analysis are discussed in Sect.~\ref{sect:atomic_data}.
In Sect.~\ref{sect:abundances} we provide details on the abundance analysis of all considered metallic elements. 
The abundance patterns for \HZ\ and \HD\ are discussed in Sect.~\ref{sect:discussion} and we conclude
in Sect.~\ref{sect:summary}.
%
\section{Spectroscopic observations}
\label{observation}
\begin{table}
\setstretch{1.2}
\caption{Spectra used for the analysis.\tablefootmark{a} }
\label{tab:obs}
\vspace{-12pt}
\begin{center}
\begin{tabular}{l l c c c}
\toprule
\toprule
Star		& Instrument	& ~Range\,(\AA)~	& ~R~	& ~S/N~	\\ 
\midrule
HD\,127493	& IUE SWP	& 1150\,$-$\,1970	& 10 000	& \phantom{1}14	\\
			& GHRS		& 1225\,$-$\,1745	& 0.07 \AA\tablefootmark{b}	& \phantom{1}40	\\
			& IUE LWR	& 1850\,$-$\,3273	& 10 000	& \phantom{1}14	\\
			& FEROS		& 3700\,$-$\,9200	& 48 000	& $180$	\\
HZ\,44		& FUSE		& \phantom{1}905\,$-$\,1188	& 19 000	& \phantom{1}30	\\
			& IUE SWP	& 1150\,$-$\,1970	& 10 000	& \phantom{1}10	\\
			& HIRES		& 3022\,$-$\,7580	& 36 000	& $142$ \\
			& ISIS		& 3700\,$-$\,5260	& 1.5 \AA\tablefootmark{b}	& $170$ \\
\bottomrule 
\end{tabular}
\end{center}
\vspace{-12pt}
\tablefoot{
\tablefoottext{a}{The signal-to-noise ratio is the average over the spectrum.}
\tablefoottext{b}{The resolution for long-slit spectrographs is given instead as $\Delta \lambda$.}
}
\end{table}

For both stars excellent archival data are available in both the optical and UV ranges.
An overview of the spectra we collected and used is given in Table \ref{tab:obs}, with additional details on the individual observations listed in Table \ref{tab:obs:detail}.
\smallskip \\
\noindent We used optical FEROS spectra to determine the atmospheric parameters of \HD\ and measure photospheric metal abundances. 
FEROS is an echelle spectrograph mounted on the MPG/ESO-2.20m telescope operated by the European Southern Observatory (ESO) in La Silla. 
It features a high resolving power of $\mathrm{R}\approx 48000$ \citep{kaufer99} and its usable spectral range, from $\sim$3700\,\AA\ to $\sim$9200\,\AA, includes all the Balmer lines as well as many He\,{\sc i}, He\,{\sc ii}, and metal lines.
The three available spectra of \HD\ were co-added to achieve a high signal-to-noise ratio (S/N) of $\gtrsim$\,100 in the 4000\,--\,6000\,\AA\ range. 
Nevertheless, the S/N decreases drastically toward both ends of the spectral range and especially below 3800\,\AA.
\smallskip\\
Both stars have been observed with the International Ultraviolet Explorer (IUE) satellite with the short-wavelength prime (SWP) camera. 
We retrieved three archival INES\footnotemark\ spectra for \HD\ and two for \HZ. For each star we co-added the individual spectra to increase the S/N. 
\footnotetext{IUE Newly-Extracted Spectra, \url{http://sdc.cab.inta-csic.es/ines/index2.html}}
They continuously cover the 1150\,--\,1980\,\AA\ range with a resolution of $\mathrm{R}\approx10000$.
Additional IUE spectra taken with the LWR camera (covering the 1850\,--\,3350\,\AA\ range) are also available for both stars. 
However these spectra have a lower quality and the S/N drops sharply at both ends of the spectra. 
Fewer lines are observed in this wavelength range but the IUE LWR spectrum of \HD\ has nevertheless been useful for the abundance analysis.\smallskip\\
\HD\ has also been observed with the Goddard High-Resolution Spectrograph (GHRS) mounted on the Hubble Space Telescope (HST). 
These spectra are publicly available in the MAST\footnotemark\ archive and cover the 1225\,--\,1745\,\AA\ range with a resolution of $\Delta\lambda \approx 0.07$\,\AA. 
The final spectrum is a combination of ten observations spanning 35\,\AA\ each and lacks coverage in the following regions: 1450.5\,--\,1532.5\,\AA, 1567.7\,--\,1623.2\,\AA, and 1658.1\,--\,1713.0\,\AA. 
Since the wavelength calibration was not perfect, we cross-correlated the individual spectra to match the synthetic spectrum of \HD. In addition, they were shifted to match the flux level of the IUE spectra.
\footnotetext{Mikulski Archive for Space Telescopes, \url{https://archive.stsci.edu/index.html}}
\smallskip\\ 
\HZ\ has been observed with the Far Ultraviolet Spectroscopic Explorer (FUSE) satellite over the spectral range between 905\,\AA\ and 1188\,\AA. 
We retrieved three calibrated observations from MAST, two taken through the LWRS ($30"\times30"$) aperture, and one through the MDRS ($4"\times20"$) aperture. 
We co-added all spectra from the eight segments in each observation. 
After inspection it turned out that the MDRS spectrum had a better quality and a better wavelength calibration, so we use this spectrum for our analysis.
\begin{figure*}
\centering
\includegraphics[width=0.98\textwidth]{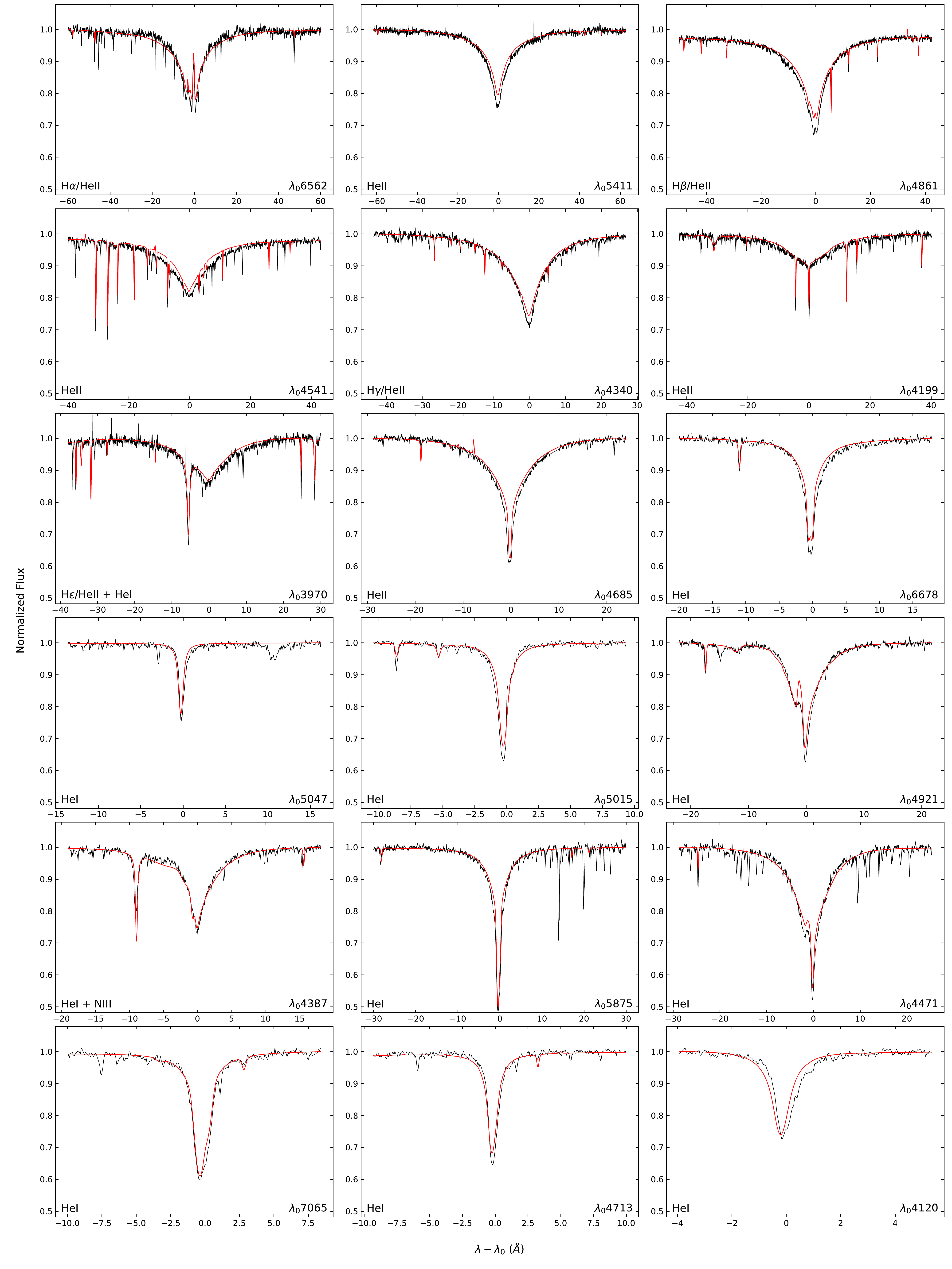}
\vspace{-10pt}
\captionof{figure}{Best fit (red) to the Balmer and helium lines selected in the normalized FEROS spectrum of \HD\ (black).}
\label{fig:atm_fit}
\end{figure*}
\begin{figure*}
\centering
\includegraphics[width=0.9\textwidth]{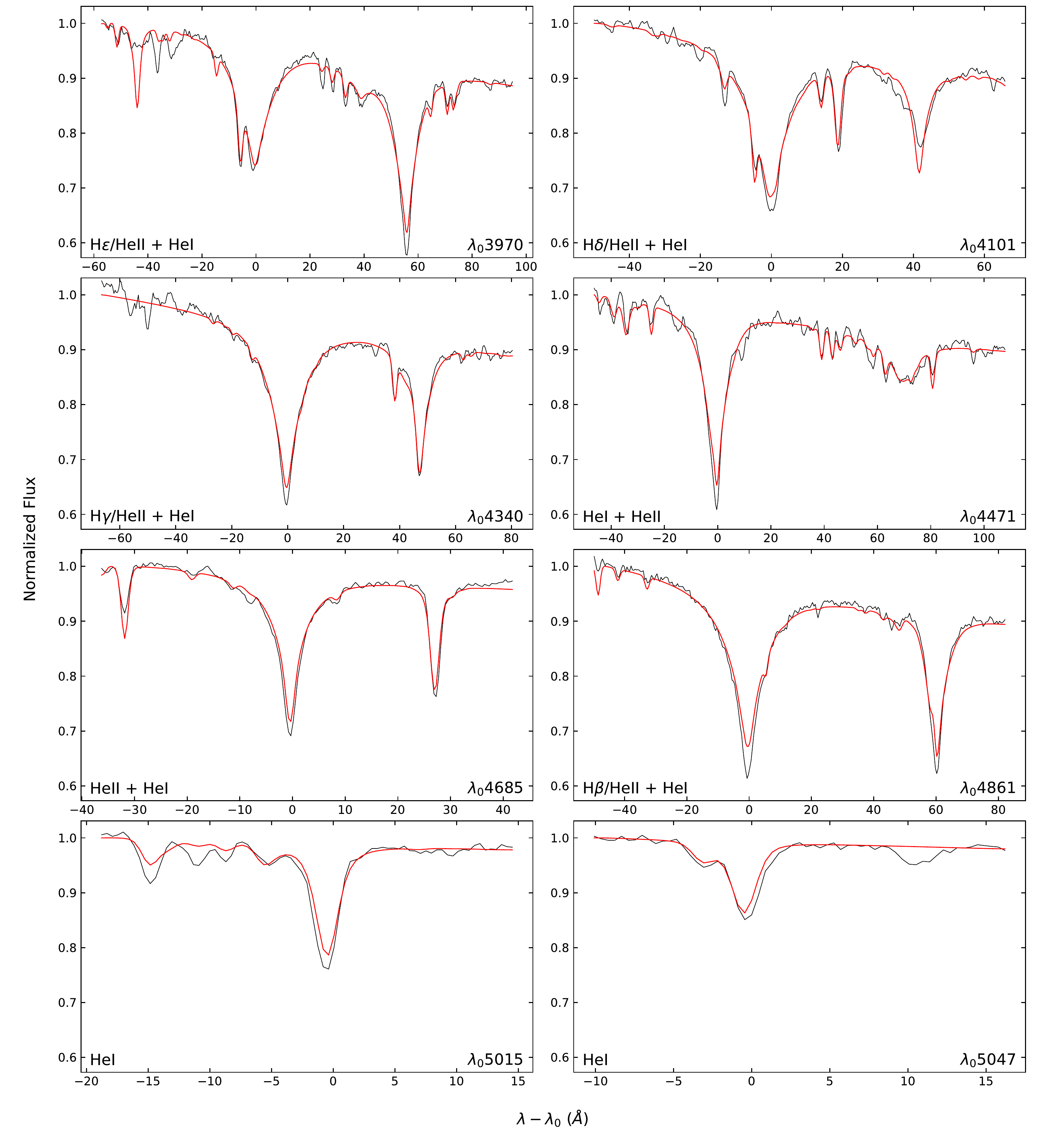}
\vspace{-10pt}
\captionof{figure}{Best fit (red) to the Balmer and helium lines selected in the flux-calibrated ISIS spectrum of \HZ\ (black).}
\label{fig:atm_fit:HZ:ISIS}
\end{figure*}
\smallskip\\
To determine the atmospheric parameters of \HZ\ we used a low resolution (1.5\,\AA), high S/N spectrum taken with the Intermediate dispersion Spectrograph and Imaging System (ISIS) mounted at the Cassegrain focus of the 4.2m William Herschel Telescope on La Palma. The spectrum covers the 3700$-$5260 \AA\ range, thus including the Balmer lines, except H$_\alpha$, as well as \ion{He}{i} and \textsc{ii} lines.  
\smallskip\\
The spectra of \HZ\ taken with the HIRES echelle spectrograph mounted on the Keck\,I telescope on Mauna Kea were most valuable for our abundance analysis.
A total of 68 extracted HIRES spectra of \HZ\ from several programs covering various wavelength ranges are available in the Keck Observatory Archive (KOA\footnotemark).
\footnotetext{Keck Observatory Archive, \url{https://koa.ipac.caltech.edu/cgi-bin/KOA/nph-KOAlogin}}
We co-added the spectra of four high S/N HIRES observations to produce the spectrum used for our abundance analysis. 
To access the ranges between 3022\,\AA\ and 3128\,\AA\ and above 5990\,\AA\ we considered two additional HIRES spectra that were used specifically for these regions. 
Additional spectra were also retrieved from the archive and used to measure radial velocities. 
Unfortunately, the normalization of HIRES spectra is difficult since the spectral orders are narrower than many broad Balmer or helium lines. 
This is not a problem for sharp metal lines, but renders the spectra next to useless for the determination of atmospheric parameters of \HZ.

%
\section{Atmospheric parameters and radial velocities}
\label{s:atmos_params}

In order to analyze the spectra of our stars we computed non-LTE model atmospheres using the \texttt{TLUSTY} and \texttt{SYNSPEC} codes developed by \cite{hubeny88} and \cite{Lanz2003}. 
A detailed description of \texttt{TLUSTY}/\texttt{SYNSPEC} has recently been published in \cite{hubeny17a,hubeny17b,hubeny17c}.
\smallskip\\
%
We derived atmospheric parameters for both stars using optical spectra (besides HIRES) and a newly constructed model atmosphere grid that includes effective temperatures from  $T_\mathrm{eff}=35\,000$\,K to 48\,000\,K in steps of 1000\,K and surface gravities from $\log g =$ 4.7 to 6.0 in steps of 0.1.
For each of these combinations, models with helium abundances from $\log n_\mathrm{He}/n_\mathrm{H}=-1.0$ to $+2.1$ in steps of 0.1 were computed. 
All models in the grid include carbon, nitrogen, and silicon in non-LTE using the abundances determined by our previous analysis of \HD\ \citep{Dorsch2018} which significantly improves the atmospheric structure compared to models that only include hydrogen and helium \citep[e.\,g.][]{schindewolf18}.
These values of C, N, and Si are also appropriate for \HZ\ as shown in the abundance analysis presented in Sect.~\ref{sect:abundances}. 
The selection of all lines we used, as well as the global best-fit model for \HD\ is shown in Fig.~\ref{fig:atm_fit}. 
Our final best fit of the ISIS spectrum of \HZ\ is shown in Fig.~\ref{fig:atm_fit:HZ:ISIS}.
The resulting parameters derived from the simultaneous fit of all selected H and He\,\textsc{i-ii} lines for both stars are reported in Table \ref{tab:atm}. 
The atmospheric parameters for \HD\ derived by \cite{hirsch09} are also listed. They were obtained with the same FEROS spectrum but different model atmospheres and are fully consistent with our results. As shown in Fig. \ref{fig:tefflogy} the atmospheric parameters of both stars fit very well the trend of helium abundance to increase with increasing effective temperatures.
We found no indication of rotation or microturbulence in either star; some optical metal lines are in fact sharper in the observations than in the models.
\begin{table*}
\setstretch{1.2}
\caption{Parameters derived from optical spectroscopy.}
\label{tab:atm}
\vspace{-2pt}
\centering
\begin{tabular}{llcccccl}
\toprule
\toprule
Name &$T_{\text{eff}}$ & $\log{g}$  & $\log{n_\mathrm{He}/n_\mathrm{H}}$ & v$_{\rm rot}\sin{i}$ & Spectrum & Ref \\ 
     &  [K]             &   [cgs]         &           &   [km\,s$^{-1}$]             &   \\
\midrule
\HZ & $39100 \pm 600$	& $5.64 \pm 0.10$	& $0.08 \pm 0.05$	& $<5$	&  ISIS	& 1	\\
\HD & $42070 \pm 180$	& $5.61 \pm 0.04$	& $0.33 \pm 0.06$	& $<10$	&  FEROS	& 1	\\
    & $42480 \pm 250$	& $5.60 \pm 0.05$	& $0.60 \pm 0.30$	& $<10$	&  FEROS	& 2	\\
\bottomrule
\end{tabular}
\tablefoot{
References:
\tablefoottext{1}{this work}
\tablefoottext{2}{\citet{hirsch09}}.
The uncertainties stated were determined using different methods. 
Uncertainties on our results are determined using the bootstrapping method.
Please to refer \citet{hirsch09} for an explanation of their uncertainties. 
}
\end{table*}
%
\smallskip\\
Radial velocities of \HZ\ in 27 HIRES spectra taken between 1995 and 2016 were measured by \citet{schork} and are listed in Table \ref{tab:rv}.
From these values an average radial velocity of $v_\mathrm{rad}=-12.7 \pm 0.4$\,km\,s$^{-1}$ was derived.
The measurements show that the radial velocity of \HZ\ does not vary on a scale of a few km\,s$^{-1}$. 
Within the radial velocity uncertainties, neither a short- nor a long-period companion is detected.
Our radial velocity measurement for \HD\ is consistent with the value derived by \cite{hirsch09} using the same  FEROS spectrum ($-17 \pm 3$\,km\,s$^{-1}$).
\begin{figure*}
    \includegraphics[width=1\textwidth]{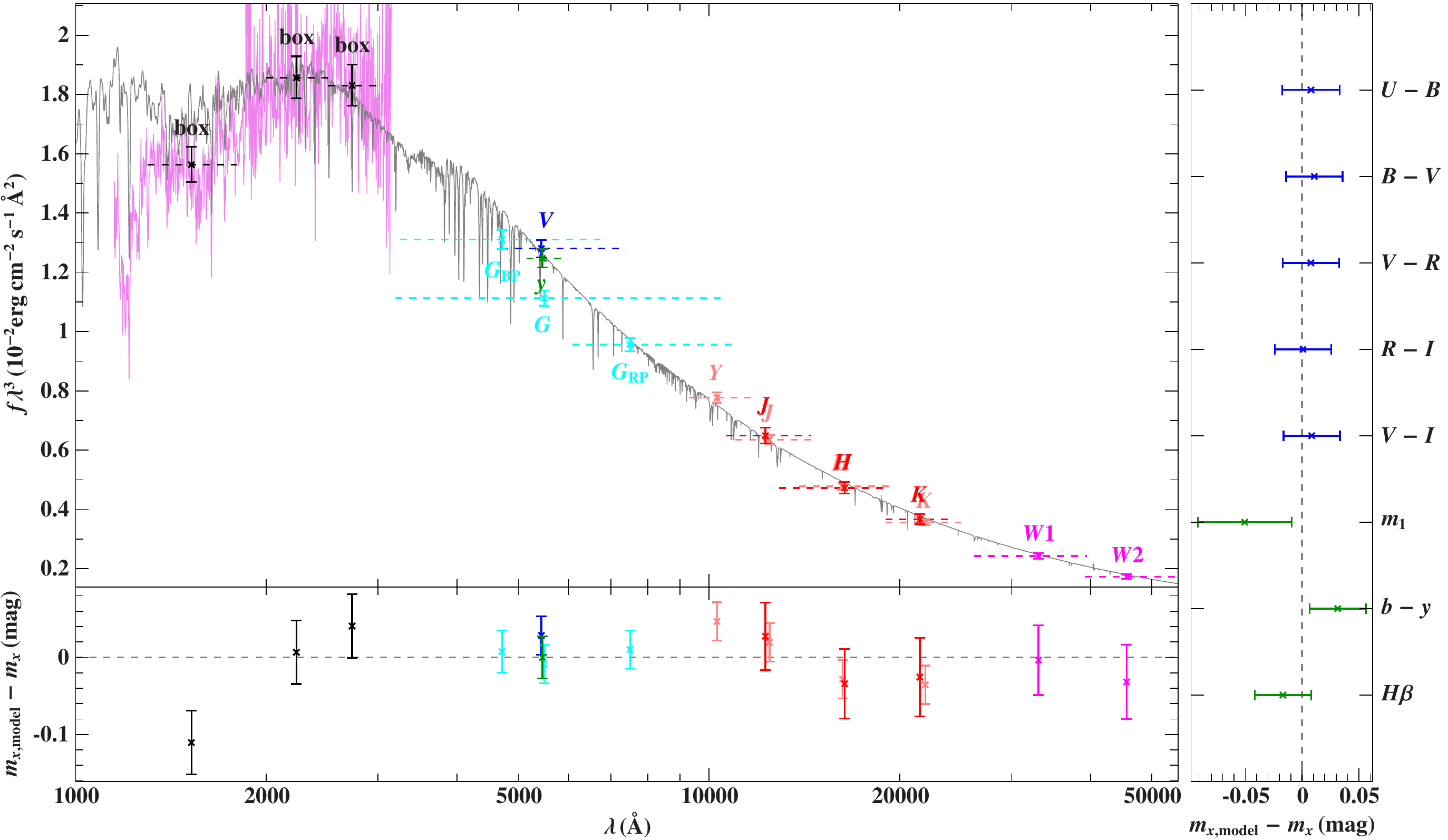}
    \vspace{2pt}
\hfill
    \label{fig:HZ44:SED}
    \includegraphics[width=1\textwidth]{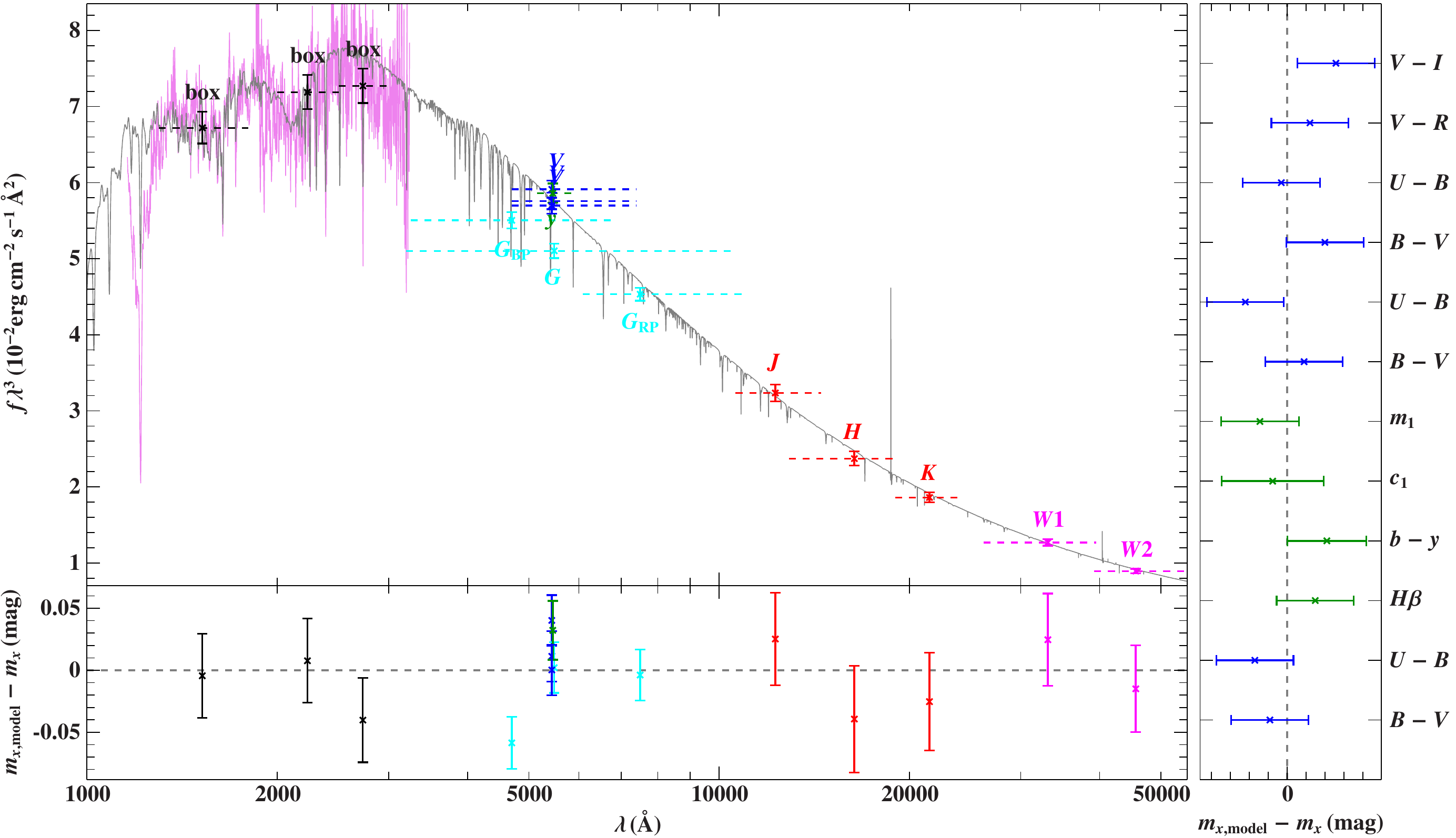}
    \captionof{figure}{
Comparison of synthetic spectra with photometric data for \HZ\ (top) and \HD\ (bottom).
The three black data points labeled ``box'' are binned fluxes from a low-dispersion IUE spectrum. 
Filter-averaged fluxes are shown as colored data points that were converted from observed magnitudes (the dashed horizontal lines indicate the respective filter widths), while the gray solid line represents a synthetic spectrum  using the atmospheric parameters given in Table~\ref{tab:atm}. 
The residual panels at the bottom and right hand side show the differences between synthetic and observed magnitudes/colors. 
The following color codes are used to identify the photometric systems: Johnson-Cousins (blue), Str\"omgren (green), \textit{Gaia} (cyan), UKIDSS (rose), 2MASS (red), WISE (magenta).
}
\label{fig:HD127493:SED}
\end{figure*}
%
\section{Stellar masses, radii, and luminosities}
\label{angular}
With the release of \textit{Gaia}\,DR2, high accuracy parallax ($\varpi$) and therefore distance measurements have become available for a large sample of hot subdwarfs. 
This allows us to derive more precise spectroscopic masses for these stars.
We collected photometric measurements from several surveys and converted them into fluxes (see Tables \ref{tab:data:HZ:phot} and \ref{tab:data:HD:phot}).
In addition, we use low-resolution, large-aperture IUE spectra that were averaged in three regions (1300--1800\,\AA, 2000--2500\,\AA, 2500--3000\,\AA) as ``box filters'' to cover the UV range.
Our photometric fitting procedure is described in detail in \cite{Heber2018}.
The $\chi ^2$ fitting procedure scales our final synthetic spectra to match the photometric data and has the solid angle $\theta = 2R/D$ and the color excess $E_{B-V}$ as free parameters.
Reddening is modeled with $R_V =3.1$ as the extinction parameter (a standard value for the diffuse ISM) and the corresponding mean extinction law from \cite{Fitzpatrick1999}.
The resulting solid angle can be combined with the \textit{Gaia} parallax distance to obtain the stellar radius, from which the stellar mass can be computed using the surface gravity derived from spectroscopy. 
The SED-fits are shown in Fig.~\ref{fig:HD127493:SED} and the derived parameters in Table \ref{tab:SED}.
Considering the non-detection of radial velocity variations and the evident lack of an IR excess, we can state that there is no indication of binarity in \HZ. 
The SED of \HD\ also shows no IR excess that would hint at a companion.
The masses determined from the SED-fits are consistent with the canonical subdwarf mass, 0.47\,$M_\odot$ \citep[and references therein]{Fontaine2012}.
\begin{table}
\setstretch{1.2}
\caption{Parallax and parameters derived from the SED fitting.}
\label{tab:SED}
\vspace{-8pt}
\begin{center}
\begin{tabular}{lcc}
\toprule
\toprule
Results& \HZ & \HD \\ 
\midrule
$\varpi$\,(mas) 				    &  $2.48  \pm 0.08$ & $5.82  \pm 0.09$ \\
$d$\,(pc) 							& $403 \pm 13$		&	$172 \pm 3$	\\
$\theta\,(10^{-11}\,\mathrm{rad})$	& $2.134 \pm 0.011$	&	$4.53 \pm 0.03$	\\
$E_{B-V}$					& $0.010 \pm 0.004$	&	$0.052 \pm 0.004$	\\
$R/R_\odot$ 						& $0.191 \pm 0.007$	&	$0.172 \pm 0.003$	\\
$M/M_\odot$ 						& $0.59 \pm 0.14$	&	$0.43 \pm 0.05$	\\
$L/L_\odot$ 						& $77 \pm 4$	&	$84 \pm 2$	\\
\bottomrule
\end{tabular}
\end{center}
\end{table}
%
\section{Atomic data}
\label{sect:atomic_data}
\begin{table*}
\begin{minipage}{1\textwidth}
\caption{Data for elements not included in the Kurucz line-list (\texttt{gfall08oct17.dat}). 
}
\label{tab:data:osc}
\begin{threeparttable}
\begin{center}
\begin{minipage}{0.5\textwidth}
\begin{tabular}[t]{lrrr}
\toprule
\toprule
 Ion	& $N_\mathrm{UV}$ & $N_\mathrm{VIS}$  & Reference	\\ 
\midrule
Ga\,\textsc{iii}		& 3	& 2	& 15, 2(28)				\\
Ga\,\textsc{iv}			& 69	& $-$ & 3				\\
Ga\,\textsc{v}			& 37	& $-$ & 3				\\
Ge\,\textsc{iii}		& 1		& $-$ & 1(25,18)		\\
Ge\,\textsc{iv}			& 7 	& 6   & 23, 2(28)		\\
Ge\,\textsc{v}			& 24	& $-$ & 3				\\
As\,\textsc{iii}		& $-$	& $-$ & 1(26,19,17)		\\
As\,\textsc{iv}			& $-$	& 6   & 2(27), 1(25)	\\
As\,\textsc{v}			& 2  	& $-$ & 1(19)			\\
Se\,\textsc{iv}$^\ast$	& 3	& $-$ & 1(22)				\\
Se\,\textsc{v}$^\ast$	& 4	& $-$ & 3					\\
Kr\,\textsc{iv}$^\ast$	& 42	& 6 & 3				\\
Kr\,\textsc{v}$^\ast$	& $-$	& $-$ & 3				\\
Sr\,\textsc{iv}$^\ast$	& 109	& 1 & 3				\\
\bottomrule
\end{tabular}
\end{minipage}
\begin{minipage}{0.5\textwidth}
\begin{tabular}[t]{lrrr}
\toprule
\toprule
 Ion	& $N_\mathrm{UV}$ & $N_\mathrm{VIS}$  & Reference	\\ 
\midrule
Sr\,\textsc{v}$^\ast$	& 23	& $-$ & 3		\\
Y\,\textsc{iii}$^\ast$	& 1	& 2   & 10, 16		\\
Zr\,\textsc{iv}			& 11	& 8   & 3		\\
Zr\,\textsc{v}			& $-$	& $-$ & 3		\\
Mo\,\textsc{iv}$^\ast$	& 92	& $-$ & 29,3	\\
Mo\,\textsc{v}$^\ast$	& 69	& $-$ & 3		\\
Mo\,\textsc{vi}$^\ast$	& 5		& $-$ & 3		\\
In\,\textsc{iii}$^\ast$	& 5	& $-$   & 4			\\
Sn\,\textsc{iii}		& $-$	& $-$ & 13		\\    
Sn\,\textsc{iv}			& 7	& $-$ & 4, 12		\\
Sb\,\textsc{iii}$^\ast$	& $-$	& $-$ & 1(17)	\\
Sb\,\textsc{iv}$^\ast$	& 1	& $-$ & 1(20)		\\
Sb\,\textsc{v}$^\ast$	& 2	& $-$ & 4			\\
Te\,\textsc{iii}$^\ast$	& $-$	& $-$ & 14		\\
\bottomrule
\end{tabular} 
\end{minipage}
\begin{minipage}{0.5\textwidth}
\begin{tabular}[t]{lrrr}
\toprule
\toprule
 Ion	& $N_\mathrm{UV}$ & $N_\mathrm{VIS}$  & Reference	\\ 
\midrule
Te\,\textsc{v}$^\ast$	& 1	& $-$   & 1(30)		\\
Te\,\textsc{vi}$^\ast$	& 2	& $-$   & 3			\\
Xe\,\textsc{iv$^\ast$}	& 5	& $-$ & 3		\\
Xe\,\textsc{v}$^\ast$	& 4	& $-$ & 3		\\
Ba\,\textsc{v}$^\ast$	& 2	& $-$ & 3		\\
Tl\,\textsc{iii}$^\ast$	& $-$	& $-$ & 5		\\
Pb\,\textsc{iii}		& 2	& $-$ & 7, 1(24)	\\
Pb\,\textsc{iv}			& 17	& 9   & 11, 5, 8, 1(24)	\\
Pb\,\textsc{v}			& 36	& $-$ & 9		\\
Bi\,\textsc{iii}$^\ast$	& $-$	& $-$ & 1(23)		\\
Bi\,\textsc{iv}$^\ast$	& $-$	& $-$ & 1(21)		\\
Bi\,\textsc{v}$^\ast$	& 1	& $-$ & 5		\\
Th\,\textsc{iv}$^\ast$	& 2	& $-$ & 6		\\
		&   		&    &  		\\
\bottomrule
\end{tabular} 
\end{minipage}
\end{center}
\tablefoot{
\small
The number of lines with predicted equivalent width greater than 5\,m\AA\ in the final model of \HZ\ (upper limits marked with $^\ast$) and in spectral ranges where observations are available for \HZ\ are listed (UV: 916$\,-\,$1980\,\AA, VIS: 3022$\,-\,$7580\,\AA). 
For references from compilations, the compilation is listed first and the individual references in parenthesis.
References:~(1) \cite{morton00}, (2) ALL, \cite{all}, 
(3) TOSS, \cite{vo:TOSS_legacy}, (4) \cite{safronova03}, 
(5) \cite{safronova04}, (6) \cite{safronova13}, 
(7) \cite{alonso-medina09}, (8) \cite{alonso-medina11}, 
(9) \cite{colon14}, (10) \cite{naslim11}, 
(11) \cite{naslim13}, (12) \cite{biswas18}, 
(13) \cite{haris12}, (14) \cite{zhang13}, 
(15) \cite{nielsen05}, (16) \cite{Redfors1991}, 
(17) \cite{Andersen1977}, (18) \cite{Andersen1979}, 
(19) \cite{Pinnington1981}, (20) \cite{Pinnington1985b}, 
(21) \cite{Pinnington1988}, (22) \cite{Bahr1982}, 
(23) \cite{Migdalek1983}, (24) \cite{Ansbacher1988}, 
(25) \cite{Curtis1992}, (26) \cite{Marcinek1993},  
(27) \cite{churilov96}, (28) \cite{oreilly98}, 
(29) \cite{kurucz18}, (30) \cite{Pinnington1985a}
}
\end{threeparttable}
\end{minipage}
\end{table*}
While atomic data and line lists for elements lighter than the iron-group are readily accessible via, for example, the Kurucz compilations and the NIST\footnotemark\ database, data for trans-iron elements are much more scarce. 
\footnotetext{National Institute of Standards and Technology, 
\url{https://physics.nist.gov/PhysRefData/ASD/lines_form.html}}
Since these elements are of special interest for the analysis of our two stars we invested particular effort into searching the literature and collecting data (energy levels, line positions, and oscillator strengths) for many trans-iron elements. 
We list in Table \ref{tab:data:osc} the ions that we took into consideration as well as the references for their atomic data. 
We also include in this table, for each ion, the number of lines visible (with a predicted equivalent width greater than 5\,m\AA) in the final model spectrum of HZ 44. 
The basis of our line list is the most recent line list published by \cite{kurucz18} and available online\footnotemark
\footnotetext{Kurucz/Linelists, \url{http://kurucz.harvard.edu/linelists/gfnew/gfall08oct17.dat}}.
The list was further extended with data listed in ALL, the Atomic Line List (v2.05b21)\footnotemark
\footnotetext{Atomic Line List (v2.05b21), \url{http://www.pa.uky.edu/~peter/newpage/}}.
In the context of their ongoing ``Stellar Laboratories'' series, \cite{vo:TOSS_legacy} have published a large collection of atomic data for elements with $\mathrm{Z}\geq 30$ on the TOSS\footnotemark  website.\footnotetext{T\"ubingen Oscillator Strengths Service, \url{http://dc.g-vo.org/TOSS}} 
While this collection was made for the analysis of hot white dwarfs with $T_\mathrm{eff}> 60\,000$\,K, it also includes atomic data for ions of stages \textsc{iv-v} that are observed in the sdOs discussed here.
Thus, additional lines were added from TOSS and other theoretical works listed in Table \ref{tab:data:osc}.
Finally the list was merged with the collection of lines from low-lying energy levels by \cite{morton00} but preferring more recent data if available.
Hyper-fine structure and isotopic line splitting are not considered because of the lack of atomic data. For subordinate lines the effect is expected to be small, but may be significant for resonance lines \citep[e.\,g.][]{Mashonkina2003} such as the Pb\,\textsc{iv}\,1313\,\AA. 
Fortunately, for the latter resonance line atomic data are available for several isotopes and we included them in the line formation calculations \citep[see][]{O'Toole2006}.
%
\section{Metal abundance analysis}
\label{sect:abundances}
\begin{table}
\begin{minipage}{1\columnwidth}
\setstretch{1.1}
\caption{Model atoms used for the final model of HZ\,44 with their number of explicit levels (L) and superlevels (SL).}
\vspace{-1pt}
\label{tab:modelatoms}
\centering
\begin{threeparttable}
\begin{center}
\begin{minipage}{0.49\columnwidth}
\begin{tabular}{lcc}
\toprule
\toprule
 Ion	& L & SL	\\ 
 \midrule
H\,\textsc{i}    & 17  & $-$\\
He\,\textsc{i}   & 24 & $-$\\
He\,\textsc{ii}  & 20 & $-$\\
C\,\textsc{ii}   & 34  & \phantom{1}5\\
C\,\textsc{iii}  & 34  & 12\\
C\,\textsc{iv}   & 35  & \phantom{1}2\\
N\,\textsc{ii}   & 32  & 10\\
N\,\textsc{iii}  & 40  & \phantom{1}9\\
N\,\textsc{iv}   & 34  & 14\\
N\,\textsc{v}    & 21  & \phantom{1}4\\
O\,\textsc{ii}   & 36  & 12\\
O\,\textsc{iii}  & 28  & 13\\
O\,\textsc{iv}   & 31  & \phantom{1}8\\
O\,\textsc{v}    & 34  & \phantom{1}6\\
O\,\textsc{vi}   & 15  & \phantom{1}5\\
Ne\,\textsc{ii}  & 23  & \phantom{1}9\\
Ne\,\textsc{iii} & 22  & 12\\
Ne\,\textsc{iv}  & 10  & \phantom{1}2\\
Mg\,\textsc{ii}  & 21  & \phantom{1}4\\
Mg\,\textsc{iii} & 37  & \phantom{1}3\\
Mg\,\textsc{iv}  & 29  & \phantom{1}5\\
Mg\,\textsc{v}   & 18  & \phantom{1}2\\
Al\,\textsc{ii}  & 20  & \phantom{1}9\\
Al\,\textsc{iii} & 19  & \phantom{1}4\vspace{5pt}\\
\bottomrule
\end{tabular}
\end{minipage}
\begin{minipage}{0.49\columnwidth}
\begin{tabular}{lcc}
\toprule
\toprule
 Ion	& L & SL	\\ 
 \midrule
Si\,\textsc{ii}  & 36  & \phantom{1}4\\
Si\,\textsc{iii} & 31  & 15\\
Si\,\textsc{iv}  & 19  & \phantom{1}4\\
P\,\textsc{iv}   & 14  & $-$\\
P\,\textsc{v}    & 13  & \phantom{1}4\\
S\,\textsc{ii}   & 23  & 10\\
S\,\textsc{iii}  & 29  & 12\\
S\,\textsc{iv}   & 33  & \phantom{1}5\\
S\,\textsc{v}    & 20  & \phantom{1}5\\
S\,\textsc{vi}   & 13  & \phantom{1}3\\
Ar\,\textsc{ii}  & 42  & 12\\
Ar\,\textsc{iii} & 27  & 17\\
Ar\,\textsc{iv}  & 39  & $-$\\
Ar\,\textsc{v}   & 25  & $-$\\
Ca\,\textsc{ii}  & 32  & $-$\\
Ca\,\textsc{iii} & 15  & \phantom{1}4\\
Ca\,\textsc{iv}  & 17  & \phantom{1}4\\
Fe\,\textsc{iii} & $-$ & 50  \\
Fe\,\textsc{iv}  & $-$ & 43  \\
Fe\,\textsc{v}   & $-$ & 42  \\
Ni\,\textsc{iii} & $-$ & 36  \\
Ni\,\textsc{iv}  & $-$ & 38  \\
Ni\,\textsc{v}   & $-$ & 48  \\
\midrule
Total & 1062 & 506\\
\bottomrule
\end{tabular}
\end{minipage}
\end{center}
\end{threeparttable}
\end{minipage}
\end{table}
Model atmospheres were calculated for each star using their atmospheric parameters as listed in Table \ref{tab:atm}.
All ions for which model atoms are available are included in non-LTE (see Table \ref{tab:modelatoms}), while the remaining elements are treated with the LTE approximation.
The next higher ionization stage of each metal listed in Table \ref{tab:modelatoms} is considered as a one-level ion.
More information on the model atoms we use can be found on the \texttt{TLUSTY} web site\footnotemark\ and in \citet{Lanz2003, Lanz2007}. The Mg\,\textsc{iii-v} and Ar\,\textsc{iv-v} model atoms are described in \cite{Latour2013}. The Ca\,\textsc{iii-iv} model atoms were constructed in a similar manner (P. Chayer, priv. comm.) while the Ca\,\textsc{ii} model atom is described in \citet{Allende2003}.
\footnotetext{\url{http://tlusty.oca.eu/Tlusty2002/tlusty-frames-data.html}}
To compute the partition functions of heavy elements (Z$>$30) in ionization stages \textsc{iv--vi} we added atomic data from NIST to \texttt{SYNSPEC}, as in \cite{chayer06}.
As a starting point, abundances in the \texttt{TLUSTY} model were set to values estimated by eye for each element.
Based on this preliminary model, a series of synthetic spectra with a range of abundances for each element were created with \texttt{SYNSPEC}.
The abundance of the elements were determined one-by-one using the downhill-simplex fitting program \texttt{SPAS} developed by \cite{hirsch09}.
This method works well for isolated lines but is not reliable for heavily blended lines, in particular in the UV region.
The abundance for these elements was estimated by manually comparing models with the observation.
Even with this method, the placement of the continuum (especially in the FUSE range) remains an important source of uncertainty. 
As noted by \cite{Pereira2006}, the true continuum in the FUSE spectral region may be well above the highest observed fluxes. 
This complicates the continuum placement since some opacity (photospheric and interstellar) is still missing in our final synthetic spectra.
Thus for some elements we could only derive upper limits. 
This includes elements having low abundances but also elements that show lines in the FUSE range only, where the aforementioned problems are most severe.
For some elements in \HD\ no abundances, or upper limits, could be derived (Cl, K, As, Se, Sb, Xe, Bi). 
This is due to insufficient spectral coverage: 
the elements in question have their strongest spectral lines in ranges where no data are available for that star (FUSE, UVA).
\\
A summary of the photospheric abundances derived for \HZ\ and \HD\ are presented at the end of this section in Fig.~\ref{fig:abupattern:nf} and in Table \ref{tab:abundances_units}.
In addition, we include in Sect.~\ref{sect:fullspectra} a comparison between the final synthetic spectrum and the observed spectrum in all wavelength ranges for both stars.
We note that for some elements, namely Ne, Ar, Cl, Sn, Tl, Pb, and Th, the uncertainty on their solar photospheric abundance \citep{asplund09} contributes significantly to the total uncertainty when computing the ratio with solar abundances.
The uncertainty stated on upper limits and by-eye abundances is defined as follows: at the upper bound the lines are judged to be clearly too strong, while they can not be distinguished from noise at the lower bound.\\
In the following subsections, we present in detail the result of our abundance analysis for each element. Light elements (C, N, O) are discussed in Sect.~\ref{sect:abu:cno}, intermediate elements (F, Ne, Na, Mg, Al, Si, P, S, Cl, Ar, K, Ca, Ti) in Sect.~\ref{sect:abu:intermediate}, iron-group elements treated in non-LTE (Fe, Ni) in Sect.~\ref{sect:abu:FeNi}, iron-group elements treated in LTE (V, Cr, Mn, Co, Cu, Zn) in  Sect.~\ref{sect:abu:irongroup}, detected trans-iron elements (Ga, Ge, As, Se, Zr, Sn, Pb) in Sect.~\ref{sect:abu:heavy:det}, and trans-iron elements with upper limits (Kr, Sr, Y, Mo, Sb, Te, Xe, Th) in Sect.~\ref{sect:abu:heavy:ul}. Finally, Sect.~\ref{sect:abu:hopeless} addresses the elements for which we could not even assess an upper limit due to the weakness of their predicted lines (Sc, In, Ba, Tl, Bi).
We then discuss the chemical portrait obtained for both stars in Sect.~\ref{sect:abu:summary}. 
In the rest of the paper we give our abundances as $\log n_\mathrm{X}/n_{\mathrm{H}}$ and use the shorter notation $\log \mathrm{X/H}$. 
Here, $n_X$ is the dimensionless number fraction.  
To put the abundances in perspective, we additionally state the corresponding number fraction relative to solar values $n_{\mathrm{X}}/n_{\mathrm{X},\odot}$. 
\subsection{Light metals (C, N, O)}
\label{sect:abu:cno}
The \textit{carbon} abundance in \HZ\ was measured using nine optical C\,\textsc{iii} lines. 
The abundance derived this way,  $\log \mathrm{C/H}=-4.31 \pm 0.13$ ($8.6\times 10^{-2}$ times solar), is consistent with the strong C\,\textsc{iii} and C\,\textsc{iv} lines observed in the UV region.
Carbon lines are weaker in the optical spectrum of \HD. 
We use the resonance doublet C\,\textsc{iv}\,$\lambda\lambda$\,1548, 1551\,\AA\ and the C\,\textsc{iii} sextuplet lines at 1175\,\AA\ to derive an abundance of $\log \mathrm{C/H}=-4.30 \pm 0.08$ ($3.9 \times 10^{-2}$ times solar).
\smallskip\\
The \textit{nitrogen} abundances measured in \HZ\ from different ionization stages/lines in the optical region are not very consistent.
Most N\,\textsc{iii} lines are well reproduced; some are too strong (e.\,g.~N\,\textsc{iii}\,$\lambda\lambda$\,4378.99, 4379.20\,\AA) while few are too weak (e.\,g.~N\,\textsc{iii}\,4003.58\,\AA). 
We identified only one strong N\,\textsc{iv} line in the optical spectrum of \HZ\ (N\,\textsc{iv}\,4057.76\,\AA) which fits the final model well.
We measure an abundance of  $\log \mathrm{N/H}=-2.39 \pm 0.21$ ($29$ times solar) 
for \HZ.
However, many strong N\,\textsc{ii} lines are too broad and shallow in the model, even assuming a rotation velocity and microturbulence of 0\,km\,s$^{-1}$ and were therefore excluded from the fit 
(e.\,g.~N\,\textsc{ii}\,$\lambda\lambda\,$3995.00, 
4041.31, 
4236.91\,\AA).
This may be related to numerical issues due to low population numbers.
In \HD\ the issues with some optical N\,\textsc{ii} lines are even more pronounced; they appear in emission in the model and were excluded from the fit. 
In addition to optical N\,\textsc{iii}-\textsc{iv} lines we used UV lines to constrain the abundance, including the N\,\textsc{v}\,$\lambda \lambda\,1238.8,1242.8$\,\AA\ resonance lines, the strong N\,\textsc{iv}\,1718\,\AA\ line, and several N\,\textsc{iii} lines. 
All ionization stages give a consistent abundance of $\log  \mathrm{N/H}=-2.09 \pm 0.17$ (26 times solar).
\smallskip\\
The abundance of \textit{oxygen} in \HZ\ was measured using optical O\,\textsc{ii} and O\,\textsc{iii} lines. 
Although these lines are weak compared to lines from other elements, they could be used to find an abundance of $\log \mathrm{O/H}=-3.77 \pm 0.13$ ($12 \times 10^{-2}$ times solar). 
The strongest observed O lines in the UV region (O\,\textsc{iii}\,1150.884\,\AA\ in FUSE and O\,\textsc{iv}\,1343.51\,\AA\ in the IUE/SWP spectrum) support this value.
O\,\textsc{iv}\,$\lambda\lambda$\,1338.62, 1343.51\,\AA\ are observed in the GHRS spectrum of \HD\ but blended with Ni lines. Since no optical lines were detected in that star we only set an upper limit of $\log \mathrm{O/H}\leq-4.3$ ($2\times 10^{-2}$ times solar) 
and note that the actual abundance is likely not significantly lower.
\subsection{Intermediate metals (F to Ti)}
\label{sect:abu:intermediate}
\begin{figure}[!ht]
\centering
    \includegraphics[width=0.95\columnwidth]{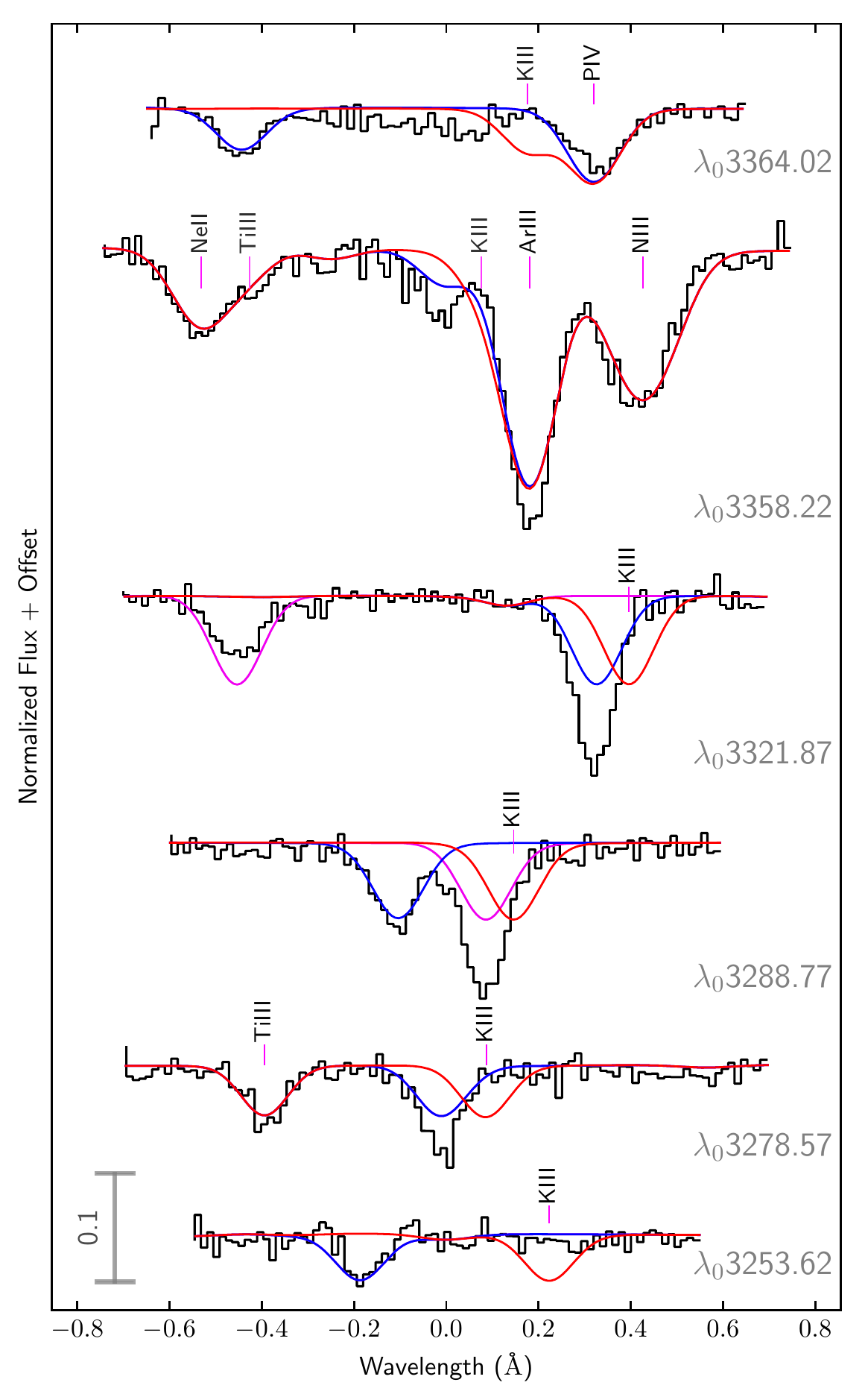}
    \vspace*{-5pt}
    \caption{The strongest K\,\textsc{iii} lines in the observed spectrum of \HZ. The model using the Kurucz wavelengths is shown in red, while a model with shifted lines is shown in blue. Alternative line shifts are shown in violet (marked with $^\ast$ in Table \ref{tab:lines:K}). All models have a potassium abundance of $\log \mathrm{K/H}=-5$.}
\label{fig:HZ:KIII}
\end{figure}

For \HZ, all elements from fluorine to titanium were analyzed. 
Due to the lack of UVA and FUSE spectra, P, Cl, Ar, K, and Ti could not be studied in \HD.
\smallskip\\
No \textit{fluorine} lines are observed in \HZ\ which allows us to provide an upper limit of $\log \mathrm{F/H}\leq-5.0\pm 0.4$ ($130$ times solar) 
based on F\,\textsc{ii}\,$\lambda\lambda$\,3503.11, 3505.63, 3847.09, 3849.99, and 3851.67\,\AA.
 None of these F\,\textsc{ii} lines are strong enough in \HD\ set a meaningful limit on the abundance.
\smallskip\\
The \textit{neon} abundance measurement for \HZ\ is based on several strong Ne\,\textsc{ii} lines, which are weaker in \HD\ due to its higher temperature. 
Most of them lie between 3300\,\AA\ and 3800\,\AA\ though some strong Ne\,\textsc{ii} lines exist at longer wavelengths. 
Fitting all accessible Ne lines results in  $\log \mathrm{Ne/H}=-3.23\pm 0.11$ ($3.3$ times solar) 
for \HZ\ and $\log \mathrm{Ne/H}=-2.38\pm 0.18$ ($3.1$ times solar) 
for \HD.
The weaker Ne\,\textsc{iii} lines are reasonably well reproduced with the abundance stated above.
\smallskip\\
\textit{Sodium} lines are weak, but clearly visible in \HZ, most notably Na\,\textsc{ii}\,$\lambda\lambda$ 3285.61, 3533.06, 3631.27, 4392.81\,\AA. 
We find $\log \mathrm{Na/H}=-4.43 \pm 0.06$ (10 times solar) 
for \HZ.
Na\,\textsc{ii}\,$\lambda\lambda$ 4392.81, 4405.12\,\AA\ allow an upper limit of $\log \mathrm{Na/H}\leq-3.4 \pm 0.3$ (48 times solar) 
to be set for \HD.
\smallskip\\
The strongest observed \textit{magnesium} lines are by far the Mg\,\textsc{ii} triplet at 4481\,\AA. 
All other optical lines are too weak to be observed in either star.
We derive $\log \mathrm{Mg/H}=-3.8 \pm 0.2$ (1.9 times solar) 
for \HZ\ and 
$\log \mathrm{Mg/H}=-3.4 \pm 0.2$ (2.1 times solar) 
for \HD, consistent with Mg\,\textsc{ii}\,2798.82\,\AA\ in the IUE LWR spectrum.
\smallskip\\
We measure the \textit{aluminum} abundance in \HZ\ to be $\log \mathrm{Al/H}=-4.86\pm 0.11$ (2.4 times solar) 
based on eleven optical Al\,\textsc{iii} lines. 
This abundance is consistent with strong  Al\,\textsc{iii} lines observed in the UV range, including the 
Al\,\textsc{iii}\,$\lambda\lambda\,$1854.72, 
1862.79\,\AA\ resonance lines.
The Al abundance in \HD, $\log \mathrm{Al/H}=-4.53\pm 0.10$ (2.2 times solar), 
is derived from Al\,\textsc{iii}\,$\lambda\lambda$\,4479.97, 4512.57, 4529.19, 5696.60\,\AA.
\smallskip\\
Both stars show strong \textit{silicon} lines in their optical and UV spectra. 
The abundance measurement for \HZ\ is based on ten optical Si\,\textsc{iii} and nine optical Si\,\textsc{iv} lines. 
The derived abundance of $\log \mathrm{Si/H}=-3.88\pm 0.11$ (2.0 times solar) 
is consistent with the resonance lines Si\,\textsc{iv}\,$\lambda\lambda\,$1394, 1403\,\AA, the very strong 
Si\,\textsc{iv}\,$\lambda\lambda\,$1066.6, 
1122.5, 
1128.3\,\AA, and 
Si\,\textsc{iii}\,1113.2\,\AA\ in the FUSE spectrum, 
as well as many more silicon lines in the UV region.
For \HD\ we used three lines in the UV (including Si\,\textsc{iv}\,$\lambda\lambda\,$1394, 1403\,\AA) for our fit in addition to four lines from the optical range.  
Both UV and optical lines give a consistent abundance of $\log \mathrm{Si/H}=-3.31\pm 0.14$ (3.2 times solar). 
\smallskip\\
The strongest observed \textit{phosphorus} lines lie in the FUSE spectral range which is only accessible for \HZ. 
This includes the 
P\,\textsc{v}\,$\lambda\lambda\,$1117.98, 
1128.01\,\AA, and  
P\,\textsc{iv}\,950.66\,\AA\ resonance lines as well as several strong P\,\textsc{iv} lines, e.\,g.~P\,\textsc{iv}\,1030.52\,\AA. 
In the optical/UVA range, there are only three observable lines: 
P\,\textsc{iv}\,$\lambda\lambda\,$3347.74, 
3364.47\,\AA\ (see Fig.~\ref{fig:HZ:KIII}) and 
P\,\textsc{iv}\,4249.66\,\AA. 
We derive an abundance of $\log \mathrm{P/H}=-5.66\pm 0.25$ (4.1 times solar) for \HZ. 
There are only two unambiguously identified P lines in \HD: P\,\textsc{iv}\,1888.52\,\AA\ and P\,\textsc{iv}\,4249.66\,\AA. They can be used to derive an abundance of $\log \mathrm{P/H}=-4.7\pm 0.4$ (16 times solar).
\smallskip\\
Besides many strong \textit{sulfur}\,\textsc{iii}-\textsc{iv} lines at optical wavelengths, we observed strong S\,\textsc{iii}-\textsc{vi} lines in the UV spectrum of \HZ\ 
(e.\,g.~S\,\textsc{iv}\,$\lambda\lambda\,$1062.66, 
1072.97, 
1623.94\,\AA). 
However, some of these lines are listed with inaccurate atomic data in the newest Kurucz line list, in  which case we preferred older values. 
The abundance measurement of
$\log \mathrm{S/H}=-3.87\pm 0.37$ (4.9 times solar) 
based on optical S\,\textsc{iii}-\textsc{iv} lines is consistent with the UV lines.
Several optical S\,\textsc{iii} lines are too weak and broad in the model 
(e.\,g.~S\,\textsc{iii}\,$\lambda\lambda\,$3497.28, 
3662.01, 
3717.77, 
3928.61, 
4361.47\,\AA) and were excluded from the fit. 
Sulfur lines are slightly weaker in \HD.
We derive an abundance of $\log \mathrm{S/H}=-3.90\pm 0.35$ (2.0 times solar) 
from UV lines, consistent with optical lines such as S\,\textsc{iv}\,$\lambda\lambda$\,4485.64, 4504.11, 5497.78\,\AA.
\smallskip\\
\textit{Chlorine} shows strong lines from low-lying levels in the FUSE spectral region (Cl\,\textsc{iv}\,$\lambda\lambda$\,973.22, 977.57, 977.90, 984.96, 985.76\,\AA). 
Although these lines are strong in \HZ, it is hard to determine abundances from them as they are rather insensitive to changes in abundance due to saturation effects.
In addition, the usual problems with lines in the FUSE spectra apply: they suffer from unidentified blends, both of stellar and interstellar origin. 
Nevertheless, the abundance of $\log \mathrm{Cl/H}=-7.0 \pm 0.7$ (0.05 times solar), 
derived from these lines is remarkably low despite the large uncertainty. 
\smallskip\\
\textit{Argon} shows many strong lines in the UVA spectrum of \HZ. 
We determine an abundance of $\log \mathrm{Ar/H}=-3.79 \pm 0.11$ (31 times solar) 
for \HZ\ based on optical/UVA lines.
Some strong Ar\,\textsc{iii} lines 
(e.\,g.~Ar\,\textsc{iii}\,$\lambda\lambda\,$3286.11, 
3302.19, 
3311.56\,\AA) were excluded from the fit since they show the same discrepancies already observed in N\,\textsc{ii} lines -- they are very narrow in the observation and too broad in the model.
Except for a very weak Ar\,\textsc{iii}\,4182.97\,\AA\ line, we could identify no optical Ar lines in \HD.
The upper limit for \HD\ derived from this line is still super-solar at $\log \mathrm{Ar/H}\leq-3.6 \pm 0.2$ (21 times solar).
The blended Ar\,\textsc{iv}\,$\lambda\lambda$\,1409.30, 1435.56\,\AA\ and Ar\,\textsc{iv}\,2641.09\,\AA\ are well reproduced at this abundance but Ar\,\textsc{v}\,1371.87\,\AA\ suggests a lower abundance.
\smallskip\\
We found no strong \textit{potassium} lines in the UV spectrum of \HZ, but some optical lines were clearly identified. 
However several lines appear to lie at shorter wavelengths than listed in the Kurucz line list.
Since the difference correlates with their LS-coupling terms, it seemed reasonable to shift them in order to match their observed position. 
Their wavelengths and configurations are listed in Table \ref{tab:lines:K}. 
Other K lines are clearly identified at wavelengths very close to their listed value (K\,\textsc{iii}\,$\lambda\lambda$\,3052.016, 3468.314, 3513.822\,\AA).
As shown in Fig. \ref{fig:HZ:KIII}, lines with a $^4$P lower term had to be shifted by approximately $-0.1$\,\AA\ whereas the shift was larger for all lines with $^2$P lower terms. 
All identified lines are reproduced reasonably well with an abundance of $\log \mathrm{K/H}=-4.92\pm0.16$ (55 times solar), when shifted to the observed position.
\begin{table}[!htbp]
\begin{center}
\caption{K\,\textsc{iii} lines with deviation between predicted and observed wavelengths.}
\label{tab:lines:K}
\begin{tabular}{clcc}
\toprule
\toprule
 $\lambda _\mathrm{Kurucz}$ (\AA) & $\lambda _\mathrm{obs}$ (\AA) & $\Delta \lambda$ (\AA)  & Configuration	\\ 
 \midrule
   3253.973 & 3253.563 & $-0.41$ & 4s\,$^2$P$_{3/2}$\,$-$\,4p\,$^2$D$^\circ_{3/2}$\\
   3278.787 & 3278.687 & $-0.10$ & 4s\,$^4$P$_{5/2}$\,$-$\,4p\,$^4$P$^\circ_{3/2}$\\
   3289.046 & 3288.796 & $-0.25$ & 4s\,$^2$P$_{3/2}$\,$-$\,4p\,$^2$D$^\circ_{5/2}$\\
            & 3288.986$^\ast$ & $-0.06$ &                                         \\
   3322.396 & 3322.326 & $-0.07$ & 4s\,$^4$P$_{5/2}$\,$-$\,4p\,$^4$P$^\circ_{5/2}$\\
            & 3321.546$^\ast$ & $-0.85$ &                                         \\
   3358.426 & 3358.346 & $-0.08$ & 4s\,$^4$P$_{3/2}$\,$-$\,4p\,$^4$P$^\circ_{1/2}$\\
   3364.326 & 3363.706 & $-0.62$ & 4s\,$^2$P$_{1/2}$\,$-$\,4p\,$^2$P$^\circ_{3/2}$\\
   3468.314 & 3468.260 & $-0.05$ & 4s\,$^4$P$_{3/2}$\,$-$\,4p\,$^4$P$^\circ_{5/2}$\\%
   3513.822 & 3513.782 & $-0.04$ & 4s\,$^4$P$_{1/2}$\,$-$\,4p\,$^4$P$^\circ_{3/2}$\\%
\bottomrule
\end{tabular} 
\end{center}
\vspace{-12pt}
\tablefoot{
The configurations are taken from NIST. The term superscript $^\circ$ indicates odd parity. The superscript $^\ast$ marks alternative identifications.}
\end{table}
\smallskip\\
While there are no usable \textit{calcium} lines in the optical spectrum of \HD, \HZ\ shows some strong Ca\,\textsc{ii} and Ca\,\textsc{iii} lines. 
The optical resonance lines 
Ca\,\textsc{ii}\,$\lambda\lambda\,$3934, 
3968\,\AA\ 
are almost entirely photospheric. 
Some strong Ca\,\textsc{iii} lines were excluded from the fit
(e.\,g.~Ca\,\textsc{iii}\,$\lambda\lambda\,$3372.68,  
3537.78\,\AA) 
since they have sharp cores and are too broad and shallow in the model. 
Non-LTE effects can not be blamed since we included Ca\,\textsc{ii} and Ca\,\textsc{iii} in non-LTE. 
We measure $\log \mathrm{Ca/H}=-3.9\pm 0.24$ (28 times solar) 
for \HZ\ and derive an upper limit of $\log \mathrm{Ca/H}\leq-3.9\pm 0.2$ (12 times solar) 
for \HD.
This upper limit is likely to be close to the actual abundance since including Ca at this abundance improves the fit for blended UV lines such as Ca\,\textsc{iii}\,1545.30\,\AA\ and Ca\,\textsc{iv}\,$\lambda\lambda$\,1647.44, 1648.62, 1655.53\,\AA.
\smallskip\\
The strongest \textit{titanium} \textsc{iii-iv} lines lie in the UVA spectral region although some lines exist at longer wavelengths.
We measure a strong enrichment in \HZ\ with $\log \mathrm{Ti/H}=-4.56 \pm 0.22$ (150 times solar). %
This abundance is consistent between strong optical and ultraviolet lines (e.\,g.~Ti\,\textsc{iv}\,$\lambda\lambda$\,1183.63, 1451.74, 1467.34, 1469.19\,\AA\ and Ti\,\textsc{iii}\,1498.70\,\AA).
Although Ti lines at wavelengths above 3800\,\AA\ are strong in \HZ, the same lines are weak in \HD. 
The few lines that can clearly be identified in \HD\ 
(Ti\,\textsc{iv}\,$\lambda\lambda\,$4397.31, 
5885.97\,\AA) 
do not give a consistent abundance. 
Other predicted lines (Ti\,\textsc{iv}\,$\lambda\lambda\,$4397.31, 5398.93, 5492.51\,\AA) are not observed.
Therefore we adopt a conservative upper limit of $\log \mathrm{Ti/H}\leq-4.40 \pm 0.25$ (94 times solar) 
for \HD\ based on optical lines. However,
Ti\,\textsc{iii-iv} lines in the IUE range would favor higher abundances.
\subsection{Fe and Ni (NLTE)}
\label{sect:abu:FeNi}
We determine \textit{iron} abundances by fitting the IUE spectrum of \HZ\ and the GHRS spectrum of \HD\ in ranges that span 10 to 20\,\AA, from 1300\,\AA\ onward (at shorter wavelengths, the amount of unidentified opacity increases).
Since Fe and Ni were fitted separately, blends are not treated exactly which may lead to overestimated abundances.
However, since abundances in the initial model were already close to the best-fit abundances, this effect is partly compensated.
Missing opacity from other sources may also introduce a bias toward higher abundances but since the observed spectrum is well-reproduced in the considered ranges, we are confident that the derived abundances are reliable within their respective uncertainties.
The average of the abundances over all ranges yields  \textcolor{black}{$\log \mathrm{Fe/H}=-4.00\pm 0.25$ (1.5 times solar) 
for \HZ\ } and $\log \mathrm{Fe/H}=-2.82 \pm 0.13$ (10 times solar)  
for \HD. 
\smallskip\\
The same procedure was applied for \textit{nickel}, resulting in $\log \mathrm{Ni/H}=-4.05 \pm 0.15$ (26 times solar) 
for \HZ\ and $\log \mathrm{Ni/H}=-3.61 \pm 0.13$ (31 times solar) 
for \HD.
\subsection{Additional iron-group abundances (LTE)}
\label{sect:abu:irongroup}
\begin{figure}
	\centering
    \includegraphics[width=1\columnwidth]{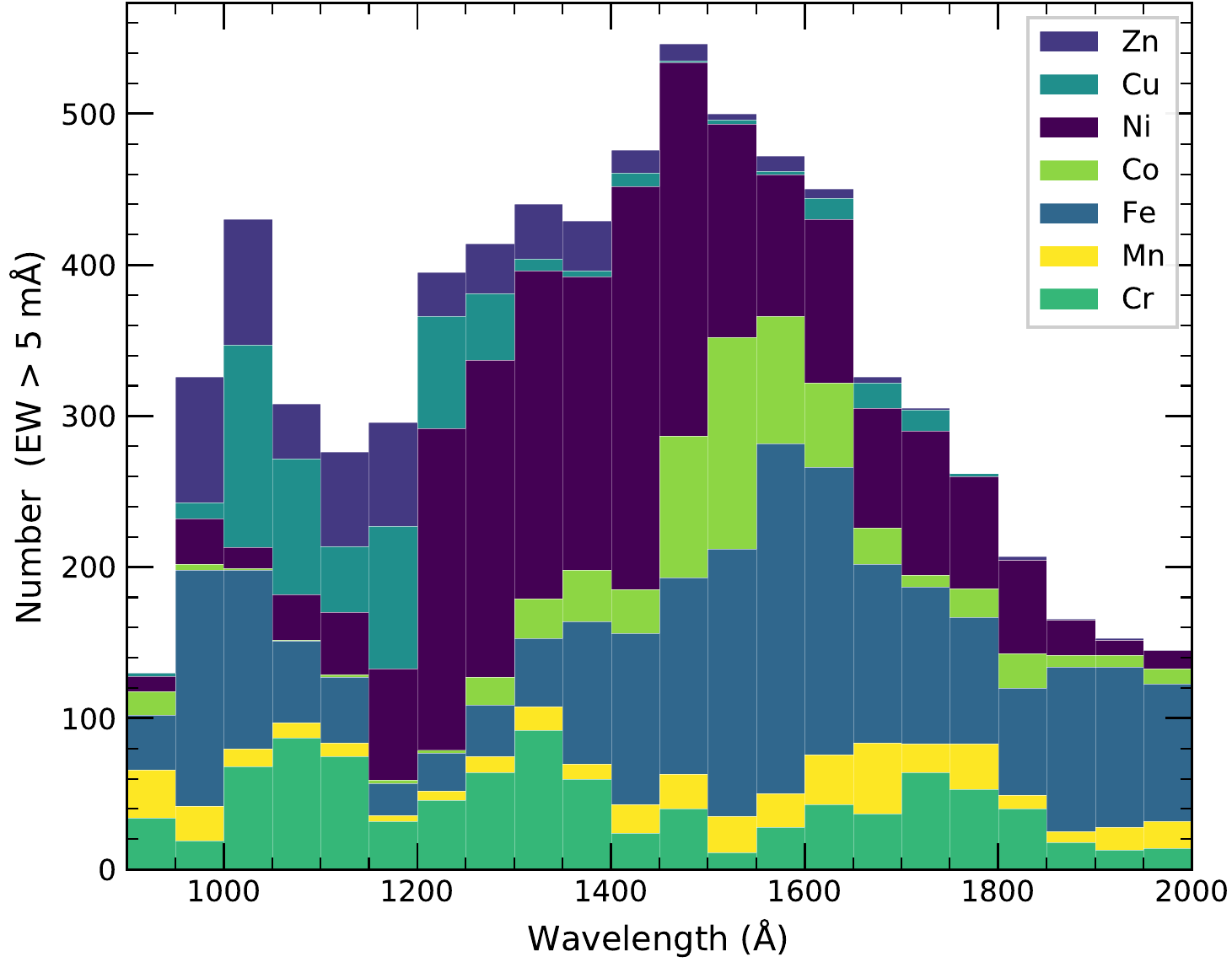}
    \vspace*{-15pt}
    \caption{Histogram of iron-peak lines with estimated equivalent widths larger than 5\,m\AA\ in the final model of HZ44. Bins are every 50\,\AA. Elements are listed in the legend in the same order as they appear in the histogram.
    }
    \label{fig:HZ44:Ironpeak}
\end{figure}
The UV spectral range is dominated by lines from iron-peak elements. 
Although most lines are from iron and nickel, opacity contributions from other iron-peak elements are also significant. 
Figure \ref{fig:HZ44:Ironpeak} shows the number of lines from iron-peak elements with estimated equivalent width larger than 5\,m\AA\ in the final model for \HZ. 
While many of these lines are observed in FUSE and IUE spectra, the opacity peak below 900\AA\ is outside of our observed spectral range.
\smallskip\\
Our models include only Fe and Ni in non-LTE. 
\smallskip\\
Many \textit{vanadium} lines in the IUE spectrum of \HZ\ would fit well with abundances of up to $\log \mathrm{V/H}=-4.8$ (e.\,g.~V\,\textsc{iv}\,$\lambda\lambda\,$1226.53, 
1308.05, 
1356.53, 
1426.65, 
1520.16, 
1522.51, 
1810.58, 
1817.69, 
1861.57\,\AA, 
and V\,\textsc{v}\,1680.20\,\AA). 
It seems unlikely that such a large amount of lines would fit the observation due to accidental alignment with unmodeled blends.
However, other lines suggest abundances below $\log \mathrm{V/H}\leq-6.0 \pm 0.4$, 
e.\,g.~V\,\textsc{iv}\,$\lambda\lambda\,$1317.56, 
1329.28, 
1355.13, 
1806.20\,\AA. 
Several lines in the FUSE spectrum seem to exclude abundances of more than $\log \mathrm{V/H}\leq-5.2\pm 0.4$, e.\,g.~V\,\textsc{iv}\,$\lambda\lambda\,$1071.06, 
1112.20, 
1123.43\,\AA, 
and V\,\textsc{iii}\,1149.95\,\AA, although they lie in regions where the continuum is poorly defined. 
We conclude that a precise abundance determination for vanadium would require a more complete model, possibly including V in non-LTE. Alternatively, it is possible that some V oscillator strengths are uncertain.
Therefore we set an upper limit of  $\log \mathrm{V/H}\leq-4.8\pm 0.4$ (893 times solar) 
for \HZ.
As for \HD, 
V\,\textsc{iv}\,$\lambda\lambda\,$1317.56, 
1329.28\,\AA\ 
seem to exclude abundances higher than $\log \mathrm{V/H}\leq-6.0\pm 0.4$ (25 times solar).
However, these lines may not be reliable since the also give a low upper limit in \HZ. We therefore adopt no upper limit for \HD.
\smallskip\\
\textit{Chromium} shows many strong lines in the ultraviolet spectrum of both stars, e.\,g.~Cr\,\textsc{iv}\,$\lambda\lambda\,$1433.89, 
1658.08,  
1825.00, 
1826.22, 
1826.88, 
1827.43\,\AA.
The overall fit is good and we adopt an abundance of $\log \mathrm{Cr/H}=-4.4\pm 0.3$ (28 times solar) 
for \HZ\ and $\log \mathrm{Cr/H}=-3.9\pm 0.3$ (76 times solar) 
for \HD.
\smallskip\\
We derive the \textit{manganese} abundance in \HZ\ from FUSE and IUE to be $\log \mathrm{Mn/H}=-4.9\pm 0.4$ (22 times solar). 
Fairly strong and unblended lines are, among many others: 
Mn\,\textsc{iii}\,$\lambda\lambda\,$917.80,  
956.47\,\AA\
and 
Mn\,\textsc{iv}\,$\lambda\lambda\,$1450.36, 
1780.00, 
1786.05\,\AA.
For \HD\, we derive an upper limit of $\log \mathrm{Mn/H}\leq-5.5\pm 0.3$ (2.5 times solar) 
from several undetected lines in the GHRS spectrum such as  Mn\,\textsc{iv}\,$\lambda\lambda\,$1244.33, 
1720.87, 
1721.49,
1724.90\,\AA.\smallskip\\
\textit{Cobalt} lines in the FUSE spectrum of \HZ\ (e.\,g.
Co\,\textsc{iii}\,$\lambda\lambda$ 944.77, 
946.54, 
946.61\,\AA)
suggest an upper limit of $\log \mathrm{Co/H}\leq-5.6\pm 0.4$.
Many Co lines in the IUE region, e.\,g.  
Co\,\textsc{iv}\,$\lambda\lambda\,$1451.43, 
1502.06, 
1502.70, 
1508.42\,\AA\
support this upper limit.
Other lines fit well with this upper limit or a slightly higher abundance:
Co\,\textsc{iv}\,$\lambda\lambda\,$1494.75, 
1500.58, 
1502.19, 
1565.91\,\AA.
Because of the unambiguous identification of Co lines and the slight discrepancy between upper limit and best-fit we adopt $\log \mathrm{Co/H}=-5.6\pm 0.5$ (12 times solar) 
with a relatively large uncertainty.
In \HD, Co\,\textsc{iv}\,$\lambda\lambda\,$1535.28, 1540.56, 1548.83, 1559.64, 1636.40\,\AA\ are resolved by GHRS and fit well with an abundance of $\log \mathrm{Co/H}=-5.1$, while Co\,\textsc{iv}\,$\lambda\lambda\,$1415.05, 1550.28, 1562.06\,\AA\ suggest an abundance no higher than $\log \mathrm{Co/H}\leq-5.3$. 
We therefore adopt an abundance of $\log \mathrm{Co/H}=-5.3\pm 0.3$ (9 times solar) 
and note that discrepancies between single lines could result from inaccuracy in line wavelengths or non-LTE effects.
\smallskip\\
Many strong \textit{copper} lines lie in the FUV spectral region. 
Cu\,\textsc{iv}\,$\lambda\lambda\,$1053.73, 1057.62\,\AA\ in the FUSE spectrum of \HZ\ are quite strong and almost free from blends.  
Other strong Cu lines are affected by unidentified blends or lie in a region where the continuum placement is not well constrained. 
Cu lines are weaker in the IUE spectrum, with a few notable exceptions:
Cu\,\textsc{iii}\,$\lambda\lambda\,$1674.59, 1684.63, 1702.11 1709.03, 1722.37\,\AA.
We determine an abundance of $\log \mathrm{Cu/H}=-5.8\pm 0.4$ (49 times solar) from the lines listed above.
Cu\,\textsc{v} lines such as Cu\,\textsc{v}\,$\lambda\lambda\,$1245.99, 1255.30, 1268.32, 1274.74, 1286.55, 1299.16\,\AA\ in the GHRS spectrum of \HD\ exclude abundances higher than $\log \mathrm{Cu/H}\leq-6.1\pm 0.4$ (11 times solar). 
\smallskip\\
The \textit{zinc} abundances in \HZ\ and \HD\ are based on strong Zn\,\textsc{iii-iv} lines that lie mostly in the IUE spectral range.
Zn\,\textsc{iv} is the dominant ion in \HD\ while \HZ\ shows about the same amount of Zn\,\textsc{iii} and Zn\,\textsc{iv} lines.
We derive $\log \mathrm{Zn/H}=-5.7\pm 0.3$ (26 times solar) for \HZ\ and  $\log \mathrm{Zn/H}=-5.3\pm 0.3$ (29 times solar) for \HD.

%
\subsection{Detected trans-iron peak elements (LTE)}
\label{sect:abu:heavy:det}
\begin{figure}
\centering
    \includegraphics[width=0.9\columnwidth]{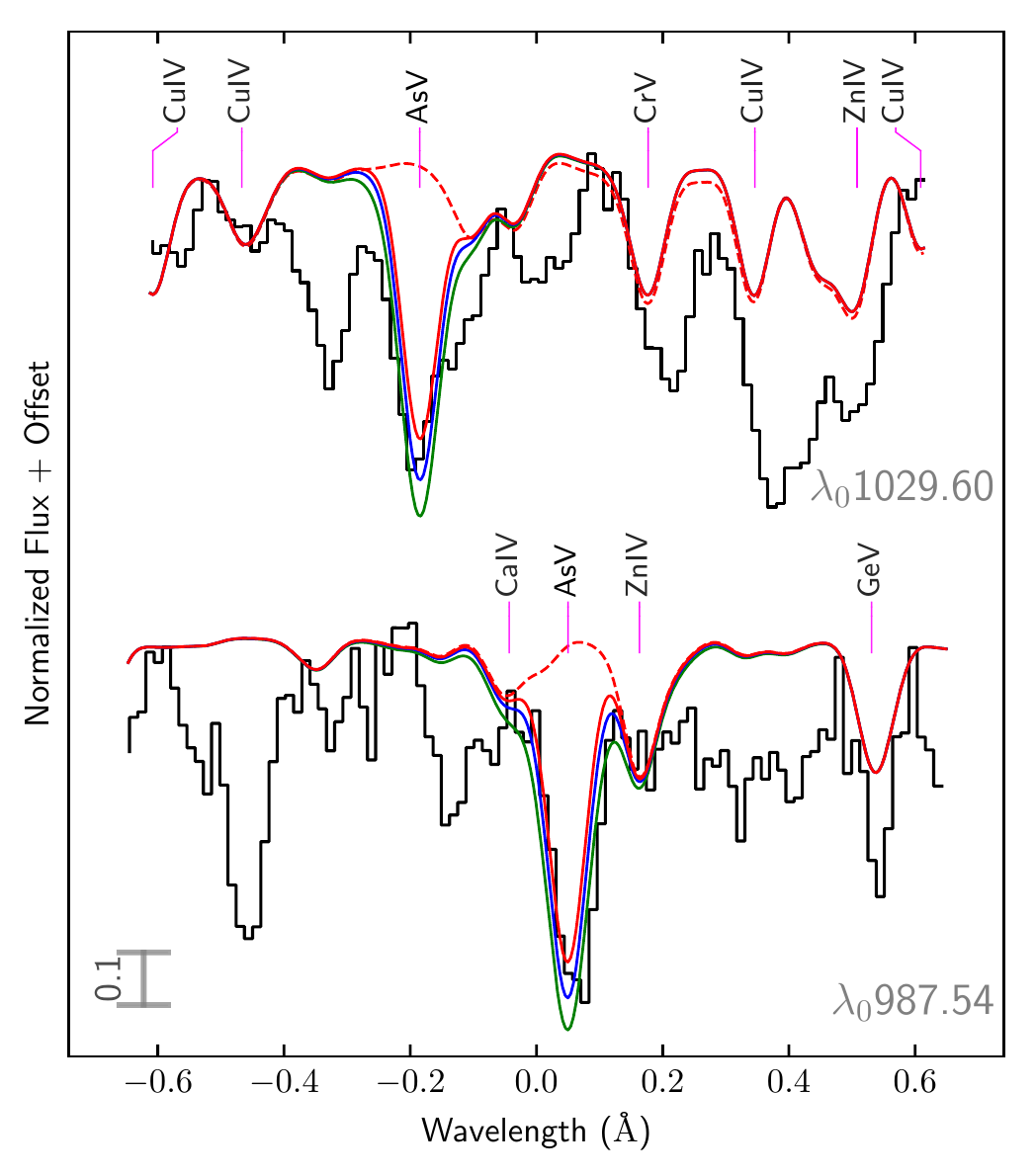}
    \vspace*{-12pt}
    \caption{As\,\textsc{v}\,987.651\,\AA\ and As\,\textsc{v}\,1029.480\,\AA\ in the FUSE spectrum of \HZ. In red: a model with $\log \mathrm{As/H}=-7.8$; dashed without As. In blue: the adopted abundance $\log \mathrm{As/H}=-7.4$. In green: $\log \mathrm{As/H}=-7.0$.}
    \label{fig:HZ:FUSE:As}
\end{figure}

We were able to measure the abundance of Ge, Ga, and Pb based on their UV lines in both \HZ\ and \HD.
In \HZ\ we could additionally derive abundances for As and Sn based on the FUSE spectrum.
\smallskip\\
In the following we will give a brief overview of the atomic data and lines used for the abundance measurement of each element.
The uncertainties on the abundances can be quite large. 
This can be the result of strong blending with unidentified lines, of the sparse atomic data available for most of these elements, and potential non-LTE effects. 
Even if atomic data are available, oscillator strengths and line wavelengths are not always well tested.
\smallskip\\
We use data from TOSS for \textit{gallium}\,\textsc{iv-v} and data from \cite{oreilly98} for Ga\,\textsc{iii} with updates for two lines from \cite{nielsen05}.
Many Ga lines are observed in the UV spectra of \HZ\ and \HD.
The strongest, isolated lines in \HZ\ include Ga\,\textsc{iv}\,$\lambda\lambda$\,1163.609, 1170.585, 1258.801, 1299.476, 1303.540, and 1347.083\,\AA.
However, a precise abundance measurement is difficult since most lines are relatively weak and blended with lines from other elements. 
Nevertheless, we measured an abundance of $\log \mathrm{Ga/H}=-6.0 \pm 0.5$ (440 times solar) 
for \HZ.
We only derive an upper limit of $\log \mathrm{Ga/H}\leq-6.4 \pm 0.4$ (80 times solar) 
for \HD\ since all Ga lines are weaker and blended.
\smallskip\\
\begin{figure}
\centering
    \includegraphics[width=0.9\columnwidth]{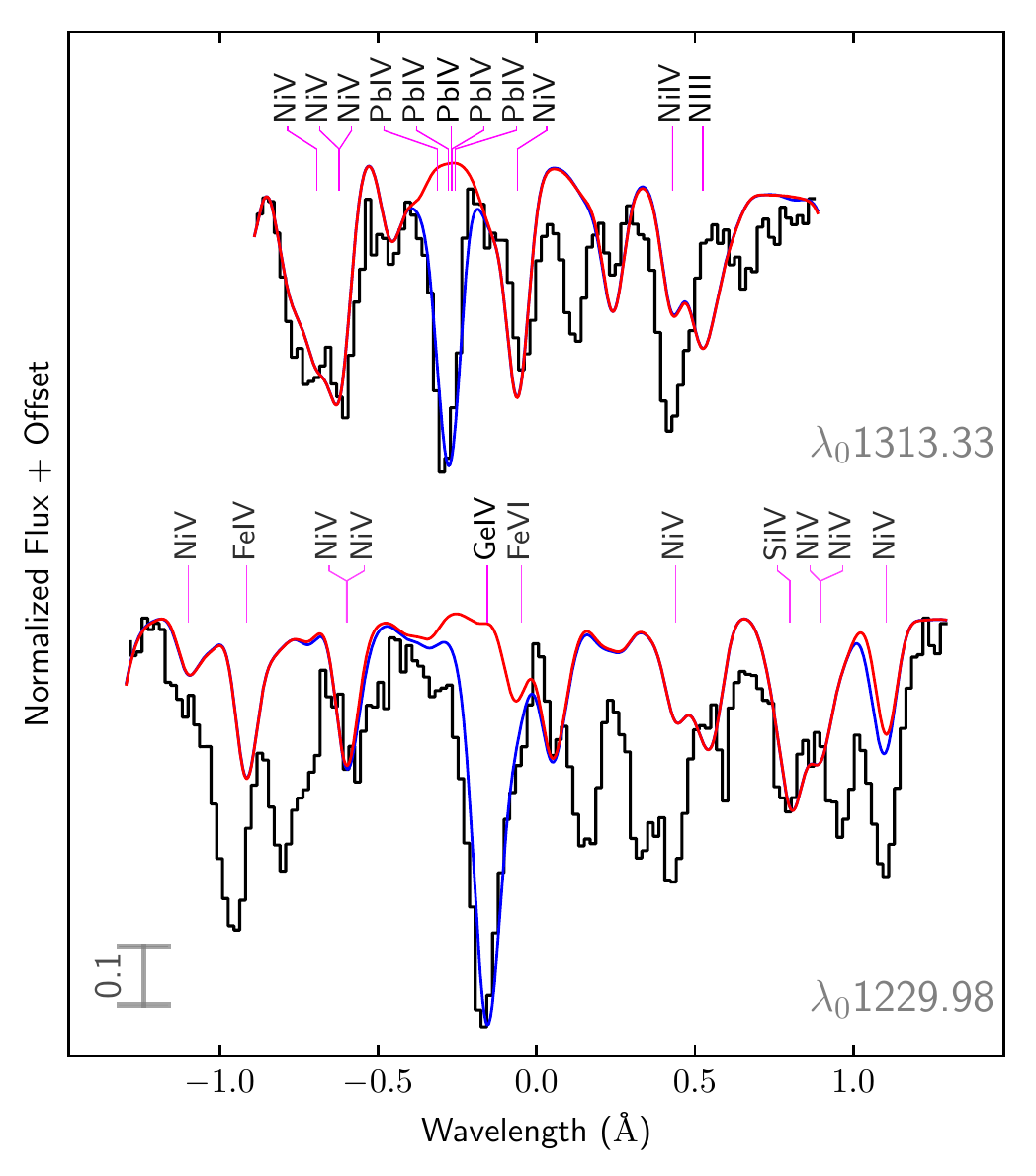}
    \vspace*{-12pt}
    \caption{Ge\,\textsc{iv} and Pb\,\textsc{iv} lines in the GHRS spectrum of \HD. In blue: a model with $\log \mathrm{Ge/H}=-5.0$, $\log \mathrm{Pb/H}=-5.7$. In red: the same model without Pb and Ge.}
    \label{fig:HD:GHRS:GePb}
\end{figure}
\textit{Germanium} shows many lines in the FUSE and IUE spectral range, including the strong resonance lines Ge\,\textsc{iv}\,$\lambda\lambda$\,1189.028, 1229.840\,\AA\ and Ge\,\textsc{iii}\,1088.463\,\AA. 
We identified lines from Ge\,\textsc{iii}-\textsc{v} in \HZ\ which can be matched at an abundance of $\log \mathrm{Ge/H}=-5.9 \pm 0.3$. (140 times solar). 
For \HD, we derive an abundance of $\log \mathrm{Ge/H}=-5.0 \pm 0.4$ (470 times solar) 
from the strong resonance lines Ge\,\textsc{iv}\,1189.028\,\AA\ (IUE) and Ge\,\textsc{iv}\,1229.840\,\AA\ (GHRS, shown in Fig. \ref{fig:HD:GHRS:GePb}).
\smallskip\\
\cite{morton00} lists oscillator strengths for ten ultraviolet \textit{arsenic}\,\textsc{iii} lines from low-lying levels, as computed by \cite{Marcinek1993}. 
Oscillator strengths for several optical As\,\textsc{iv} lines are listed in ALL, originally from \cite{churilov96}. 
The only ultraviolet As\,\textsc{iv} line listed in \cite{morton00} is the resonance line As\,\textsc{iv}\,1299.28\,\AA\ but the oscillator strength provided by \cite{Curtis1992} is low ($f=1.94\cdot10^{-3}$).
Atomic data for the two resonance lines As\,\textsc{v}\,987.651\,\AA\ and As\,\textsc{v}\,1029.480\,\AA\ is provided by \cite{Pinnington1981}, as listed in \cite{morton00}. 
These  As\,\textsc{v} oscillator strengths have previously been used for the As abundance measurement in DO white dwarfs by \cite{Chayer2015} and \cite{Rauch2016a}.
\cite{morton00} also lists a third resonance line, As\,\textsc{v}\,1001.211\,\AA. This line is not observed in the spectrum of \HZ\ and was disregarded by both \cite{Chayer2015} and \cite{Rauch2016a}. 
It is only mentioned in \cite{FroeseFischer1977} (and may have been confused with the $^2$S$_{1/2}-^2$P$_{1/2}$ transition line As\,\textsc{v}\,1029.480\,\AA).
Neither NIST \citep{Moore1971} nor \cite{Joshi1986} list an energy level that would be consistent with an As\,\textsc{v} resonance line at 1001.211\,\AA, so we decided to exclude it as well.
As\,\textsc{iii} lines are weak in \HZ\ and As\,\textsc{iii}\,$\lambda\lambda$\,927.540, 944.726\,\AA\ exclude abundances higher than $\log \mathrm{As/H}\leq-6.4 \pm 0.4$ (960 times solar).
As\,\textsc{iv}\,1299.28\,\AA\ would fit an otherwise unidentified line at an abundance of $\log \mathrm{As/H}=-5.6$ which is excluded by other lines.
Figure \ref{fig:HZ:FUSE:As} shows the strongest observed As lines in \HZ, As\,\textsc{v}\,987.651\,\AA\ and As\,\textsc{v}\,1029.480\,\AA. 
We use these lines to derive an abundance of $\log \mathrm{As/H}=-7.4 \pm 0.4$ (100 times solar). 
As\,\textsc{iv}\,1299.28\,\AA\ is also visible in the GHRS spectrum of \HD\ and fits the observation at an abundance of $\log \mathrm{As/H}=-4.9 \pm 0.4$ (1300 times solar).
Due to the discrepancy observed in \HZ, we do not consider this an abundance measurement.
\smallskip\\
Also as part of their ``Stellar Laboratories'' series \cite{Rauch2017SL9} have measured the abundance of \textit{selenium} in the peculiar DO white dwarf \RE. 
We use their oscillator strengths for Se\,\textsc{v} and results from \cite{Bahr1982} for Se\,\textsc{iv}, as listed in \cite{morton00}. 
Se\,\textsc{iv} $\lambda\lambda\,$\,959.590, 
996.710\,\AA\ 
and Se\,\textsc{v}\,1094.691\,\AA\ fit strong, otherwise unidentified lines at $\log \mathrm{Se/H}=-6.3\pm 0.6$.
However, Se\,\textsc{iv}\,984.341\,\AA\ seems to exclude abundances higher than $\log \mathrm{Se/H}\leq-7.6\pm 0.4$.
Like As\,\textsc{v}\,1001.211\,\AA\ this line may not be real; its lower level is not listed in NIST nor in the newest reference on Se\,\textsc{iv} energy levels we found, \cite{Pakalka2018}.
Therefore, we adopt an abundance of $\log \mathrm{Se/H}=-6.3\pm 0.6$ (110 times solar) for \HZ.
\smallskip\\
\begin{figure}[!htbp]
\centering
    \includegraphics[width=0.95\columnwidth]{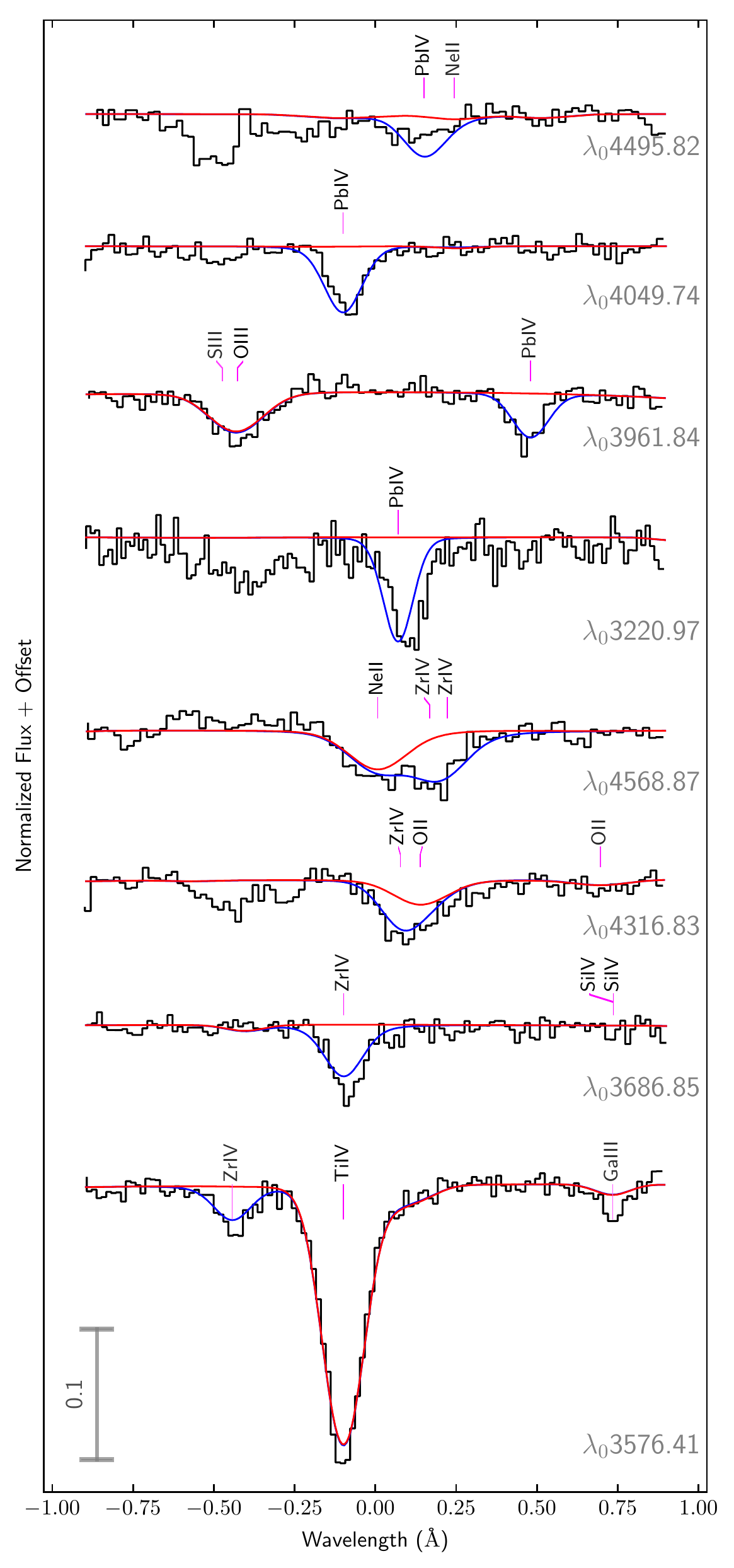}
    \vspace*{-7pt}
    \caption{A selection of  Zr\,\textsc{iv} and Pb\,\textsc{iv} lines in the HIRES spectrum of \HZ. In blue: a model with $\log \mathrm{Zr/H}=-5.9$ and $\log \mathrm{Pb/H}=-5.9$. In red: the same model, but without Zr or Pb.}
\label{fig:HZ:ZrIVPbIV}
\end{figure}
\begin{figure}[!htbp]
\centering
    \includegraphics[width=0.9\columnwidth]{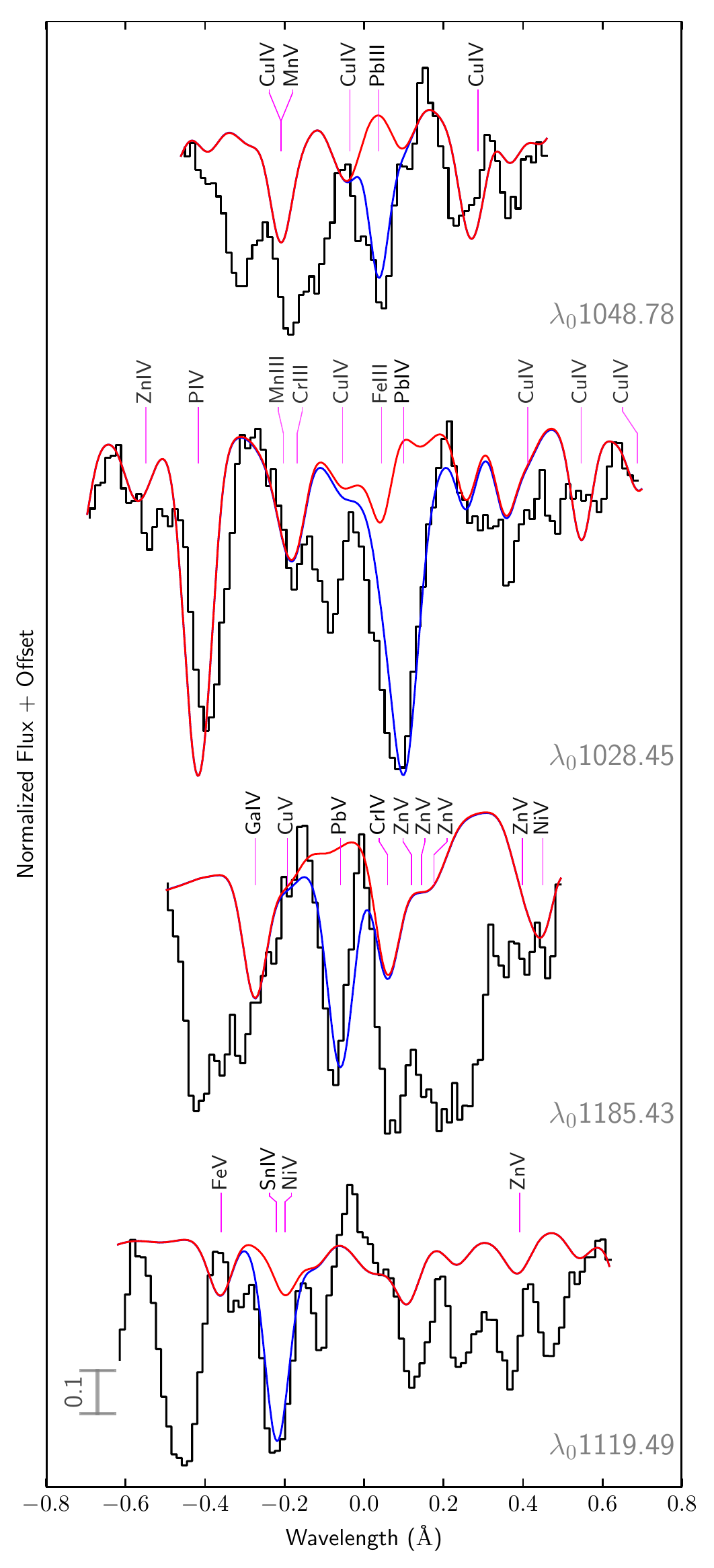}
    \vspace*{-7pt} 
    \caption{Pb\,\textsc{iii}, Pb\,\textsc{iv}, Pb\,\textsc{v}, and  Sn\,\textsc{iv} lines in the FUSE spectrum of \HZ. In blue: a model with $\log \mathrm{Pb/H}=-5.9$ and $\log \mathrm{Sn/H}=-6.9$. In red: the same model without Pb and Sn.}
    \label{fig:HZ:FUSE:Pb}
\end{figure}
\HZ\ is one of the very few hot subdwarf stars showing \textit{zirconium} in its optical spectrum. 
We used the atomic data from \cite{Rauch2017} for our analysis.
We fitted four distinct Zr\,\textsc{iv} lines in the HIRES spectrum of \HZ\ 
(Zr\,\textsc{iv}\,$\lambda\lambda\,$ 
3576.107, 
3686.902, 
4317.077, 
4569.218, 4569.272\,\AA, see Fig. \ref{fig:HZ:ZrIVPbIV}) 
and found an abundance of $\log \mathrm{Zr/H}=-5.92 \pm 0.19$ (1500 times solar). 
The UV spectra of \HZ\ also show Zr\,\textsc{iv-v} lines, and although none of them are strong or isolated enough to independently measure the abundance, they are consistent with abundance derived from the optical lines.
The doublet Zr\,\textsc{iv}\,$\lambda\lambda$\,4569.218, 4569.272\,\AA\ is visible in \HD\ and would fit well with an abundance of $\log \mathrm{Zr/H}=-5.5$ (1700 times solar).
Since no other lines could clearly be identified as Zr we adopt an upper limit of $\log \mathrm{Zr/H}\leq-5.4\pm 0.2$.
\smallskip\\
\textit{Tin}  is one of the elements that were identified in \HZ\ by \cite{O'Toole2004}. 
We used atomic data from \cite{safronova03}, supplemented with data from \cite{biswas18} for Sn\,\textsc{iv} and results from \cite{haris12} for Sn\,\textsc{iii}.
We derived the Sn abundance in \HZ\ to be $\log \mathrm{Sn/H}=-6.9 \pm 0.4$ (550 times solar), 
based on the strong Sn\,\textsc{iv}\,1119.338\,\AA\ line which is almost free from blends (see Fig.~\ref{fig:HZ:FUSE:Pb}). 
The other strong, but blended Sn\,\textsc{iv}\,$\lambda\lambda$\,1019.720, 1044.487, 1314.539\,\AA, and Sn\,\textsc{iv}\,1437.527\,\AA\ (blended with Co\,\textsc{iv}\,1437.488\,\AA) lines support this measurement.
Even if the FUSE continuum is estimated too high in our models, Sn lines in the IUE spectrum, where the model is more complete, set the upper limit to $\log \mathrm{Sn/H}\leq-6.5 \pm 0.4$ (1400 times solar). 
Sn\,\textsc{iv}\,1314.539\,\AA\ excludes abundances higher than $\log \mathrm{Sn/H}\leq-6.6 \pm 0.4$ (480 times solar) for \HD.
\smallskip\\
\noindent We collected atomic data for \textit{lead}\,\textsc{iii}-\textsc{v} from several sources. 
Pb\,\textsc{iii} oscillator strengths are from \cite{alonso-medina09}, with the exception of the resonance lines Pb\,\textsc{iii}\,1048.877\,\AA\ and  Pb\,\textsc{iii}\,1553.021\,\AA\ which are based on lifetime measurements by \cite{Ansbacher1988} as listed in \cite{morton00}. 
For Pb\,\textsc{iv}, we use oscillator strengths from \cite{safronova04} with additional lines from \cite{alonso-medina11} and one line (Pb\,\textsc{iv}\,4496.15\,\AA) from \cite{naslim13}.
Data for Pb\,\textsc{v} is provided by \cite{colon14}.
While this collection is far from complete, many Pb lines could be identified, including not only strong Pb\,\textsc{iii-v} lines in the ultraviolet spectrum of \HZ\ but also five Pb\,\textsc{iv} lines in its optical spectrum (Pb\,\textsc{iv}\,$\lambda\lambda$\,3052.56, 
3221.17, 
3962.48, 
4049.80, 
4496.15\,\AA, see Fig.~\ref{fig:HZ:ZrIVPbIV}).

Fitting all identified Pb lines in the HIRES spectrum except Pb\,\textsc{iv}\,3052.56\,\AA\ (S/N too low)
results in $\log \mathrm{Pb/H}=-5.89 \pm 0.09$ (11000 times solar). 
This is remarkably consistent with Pb lines observed in the UV region, including lines from Pb\,\textsc{iii} and Pb\,\textsc{v}. 
As far as we know, this is the first time Pb\,\textsc{v} lines were modeled in any star.
The strongest Pb lines per ionization stage observed in the FUSE spectrum of \HZ\ are shown in Fig.~\ref{fig:HZ:FUSE:Pb}. 
The Pb abundance measurement in \HD\ is based mostly on Pb\,\textsc{iv}\,1313\,\AA\ (see Fig.~\ref{fig:HD:GHRS:GePb}), assuming a solar isotopic ratio as in \cite{O'Toole2006}. 
We derive an abundance of $\log \mathrm{Pb/H}=-5.65 \pm 0.40$ (8400 times solar), 
consistent with Pb\,\textsc{v}\,$\lambda\lambda$\,1233.50, 1248.46\,\AA\ in the GHRS spectrum. 
%
\subsection{Trans-iron elements with upper limits}
\label{sect:abu:heavy:ul}
The abundance measurements for Se, Kr, Sr, Y, Mo, Sb, Te, Xe, and Th turned out to be inconclusive because too few lines were found and/or their relative line strengths were at variance with model predictions. Instead we derived upper limits for these elements.
\smallskip\\
\textit{Krypton} and \textit{strontium} belong to the group of elements that have been studied in white dwarfs by \cite{Rauch2016b, Rauch2017SL9}.
Despite the large number of Kr\,\textsc{iv}-\textsc{v} lines in the TOSS line list, none of them are strong enough in the final synthetic spectrum of \HZ\ to be identified in the observation. 
We derive an upper limit of $\log \mathrm{Kr/H}\leq-5.2\pm 0.6$ (1700 times solar) 
from four undetected Kr\,\textsc{iv} lines, the strongest being Kr\,\textsc{iv}\,1538.211\,\AA. 
Kr\,\textsc{iv}\,999.388\,\AA\ would fit well with an abundance of $\log \mathrm{Kr/H}=-5$ but is likely blended.
Kr\,\textsc{iv}\,$\lambda\lambda$\,1400.898, 1538.211, 1558.514\,\AA\ and Kr\,\textsc{v}\,1293.917\,\AA\ in GHRS spectrum of \HD\ exclude abundances higher than $\log \mathrm{Kr/H}\leq-4.8 \pm 0.4$ (1900 times solar).
\smallskip\\
The situation is similar for \textit{strontium} in \HZ.  
The undetected Sr\,\textsc{v}\,962.378\,\AA, Sr\,\textsc{iv}\,1244.137\,\AA, and Sr\,\textsc{iv}\,1244.763\,\AA\ lines exclude abundances higher than $\log \mathrm{Sr/H}\leq-5.1 \pm 0.6$ (5100 times solar). 
Sr\,\textsc{iv}\,1331.129\,\AA\ would fit well with $\log \mathrm{Sr/H}=-5.0$.
Sr\,\textsc{iv}\,$\lambda\lambda$\,1244.137, 1244.888, 1268.622, 1275.354, 1729.533\,\AA\ in GHRS exclude abundances higher than $\log \mathrm{Sr/H}\leq-4.9 \pm 0.3$ (3600 times solar) in \HD.
\smallskip\\
\cite{naslim11} have observed \textit{yttrium} in the iHe hot subdwarf LS\,IV$-14^\circ116$. 
So far it has been observed to be extremely enriched in two additional iHe-sds: HE\,2359$-$2844 \citep{naslim13} and \UVO\ \citep{Jeffery2017}.
We used their oscillator strengths for Y\,\textsc{iii}\,4039.602\,\AA\ and Y\,\textsc{iii}\,4040.112\,\AA\ to search for Y in \HZ\ and \HD. 
Both lines are predicted to be weak in the models and were not detected in the spectra of either star.
We derive an upper limit of $\log \mathrm{Y/H}\leq-5.3 \pm 0.2$ (14800 times solar) for \HZ.
Due to the lower S/N of the FEROS spectrum and the higher temperature in \HD, the upper limit derived from the same lines in that star is even higher: $\log \mathrm{Y/H}\leq-4.7 \pm 0.3$ (26000 times solar). 
None of the ultraviolet Y\,\textsc{iii} lines for which \cite{Redfors1991} computed oscillator strengths are strong enough to improve on this threshold. 
Unfortunately, we found no oscillator strength measurements for the resonance lines Y\,\textsc{iii}\,1000.563\,\AA\ and Y\,\textsc{iii}\,1006.587\,\AA\ that are listed in \cite{morton00}.
A more complete analysis of Y in subdwarf stars would benefit from oscillator strengths for Y\,\textsc{iv} which is the dominant ionization stage at effective temperatures around 40\,000\,K.
\smallskip\\
\cite{Rauch2016a} have observed \textit{molybdenum} in \RE. 
We use their atomic data for Mo\,\textsc{v} and atomic data from the Kurucz line list for Mo\,\textsc{iv} to search for Mo in \HZ.
Mo\,\textsc{iv}\,$\lambda\lambda\,$965.485, 
966.638\,\AA\ and 
Mo\,\textsc{v}\,$\lambda\lambda\,$939.248, 
1127.101\,\AA\ 
are well reproduced with $\log \mathrm{Mo/H}=-6.4$ (2500 times solar). 
Since these lines are relatively weak and our model is missing opacity in the FUSE region, we adopt an upper limit of $\log \mathrm{Mo/H}\leq-6.2 \pm 0.4$ (4000 times solar).
Several lines in the IUE spectrum return a non-detection compatible with this upper limit, for example Mo\,\textsc{v}\,$\lambda\lambda\,$\,1586.898, 
1774.317\,\AA.
As for \HD\,
Mo\,\textsc{v}\,$\lambda\lambda\,$\,1586.898, 
1590.414, 
1653.541
1661.215
1774.317\,\AA\
exclude abundances higher than $\log \mathrm{Mo/H}\leq-6.1\pm0.4$ (2200 times solar).
\smallskip\\
\cite{Werner2018} have recently measured photospheric abundances of \textit{antimony} in two DO white dwarfs: \RE\ and PG\,0109+111. 
Both stars are chemically peculiar, with strong enrichment of trans-iron elements.
Despite blends with unidentified lines in FUSE and the low resolution of IUE, we were able to set the upper limit on the Sb abundance in \HZ\ to $\log \mathrm{Sb/H}\leq-8.0\pm0.5$ (470 times solar), 
which is well below the extreme enrichment observed in the two aforementioned white dwarfs. 
This upper limit is based on three lines: Sb\,\textsc{iv}\,1042.190\,\AA\ (blended with Cr\,\textsc{iv}), Sb\,\textsc{v}\,1104.23\,\AA, and Sb\,\textsc{v}\,1226.001\,\AA. 
Sb\,\textsc{v}\,1104.23\,\AA\ even fits well at this abundance.
\begin{figure}
\centering
    \includegraphics[width=0.9\columnwidth]{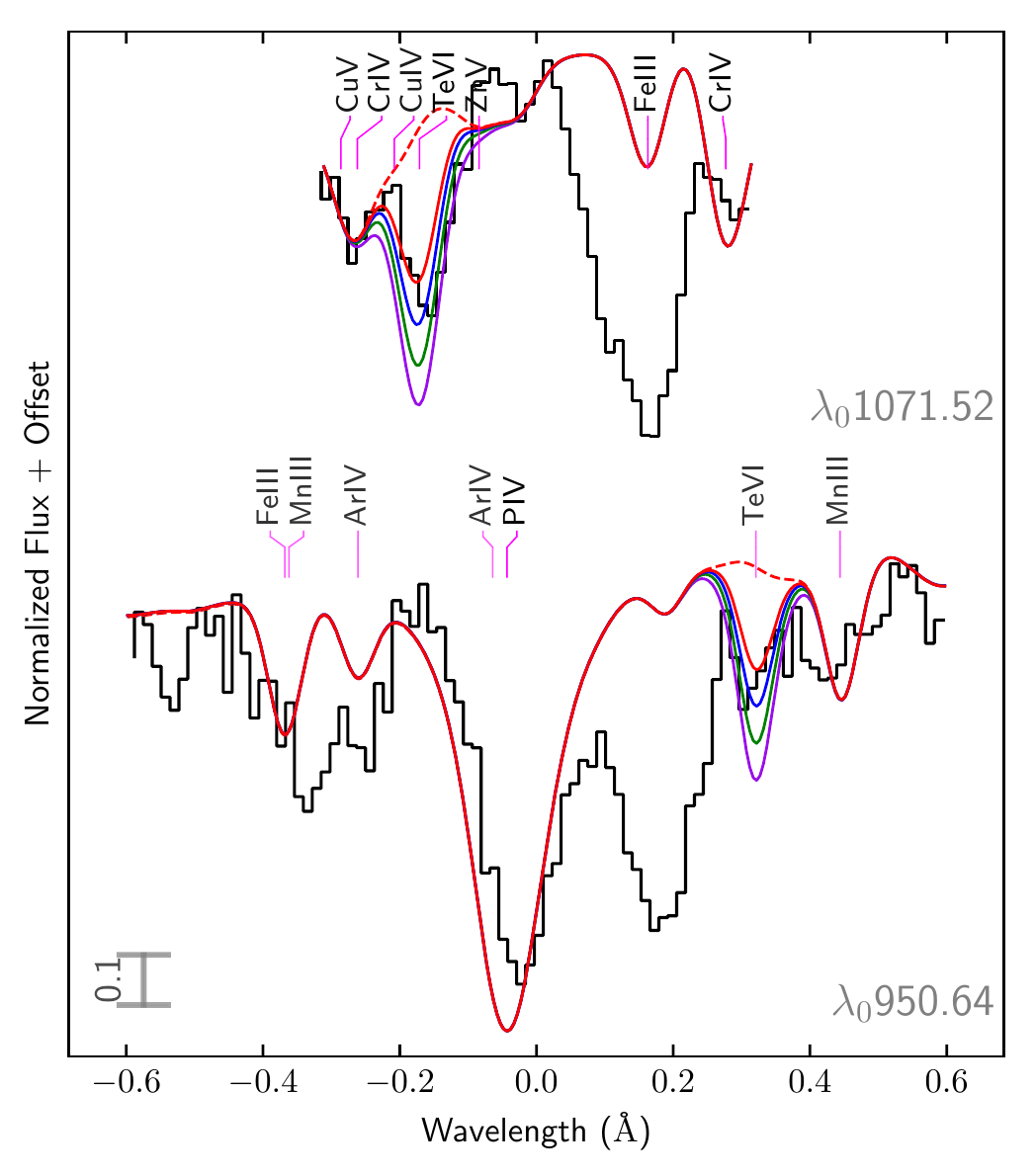}
    \vspace*{-12pt}
    \caption{Te\,\textsc{vi} resonance lines in the FUSE spectrum of \HZ. We show a model with $\log \mathrm{Te/H}=-7.9$ (red) and without Te (dashed). Models including higher Te abundances are further shown: $\log \mathrm{Te/H}=-7.5$ (blue), $\log \mathrm{Te/H}=-7.1$ (green), and $\log \mathrm{Te/H}=-6.7$ (pink).}
    \label{fig:HZ:FUSE:Te}
\end{figure}
\smallskip\\
\cite{zhang13} have computed oscillator strengths for \textit{tellurium}\,\textsc{ii}-\textsc{iii}, including UV and optical lines.
However, since the Te\,\textsc{iii} population numbers are low in \HZ, none of these lines are visible.
\citet{Rauch2017SL9} provided oscillator strengths for Te\,\textsc{vi} lines, including the resonance lines Te\,\textsc{vi}\,$\lambda\lambda$\,951.021, 1071.414\,\AA, which are visible in \HZ\ (see Fig.~\ref{fig:HZ:FUSE:Te}).
They are best reproduced with an abundance of $\log \mathrm{Te/H}=-7.5\pm0.4$ (100 times solar).
Due to the weakness of the lines and their position in the FUSE spectrum we only adopt an upper limit of $\log \mathrm{Te/H}=-7.1\pm0.4$ (250 times solar).
The Te\,\textsc{v}\,1281.670\,\AA\ resonance line listed in \citet{morton00,Pinnington1985a} also supports this upper limit.
Te\,\textsc{v}\,1281.670\,\AA\ is also visible in the GHRS spectrum of \HD, but blended with a weaker Co\,\textsc{v} line. We derive an upper limit of $\log \mathrm{Te/H}=-5.8\pm0.4$ (2200 times solar).
\smallskip\\
We use oscillator strengths for \textit{xenon}\,\textsc{iv-v} from \cite{Rauch2017} provided on the TOSS website in order to look for Xe in \HZ.
Most Xe lines in the FUSE spectrum of \HZ\ are blended with unidentified lines.
The resonance lines Xe\,\textsc{iv}\,$\lambda\lambda$\,935.251, 1003.373\,\AA\ and Xe\,\textsc{v}\,$\lambda\lambda$\,936.284, 945.248\,\AA\ allow us to set an upper limit of $\log \mathrm{Xe/H}\leq-7.3\pm 0.4$ (140 times solar). 
The actual Xe abundance in \HZ\ might be close to our upper limit, since the additional opacity improves the fit for many lines. 
However, the only isolated line that could be identified is Xe\,\textsc{v}\,936.284\,\AA. 
\smallskip\\
\textit{Thorium} is the heaviest element for which we found atomic data. 
It is of particular interest since it is not produced through the s-process and can be used for age determination because of its radioactivity.
\cite{safronova13} have computed atomic properties of 24 low-lying states of the Th\,\textsc{iv} ion. 
Atomic data for Th\,\textsc{iii} are published in \cite{Safronova2014}, but could not be used here since all transitions with calculated oscillator strengths lie in the infrared region.
Even with a relatively low abundance of $\log \mathrm{Th/H}=-8.5$ (1400 times solar), 
our models predict several Th\,\textsc{iv} lines with estimated equivalent widths up to 30\,m\AA\ in the UV range. 
In particular, the non-detection of Th\,\textsc{iv}\,983.140\,\AA\ (which falls conveniently on one of the few points of pseudo-continuum in the FUSE spectrum of \HZ), Th\,\textsc{iv}\,1140.612\,\AA\ (in the wing of a well-modeled Si\,\textsc{iii} line), and Th\,\textsc{iv}\,1682.213\,\AA\ allow us to set the upper limit for the photospheric Th abundance in \HZ\ to $\log \mathrm{Th/H}\leq-8.0$ (4600 times solar). 
We derive an upper limit of $\log \mathrm{Th/H}\leq-7.8\pm 0.3$ (3200 times solar) for \HD\ based on the Th\,\textsc{iv}\,$\lambda\lambda$\,4413.576, 5420.380, 5841.397, 6019.151\,\AA\ lines in the FEROS spectrum.
%
\subsection{An unsuccessful search for additional trans-iron elements}
\label{sect:abu:hopeless}
We also searched for predicted lines of Sc, In, Ba, Tl, and Bi in the ultraviolet and optical spectra of \HZ.
However, no lines from these elements could be identified and no meaningful upper limit could be derived either.
\smallskip\\
The strongest \textit{scandium} lines in the model of \HZ\ are the resonance line Sc\,\textsc{iii}\,1610.194\,\AA\ and Sc\,\textsc{iii}\,1603.064\,\AA. 
Both lines are weak in the model, even at a high abundance of $-5$\,dex relative to hydrogen and are blended with both modeled and unidentified lines, so no meaningful upper limit could be determined.
\smallskip\\
\cite{safronova03} have calculated atomic properties along the silver isoelectronic sequence, including In\,\textsc{iii}, Sn\,\textsc{iv}, and Sb\,\textsc{v}. 
They predict \textit{indium}\,\textsc{iii} lines from low-lying levels with high oscillator strengths in the IUE spectral region. 
However, the population numbers for In\,\textsc{iii} are too low to set a meaningful upper limit for both \HZ\ and \HD.
\smallskip\\
\textit{Barium} was observed in RE\,0503$-$289 by \cite{Rauch2014} who also provide atomic data for Ba\,\textsc{v}.
The predicted Ba\,\textsc{v} lines are so weak in the model of \HZ\ that $\log \mathrm{Ba/H}=-5$ (32000 times solar) 
is required before the strongest predicted lines in synthetic spectrum of \HZ\ 
(Ba\,\textsc{v}\,$\lambda\lambda\,$1097.415, 
1103.140\,\AA) 
reach equivalent widths of 0.1\,m\AA. 
It was therefore not possible to set a meaningful upper limit. 
\smallskip\\
The strongest observable \textit{bismuth} line in the spectrum of \HZ\ is by far Bi\,\textsc{v}\,1139.549\,\AA. 
Since it is blended with an unidentified line, we can only derive an upper limit of $\log \mathrm{Bi/H}\leq-8.5$ (340 times solar).
We consider this upper limit preliminary since it is based on a single, blended line in the FUSE spectrum.\smallskip\\
Oscillator strengths along the gold isoelectronic sequence were computed by \cite{safronova04}, including not only Pb\,\textsc{iv} but also \textit{thallium}\,\textsc{iii}. 
Similar to In\,\textsc{iii}, the population number for Tl\,\textsc{iii} is too low to set a meaningful upper limit in both \HZ\ and \HD.
%

\subsection{Chemical composition summary}
\label{sect:abu:summary}
\begin{figure*}
\begin{center}
\includegraphics[width=1\textwidth]{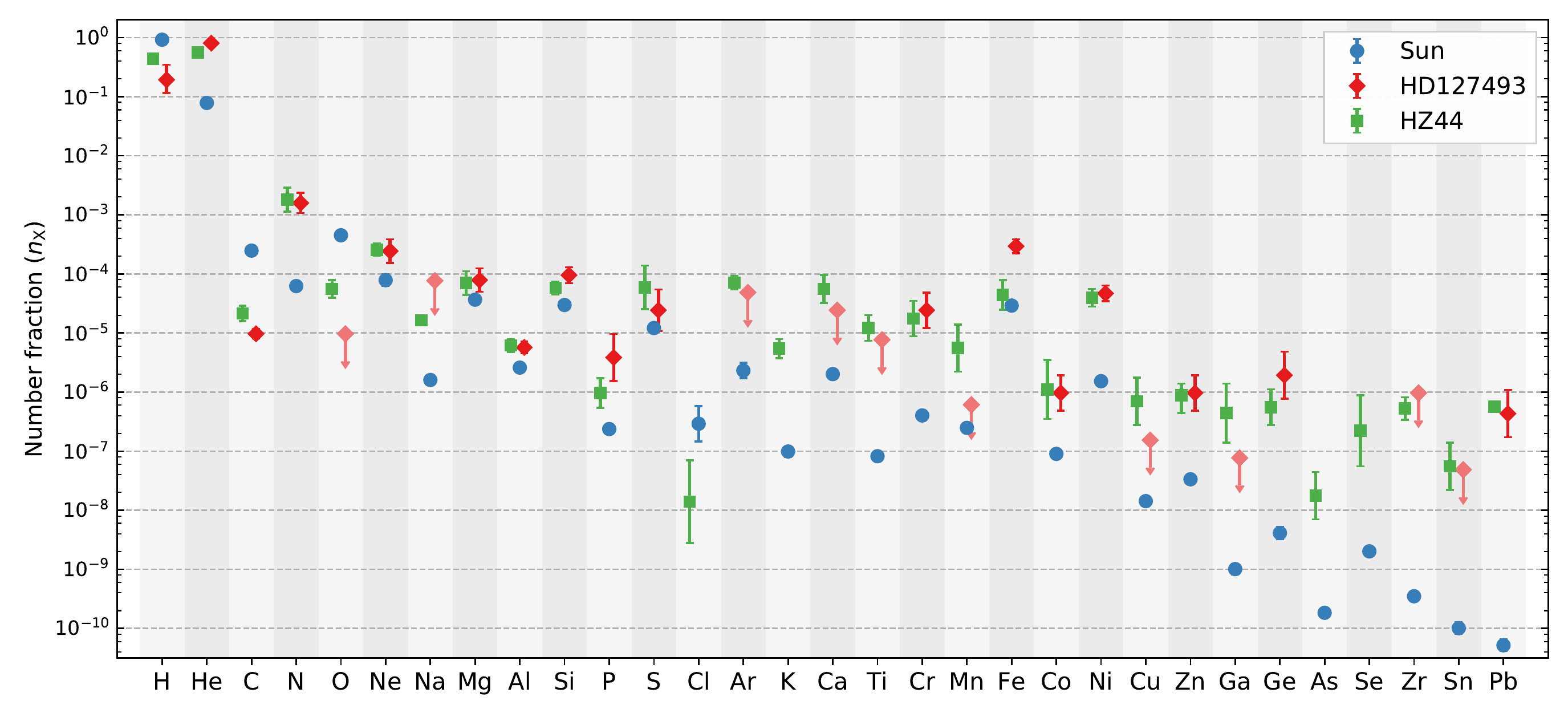}
\vspace{-12pt}
\captionof{figure}{
Abundance patterns of \HD\ and \HZ\ compared to that of the Sun (by number fraction). Only elements with an abundance measurement in at least one of the star are shown. Upper limits are marked with an arrow and less saturated colors.}
\label{fig:abupattern:nf}
\end{center}
\end{figure*}

\begin{figure*}
\begin{center}
\includegraphics[width=1\textwidth]{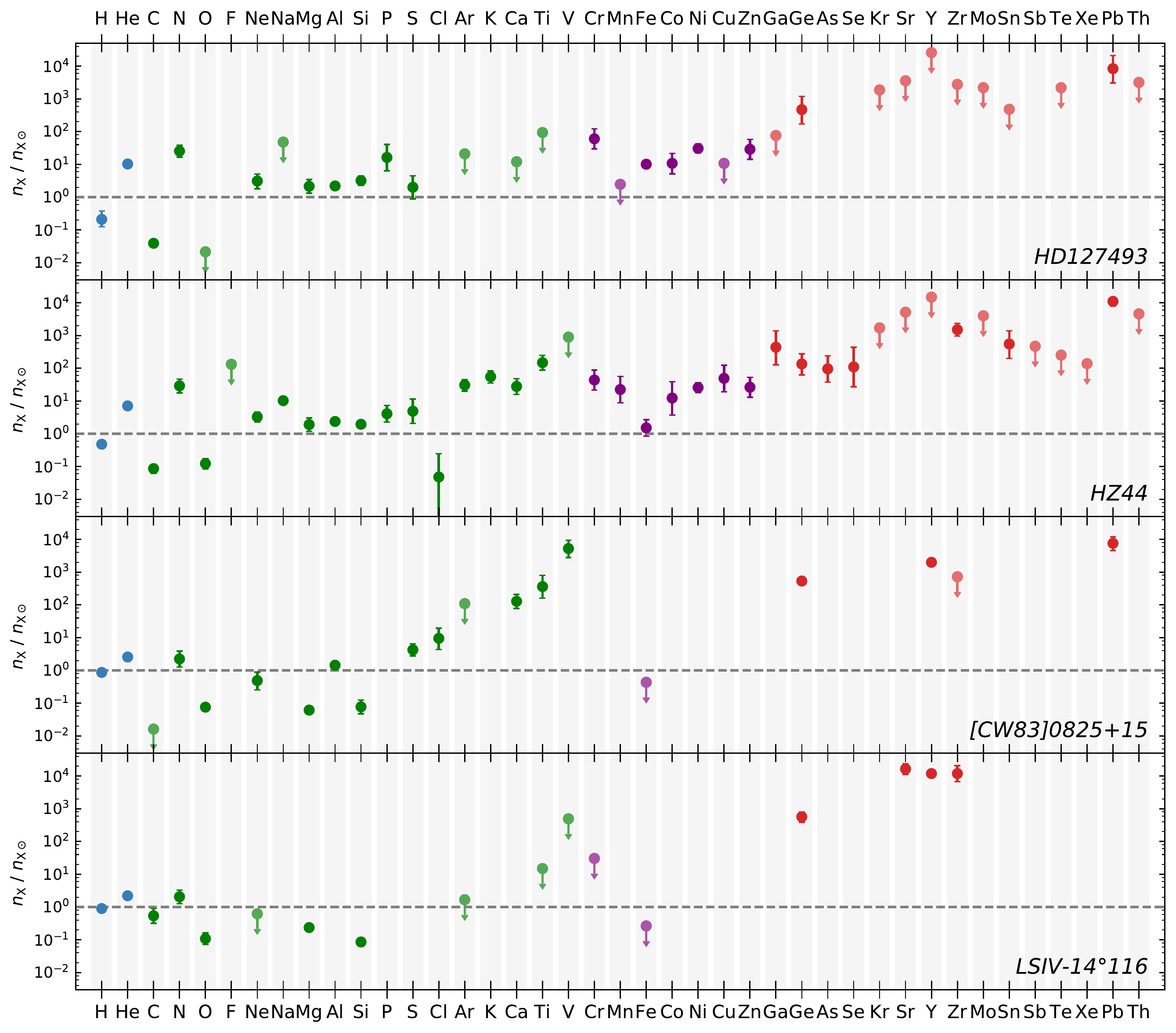}
\captionof{figure}{
Abundance pattern of \HD\ and \HZ\ with respect to solar composition. 
Results for the heavy-metal subdwarfs [CW83]\,0825+15 \citep{Jeffery2017} and \lsIV\ \citep{naslim11} are shown for comparison.
Light elements ($23\leq\mathrm{Z}$) are marked by green symbols, iron-peak elements ($24\leq\mathrm{Z}\leq30$) in purple, and heavier elements ($\mathrm{Z}\geq31$) in red. Upper limits are marked with an arrow and less saturated colors. 
\vspace{00pt}}
\label{fig:abupattern}
\end{center}
\end{figure*}

Figures \ref{fig:abupattern:nf} and \ref{fig:abupattern} as well as Table \ref{tab:abundances_units} show our final abundance values for \HZ\ and \HD. The abundance patterns are remarkably similar in both stars despite the $\sim$2500\,K difference in their effective temperature. The overall resemblance between the chemistry of both stars is especially visible when comparing their abundances in number fractions (Fig~\ref{fig:abupattern}). 
In addition, some trans-iron elements are present in the atmosphere with very similar abundances. For example, in \HZ\ Cu, Zn, Ga, Ge, Zr, and Pb have the same number fraction of $\sim10^{-6.2}$, whereas the solar abundances show a strong decrease with increasing atomic mass. 
As and Sn are significantly less abundant than the other trans-iron elements.\\
Both stars show a strong CNO cycle pattern, most obvious in Fig. \ref{fig:abupattern}, with nitrogen being enriched while carbon and oxygen are depleted with respect to solar values. 
Ne is mildly enriched (by a factor of 3) in both stars compared to solar.
With the exception of Cl, all elements with $11 \leq \mathrm{Z} \leq 20$ are more abundant in \HZ\ than in the Sun. 
The abundance of Mg, Al, Si, and S is similar in both stars.
With a measured abundance of $148^{+98} _{-61}$ times solar, the Ti\,\textsc{iv} lines are very strong in the UVA spectrum of \HZ. 
In contrast, the Ti lines covered by the FEROS spectrum of \HD\ are weak and set an upper limit for Ti to $94^{+105}$ times solar.
Co and Ni have very similar abundances in both stars: they are about 30 times the solar values. 
Mn and Cu could not be detected in \HD, which indicated that they are less abundant than in \HZ.
As seen in many other hot subdwarf stars, Fe is the least enriched element among the iron group in \HZ\ ($\sim$1.5 times solar).
Fe is more enriched ($\sim$10 times solar) in \HD. 
The Zn abundance in both stars is similar to that of Ni, between 25 and 30 times solar. 
While the Ge abundances in \HD\ ($\sim$470 times solar) and \HZ\ ($\sim$140 times solar) are similar considering uncertainties, the Ga abundance in \HD\ ($\lesssim$\,75 times solar) is lower than in \HZ\ ($\sim$440 times solar).
The As abundance exceeds that of the Sun by a factor of about 100 in \HZ. 
Interestingly, \HZ\ is one of the few hot subdwarf stars showing Zr\,\textsc{iv} and Pb\,\textsc{iv} lines in their optical spectrum.
As far as we know Zr\,\textsc{iv} has been identified in the optical spectrum of only two iHe hot subdwarfs, LS\,IV$-$14$^\circ$\,116 and HE\,2359-2844 \citep{naslim11,naslim13}.
Zr is enriched at about 1500 times solar in \HZ, which is not excluded also in \HD. 
Interestingly, the measured Pb abundance in both stars is almost identical; they are enhanced by a factor of 10000.
The enrichment is significantly lower for other heavy elements in \HZ. 
In particular Xe and Te exceed the solar abundance by a factor of less than 500 in \HZ.

\section{Discussion}
\label{sect:discussion}
%
The Carnegie Yearbook No.~55, for 1955/56 reports the discovery of a new sdO star, BD+25$^\circ4655$, to be similar to \HD\ and \HZ\ and quotes a very foresighted conclusion (probably by Muench): that the spectrum of the newly discovered sdO ``is extremely rich in faint sharp lines, those of N\,\textsc{ii}, N\,\textsc{iii}, and Ne\,\textsc{ii} being especially conspicious. The complete absence of lines of oxygen and carbon, in any stage of ionization, suggests that the surface material of this kind of star has undergone nuclear processes which transformed the carbon into nitrogen and the oxygen to neon.'' The abundance of those elements derived in previous quantitative spectral analyses as well as in ours substantiate this statement. The strong overabundances of heavy elements found for \HZ, \HD, and a few other iHe hot subdwarfs stars are generally believed to be caused by atmospheric diffusion processes (radiative levitation). However, it is not plausible to assume that diffusion creates an abundance pattern of C, N, O, and Ne, that mimics the nucleosynthesis pattern so well. 
In the following subsections, we first discuss the evidence for diffusion processes and compare the abundance patterns of our two stars to that of other iHe hot subdwarfs. Then we revisit the nuclear synthesis aspect and discuss implications for the evolutionary status. 
\subsection{Diffusion} 
\label{sect:diffusion}
Diffusion refers to the equilibrium between gravitational settling and radiative levitation.
While heavy elements are pulled downwards by gravity, their important line opacities in the UV region, where the photospheric flux distribution peaks, lead to opposing forces due to radiation pressure.
This force is limited by the saturation of spectral lines at high abundances. 
Once an equilibrium of both forces has been established, the elemental abundances should be fixed.
\smallskip\\
Models that account for gravitational settling and radiative levitation only fail to reproduce the observed abundances pattern of sdB stars \citep[see][for a discussion]{heber16}. Additional processes have to be taken into account. Stellar winds and turbulent mixing have been suggested. 
\cite{Michaud2011} have studied the effects of non-equilibrium diffusion and radiative levitation on element abundances up to Ni for sdB stars on the horizontal branch (up to $T_\mathrm{eff} \approx 37\,000$\,K), but not for sdOs. 
To match the iron abundances observed in sdBs by \cite{Geier2010}, and later \cite{Geier2013}, they required some process to dampen the effect of radiative levitation. 
\cite{Michaud2011} adopted a turbulent surface mixing zone during the HB evolution that includes the outer $\sim$\,$10^{-7.5}\,\mathrm{M}_\odot$ in the envelope. 
Similarly to the sdOs discussed in this paper, the photospheric iron abundance in sdBs is approximately solar.
This low Fe enhancement is a result of its high absolute abundance in the photosphere and the consequent line saturation (see Fig.~\ref{fig:abupattern:nf}). 
Since heavy elements, such as Zr and Pb, are initially less abundant in absolute terms, a stronger enrichment due to radiative levitation is expected.
The models for the hottest stars ($T_\mathrm{eff}$ = 35$-$37 kK) in \citet{Michaud2011} predict abundances that are lower than what is observed in \HZ\ and \HD. For example, N, Ne, Al, Si, and Mg are predicted to be depleted with respect to the solar values. Thus additional processes are required to explain the abundance pattern of our two sdOs; nucleosynthesis during the formation of the stars, weak stellar winds \citep{unglaub08,Hu2011} and a possible atmospheric surface convection zone \citep{groth85,unglaub10} might well be involved.
\smallskip\\
The models by \cite{Michaud2011} can not reproduce the He-enrichment and CNO-cycle pattern observed in some sdBs (and the sdOs discussed here) since they use approximated methods to evolve their models through the He-flash. 
\cite{Byrne2018} have preformed similar calculations for post common envelope sdBs from the top of the RGB to the zero age HB with a more self-consistent treatment of the He-flash. 
They produced He-rich atmospheres in their delayed He-flash models and predict C and N to be enriched and O to be depleted for sdBs on the zero-age horizontal branch (ZAHB). The abundances of other elements are similar to those of \cite{Michaud2011} but both models are not especially well-suited for the hotter stars discussed here.
Detailed sdO evolutionary models (e.\,g.~through the HeWD-merger channel) including diffusion of heavy elements beyond the iron group would be required to explain the observed abundance pattern.
Unfortunately, the atomic data required for modeling diffusion of elements heavier than Ni is still lacking.
%
\subsection{Comparison  with other iHe hot subdwarfs} 
\label{sect:diffusion:examples}
In Fig. \ref{fig:abupattern} we compared the abundance pattern of \HZ\ and \HD\ with literature abundances of two other iHe subdwarf stars:
\UVO\ and \lsIV.
\smallskip\\
\UVO\ is the closest match to \HZ\ and \HD\ in terms of of atmospheric properties with $T_\mathrm{eff}=38\,900$\,K and $\log g=5.97$ \citep{Jeffery2017}. Although being less He-rich ($\log\,n_\mathrm{He}/n_\mathrm{H} = -0.6$), it is the only known heavy-metal iHe hot subdwarf to be C-deficient, like the two stars we analysed here. 
Its abundance pattern is also similar to \HZ\ and \HD\ in that the CNO-cycle pattern is evident and lead is equally enriched.
However, the abundances of some specific elements differ significantly: 
Mg and Si are less abundant by $\sim$1\,dex while Cl is about 2\,dex more abundant in \UVO.
\smallskip\\
\lsIV\ was the first heavy-metal hot subdwarf to be recognized as so and is considered as the prototype of the class, with its extreme enrichment in Sr, Y and Zr \citep{naslim11}
With $T_\mathrm{eff}=34\,950$\,K, $\log g=5.93$, and $\log n_\mathrm{He}/n_\mathrm{H} = -0.62$ \citep{Green2011}, the star is cooler and less helium-rich than \HZ\ and \HD. Similarly to the two other heavy-metal subdwarfs, HE2359-2844 and HE1256-2738, its C abundance is higher than in \HZ\ and \HD\ (see also Fig.~\ref{fig:tefflogy}).
Its Sr and Zr enrichment is stronger than in \HZ\ and \HD\ (and \UVO).
\smallskip\\
Although the abundances of \HZ\ and \HD\ are remarkably similar, the patterns observed in other heavy-metal subdwarfs appear to be different. However, it is difficult to draw firm conclusions when abundances are known only for a much more limited subset of elements in the other stars. In the case of
\lsIV\ and \UVO, the lack of UV data strongly restricts their chemical portrait. 
Along with \UVO, our two stars \HZ\ and \HD\ are the only known heavy-metal subwarfs to be enriched in nitrogen, but depleted in carbon and oxygen. 
The three other known heavy metal subdwarfs have higher C-abundances, similarly to the group of CN-rich eHe subdwarfs that is observed at higher temperatures.
Whether the differences in the the abundances of carbon and nitrogen in iHe subdwarfs are related to stellar evolution or the effects of diffusion remains unclear.

\subsection{Nuclear synthesis and evolutionary status} 
\label{sect:evolution}
The formation of hot subdwarfs with intermediate He abundances (10\%--90\% by number) through merging He-WDs with low-mass MS stars was investigated by \cite{zhang17}.
In these models, subdwarfs with intermediate He-rich atmospheres represent a short ($<5\,$Myr) phase after the He-flash is ignited during the merger.
The initially He-rich atmosphere of the merger remnant transforms into a H-rich one as the heavier He diffuses downward (gravitational settling) until the atmosphere is H-rich when the ZAHB is reached.
The same process is predicted in accretion-based HeWD+HeWD mergers \citep{zhang12a} that can also reproduce the He abundance in iHe-sds.
\smallskip\\
A known problem is that merger calculations predict a fast surface rotation.
\cite{Schwab2018} has calculated post-HeWD+HeWD merger models with initial conditions taken from hydrodynamic merger calculations and found that merger products have $v_\mathrm{rot} \succeq 30$\,km\,s$^{-1}$ once they appear as hot subdwarfs.
This rotation is usually not observed in single sdBs \citep{Geier2012} and \cite{hirsch09} found N-rich He-sdOs (such as \HZ\ and \HD) to have $v_\mathrm{rot}$ similar to sdBs.
For individual stars, this can be explained by a small inclination $i$ (which leads to a small $v_\mathrm{rot}\sin i$). 
However, with increasing evidence for slowly rotating (intermediate) He-sdOs, it seems likely that additional physics is needed to match the observations \citep{Schwab2018}.
Alternatively, slowly rotating hot subdwarfs may be created through a different process altogether.
\smallskip\\
The observation of the CNO cycle pattern in \HD\ and \HZ\ indicates that the CNO process must have been efficient in a H-burning shell or mixed from a sufficiently hot core in the stars' progenitor.
In fact, the slow HeWD+HeWD merger model by  \cite{zhang12a} is able to reproduce the CNO pattern observed in \HZ\ and \HD\ well except for somewhat higher predicted O-abundances.
This may be an indication that O has been processed to Ne through the $\alpha$ capture $\isotope[16][8]{O}(\alpha,\gamma)\isotope[20][10]{Ne}$. 
In HeWD+MS merger models presented by \cite{zhang17}, temperatures high enough for $\isotope[18][8]{O}(\alpha,\gamma)\isotope[22][10]{Ne}$ burning are reached following the first He-flash, even if the processed material is not always mixed to the surface.
That the He, C, N, O, and Ne abundances in some He-sdOs, and in the two stars analysed here, can be explained by nuclear synthesis might indicate that these light elements are less affected by diffusion in this type of stars.
\smallskip\\
An alternative explanation to diffusion for the extreme enrichment of heavy element could be that they were created in the stars' progenitor.
Heavy elements like Zr and Pb are produced mainly in the s-process, which is thought to be efficient in asymptotic giant branch (AGB) stars. 
While most hot subdwarfs do not evolve through the AGB phase, low-mass post-AGB tracks are crossing the log $g$ $-T_\mathrm{eff}$ diagram in the region populated by luminous hot subdwarfs \citep{napiwotzki08}. 
Therefore such an evolutionary channel might be responsible for a small fraction of the hot subdwarfs. 
However, diffusion calculations for these elements are required before conclusions on possible AGB progenitors of heavy-metal enriched iHe-sds can be made.
%
\section{Conclusion}
\label{sect:summary}
We have performed a detailed spectroscopic analysis of the two intermediate He-sdOs \HZ\ and \HD.
SED-fits combined with parallax distances for both stars result in masses that are consistent with the canonical subdwarf mass of 0.47\,M$_\odot$ within 1-$\sigma$ uncertainty.
No indication of binarity was found for either star.
Our main focus was the determination of photospheric metal abundances, including heavy elements.
We found the abundance pattern in both stars to be very similar.
They show a typical CNO-cycle pattern and slight enrichment of intermediate-mass elements ($\mathrm{Z}\leq30$, except Cl) compared to solar values.
Heavier elements such as Ga, Ge, and As were found to be enriched in the order of 100 times solar.
Most interestingly, the abundances of Zr and Pb were measured from optical lines and confirmed with UV transitions in \HZ, and turned out to be more than 1000 and 10000 times solar, respectively.
\HD\ shows no optical Zr or Pb lines, but we derived a Pb enrichment of about 8000 times solar from Pb\,\textsc{iv-v} lines in its HST/GHRS UV spectrum. \ion{Pb}{v} lines were modeled for the first time in a stellar photosphere and their predicted strength reproduced well the observations of both stars.  
We also determined upper limits for several additional heavy elements.
Some of them, for example Xe and Te, have a moderate enrichment ($\la$ 500 times solar) in \HZ.
\smallskip\\
In order to improve the accuracy of abundance measurements, additional atomic data are much-needed, in particular for the heavy elements.
Many lines in both the optical and ultra-violet spectra still remain unidentified. 
This is especially evident in the FUSE spectrum of \HZ, where not only interstellar but also many photospheric lines are missing from our models.
Some of those lines likely belong to ionized heavy elements for which no atomic data, or only a limited subset, are available.
\smallskip\\
Interestingly, pulsations were observed in three other iHe subdwarfs, namely \UVO\ \citep{Jeffery2017}, \lsIV\
\citep{Ahmad2005}, and Feige\,46 \citep{Latour2019}. This would make it worth looking for photometric variability in \HZ\ and \HD\ as well.\\
As of now, we are not able to fully explain the observed abundance pattern in intermediate He-sdOs.
Evolutionary simulations for sdOs including diffusion for heavy elements and mixing during hot flasher/merger evolution would be required to interpret the abundance pattern. Even though we obtained a quite exhaustive chemical portrait for the two stars analysed here, this is generally not the case for the other iHe hot subdwarfs. More complete set of abundances for additional stars are also necessary to properly investigate these intriguing patterns. 
\smallskip\\
The efficiency of radiative support on heavy elements in hot subdwarfs might be linked to their helium abundance, given that the intermediate helium-rich hot subdwarfs seem to favorably display extreme enhancements. 
Hydrogen-rich sdB stars were found to be enriched in some heavy elements as well \citep{O'Toole2006,blanchette2008}, but their enrichment in Pb for example is significantly lower than that observed in the heavy-metal iHe subdwarfs. 
At the other end of the helium abundance spectrum, abundance analyses of He-sdOs are more limited, especially concerning heavy metals.
The only He-rich sdO for which heavy metal abundances have been derived, BD+39\,3226, turned out the be less than 2 dex enhanced in Zr and Pb \citep{Chayer2014}. 
It would be most interesting to determine abundances of heavy elements in additional He-rich stars.
The He-sdOs recently analyzed by \cite{schindewolf18} would be well-suited to confirm (or not) this milder enrichment in heavy metals. 
Their atmospheric parameters, as well as their abundances of lighter elements are well constrained, and excellent UV data are available. 
The current set of hot subdwarfs for which abundances of heavy elements are known do not allow us to rule out the possibility that the effective temperature also plays a role in favoring the radiative support of particular elements. Once again abundances for a larger sample of stars across the $T_\mathrm{eff}$ range where the extreme overabundances are observed ($\sim$ 34$-$43 kK), also including hydrogen-rich stars such as the two hottest objects from \citet{O'Toole2006}, will be necessary in order to investigate the relation between $T_\mathrm{eff}$ and the (over)abundances of particular elements.
\smallskip\\
%
\begin{acknowledgements}
We thank Andreas Irrgang and Simon Kreuzer for the development of the SED fitting tool, Monika Schork for measuring the radial velocities of \HZ\ from HIRES spectra, and Markus Schindewolf for providing preliminary atmospheric parameters of HZ\,44. M.L.\ acknowledges funding from the Deutsche Forschungsgemeinschaft (grant DR 281/35-1).
Based on observations made with the NASA/ESA Hubble Space Telescope, obtained from the data archive (prop. ID GO5305) at the Space Telescope Science Institute. 
STScI is operated by the Association of Universities for Research in Astronomy, Inc. under NASA contract NAS 5-26555. 
Support for MAST for non-HST data is provided by the NASA Office of Space Science via grant NNX09AF08G and by other grants and contracts. 
Based on INES data from the IUE satellite. 
Based on observations made with ESO Telescopes at the La Silla Paranal Observatory under programme ID  074.B-0455(A).
This research has made use of the Keck Observatory Archive (KOA), which is operated by the W. M. Keck Observatory and the NASA Exoplanet Science Institute (NExScI), under contract with the National Aeronautics and Space Administration.
This work has made use of data from the European Space Agency (ESA) mission {\it Gaia} (\url{https://www.cosmos.esa.int/gaia}), processed by the {\it Gaia} Data Processing and Analysis Consortium (DPAC, \url{https://www.cosmos.esa.int/web/gaia/dpac/consortium}). Funding for the DPAC has been provided by national institutions, in particular the institutions participating in the {\it Gaia} Multilateral Agreement.
The TOSS service (\url{http://dc.g-vo.org/TOSS}) used for this paper was constructed as part of the activities of the German Astrophysical Virtual Observatory.
We acknowledge the use of the Atomic Line List (\url{http://www.pa.uky.edu/~peter/newpage/}). We also thank the referee, C. Moni Bidin,
for his helpful comments.

\end{acknowledgements}
%
\bibliographystyle{aa} 
\bibliography{hzhd.bib} 
%
\begin{appendix}
\section{Additional material}
\subsection{Spectroscopic and photometric data}
\label{a:data:spectra}
\begin{table*}
\setstretch{1.2}
\captionof{table}{List of spectra used in our analysis.}
\label{tab:obs:detail}
\vspace*{-5pt}
\begin{center}
\begin{tabular}{l l l l r c}
\toprule
\toprule
Star & Instrument	&	Dataset	&	 Range (\AA) 	& Exp. (s) &	R	\\ 
\midrule
\HD	&	FEROS		&	ADP.2016-09-21T07:07:18.680		&	$3527.9-9217.7$	&	$\phantom{0}600$	&	48000\\
	&				&	ADP.2016-09-21T07:07:18.736		&	$3527.9-9217.7$	&	$\phantom{0}300$	&	\\
	&				&	ADP.2016-09-21T07:07:18.686		&	$3527.9-9217.7$	&	$\phantom{0}300$	&	\\ 
	&	IUE	LWR		&	LWR03587HS	&	$1850.0-3350.0$	&	$5400$	&	10000\\
	&				&	LWR04198HL	&	$1850.0-3350.0$	&	$4450$	&	\\
	&				&	LWR06702HL	&	$1850.0-3350.0$	&	$5400$	&	\\
	&				&	LWR07211HL	&	$1850.0-3350.0$	&	$2950$	&	\\
	&	GHRS G160M	&	Z2H60107T	&	$1222.6-1258.8$	&	$\phantom{0}462$	&	0.07\AA\\
	&				&	Z2H60109T	&	$1254.9-1291.0$	&	$\phantom{0}462$	&	\\
	&				&	Z2H6010BT	&	$1285.6-1321.6$	&	$\phantom{0}462	$	&	\\
	&				&	Z2H6010DT	&	$1317.7-1353.6$	&	$\phantom{0}462	$	&	\\
	&				&	Z2H6010FT	&	$1349.7-1385.5$	&	$\phantom{0}517	$	&	\\	
	&				&	Z2H6010HT	&	$1383.0-1418.8$	&	$\phantom{0}598	$	&	\\	
	&				&	Z2H6010JT	&	$1414.9-1450.5$	&	$\phantom{0}517	$	&	\\	
	&				&	Z2H6010LT	&	$1532.5-1567.7$	&	$\phantom{0}653	$	&	\\	
	&				&	Z2H6010OT	&	$1623.2-1658.1$	&	$\phantom{0}462	$	&	\\	
	&				&	Z2H6010QT	&	$1713.0-1747.6$	&	$\phantom{0}462	$	&	\\ 
	&	IUE	SWP		&	SWP04071HS	&	$1150.0-1980.0$	&	$5880$	&	10000\\
	&				&	SWP04860HL	&	$1150.0-1980.0$	&	$3000$	&	\\
	&				&	SWP07695HL	&	$1150.0-1980.0$	&	$4500$	&	\\ 
	&				&	SWP08276HL	&	$1150.0-1980.0$	&	$3930$	&	\\ 
\HZ	&	HIRES		&	HI.20050810.20686	&	$3214\phantom{.0}-5990\phantom{.0}$	&	$600$	&	36000\\ 
	&	    		&	HI.20050810.21381	&	$3890\phantom{.0}-6732\phantom{.0}$	&	$600$	&	 \\ 
	&	    		&	HI.20050812.21565	&	$3214\phantom{.0}-5990\phantom{.0}$	&	$500$	&	 \\ 
	&	    		&	HI.20070504.38715	&	$3022\phantom{.0}-5800\phantom{.0}$	&	$900$	&	 \\ 
	&	    		&	HI.20160203.58141	&	$4716\phantom{.0}-7580\phantom{.0}$	&	$600$	&	 \\ 
	&	    		&	HI.20160401.55323	&	$3128\phantom{.0}-5947\phantom{.0}$	&	$600$	&	 \\ 
	&	IUE	SWP		&	SWP16294HL	&	$1150.0-1980.0$	&	$9600$	&	10000\\
	&				&	SWP17350HL	&	$1150.0-1980.0$	&	$14820$	&	\\ 
	&	FUSE MDRS 	&	p3020401000	&	$ \phantom{0}904.3-1188.4$	&	$5919$	&	19000\\
	&	FUSE LWRS	&	m1080401000	&	$ \phantom{0}904.3-1188.4$	&	$4679$	&	17000\\
	&	FUSE LWRS	&	s5051901000	&	$ \phantom{0}904.3-1188.4$	&	$3937$	&	\\
\bottomrule 
\end{tabular}
\end{center}
\end{table*}
\begin{table*}
\setstretch{1.2}
\captionof{table}{Radial velocity measurements for \HZ\ and HIRES spectra used.}
\label{tab:rv}
\vspace*{-10pt}
\begin{center}
\begin{tabular}{c c c}
\toprule
\toprule
Time (\texttt{YYYY-MM-DD hh:mm})& Number of considered lines	&	$v_\mathrm{rad}$ (km\,s$^{-1}$)	\\
\midrule
1995-07-02 05:26  &  10   &  $12.4 \pm 0.8$ \\
1995-07-02 05:33  &  10   &  $12.9 \pm 0.7$ \\
1995-07-02 05:40  &  10   &  $13.0 \pm 0.6$ \\
1996-05-26 07:50  &  10   &  $12.8 \pm 0.6$ \\
1997-07-13 05:54  &  19   &  $13.0 \pm 0.6$ \\
1998-05-19 05:33  &  25   &  $11.9 \pm 0.7$ \\
1999-02-14 15:45  &  20   &  $12.6 \pm 0.5$ \\
2000-02-05 16:10  &  5    &  $12.0 \pm 1.0$ \\
2001-03-02 16:04  &  23   &  $12.2 \pm 0.6$ \\
2002-02-03 16:18  &  10   &  $12.9 \pm 0.7$ \\
2002-08-04 05:26  &  18   &  $12.7 \pm 0.9$ \\
2005-08-10 05:45  &  27   &  $12.6 \pm 0.9$ \\
2005-08-12 05:59  &  27   &  $11.9 \pm 0.6$ \\
2005-08-12 08:10  &  22   &  $12.7 \pm 0.6$ \\
2006-06-18 05:37  &  29   &  $13.2 \pm 0.7$ \\
2006-06-18 05:39  &  27   &  $13.3 \pm 0.8$ \\
2006-06-18 05:42  &  26   &  $13.2 \pm 0.8$ \\
2007-05-04 10:45  &  17   &  $13.2 \pm 0.5$ \\
2008-07-11 05:53  &  10   &  $12.7 \pm 0.9$ \\
2008-07-11 06:00  &  10   &  $12.4 \pm 0.8$ \\
2009-07-13 06:56  &  19   &  $12.9 \pm 0.7$ \\
2012-01-04 16:14  &  17   &  $12.4 \pm 0.5$ \\
2012-01-04 16:17  &  17   &  $12.3 \pm 0.5$ \\
2013-05-05 05:21  &  22   &  $13.0 \pm 0.7$ \\
2013-05-07 05:24  &  25   &  $12.7 \pm 0.7$ \\
2015-04-09 15:34  &  12   &  $12.0 \pm 0.6$ \\
2015-04-11 15:24  &  19   &  $12.4 \pm 0.6$ \\
\bottomrule 
\end{tabular}
\end{center}
\end{table*}

%

\begin{table*}
\setstretch{1.2}
\captionof{table}{Photometric data used for the SED-fit of \HZ.}
\label{tab:data:HZ:phot}
\vspace*{-5pt}
\begin{center}
\begin{tabular}{l l r r r c c}
\toprule
\toprule
    System & Passband & Magnitude & Uncertainty & Type & Reference\\ 
\midrule
     2MASS &          H  &   12.569  & 0.023  & magnitude  & \cite[2MASS: II/246/out]{Cutri2003_2MASS}\\ 
     2MASS &          J  &   12.386  & 0.022  & magnitude  & \cite[2MASS: II/246/out]{Cutri2003_2MASS}\\ 
     2MASS &          K  &   12.672  & 0.027  & magnitude  & \cite[2MASS: II/246/out]{Cutri2003_2MASS}\\ 
Stroemgren &  H$_\beta$  &    2.617  &        &     color  & \cite[J/A+A/580/A23/catalog]{Paunzen2015_Stromgen}\\ 
Stroemgren &        b$-$y  &  $-0.151$ &        &     color  & \cite[J/A+A/580/A23/catalog]{Paunzen2015_Stromgen}\\ 
Stroemgren &         m1  &   0.104   & 0.020  &     color  & \cite[J/A+A/580/A23/catalog]{Paunzen2015_Stromgen}\\ 
Stroemgren &          y  &   11.715  & 0.007  & magnitude  & \cite[J/A+A/580/A23/catalog]{Paunzen2015_Stromgen}\\ 
    UKIDSS &          H  &   12.560  & 0.002  & magnitude  & \cite[UKIDSS DR9: II/319/las9]{Lawrence2013_UKIDSS}\\ 
    UKIDSS &          J  &   12.400  & 0.001  & magnitude  & \cite[UKIDSS DR9: II/319/las9]{Lawrence2013_UKIDSS}\\  
    UKIDSS &          K  &   12.687  & 0.002  & magnitude  & \cite[UKIDSS DR9: II/319/las9]{Lawrence2013_UKIDSS}\\ 
    UKIDSS &          Y  &   12.276  & 0.001  & magnitude  & \cite[UKIDSS DR9: II/319/las9]{Lawrence2013_UKIDSS}\\ 
      WISE &         W1  &   12.750  & 0.023  & magnitude  & \cite[AllWISE: II/328/allwise]{Cutri2012_WISE}\\ 
      WISE &         W2  &   12.830  & 0.025  & magnitude  & \cite[AllWISE: II/328/allwise]{Cutri2012_WISE}\\ 
   IUE box & $1300-1800$\,\AA\  &    7.903  & 0.020  & magnitude  & \cite[VI/110/inescat, SWP03432LL]{Wamsteker2000_INES}\\ 
   IUE box & $2000-2500$\,\AA\  &    8.609  & 0.020  & magnitude  & \cite[VI/110/inescat, LWR03017LL]{Wamsteker2000_INES}\\ 
   IUE box & $2500-3000$\,\AA\  &    9.047  & 0.020  & magnitude  & \cite[VI/110/inescat, LWR03017LL]{Wamsteker2000_INES}\\ 
\textit{Gaia}       &     G       &   11.6350 & 0.001  & magnitude  & \cite[I/345/gaia2]{Gaia2018_VizieR}           \\  
\textit{Gaia}       &   GBP       &   11.3913 & 0.007  & magnitude  & \cite[I/345/gaia2]{Gaia2018_VizieR}\\ 
\textit{Gaia}       &   GRP       &   11.9377 & 0.001  & magnitude  & \cite[I/345/gaia2]{Gaia2018_VizieR}\\ 
Johnson    &     V$-$I     &   $-0.322$  & 0.002  &     color  & \cite[J/AJ/133/768/table4]{Landolt2007_Johnson}\\ 
Johnson    &     R$-$I     &   $-0.181$  & 0.001  &     color  & \cite[J/AJ/133/768/table4]{Landolt2007_Johnson}\\ 
Johnson    &    V$-$R      &   $-0.141$  & 0.001  &     color  & \cite[J/AJ/133/768/table4]{Landolt2007_Johnson}\\ 
Johnson    &     B$-$V     &   $-0.291$  & 0.001  &     color  & \cite[J/AJ/133/768/table4]{Landolt2007_Johnson}\\ 
Johnson    &     U$-$B     &   $-1.196$  & 0.003  &     color  & \cite[J/AJ/133/768/table4]{Landolt2007_Johnson}\\ 
Johnson    &       V     &    11.673   & 0.002  & magnitude  & \cite[J/AJ/133/768/table4]{Landolt2007_Johnson}\\ 
\bottomrule 
\end{tabular}
\end{center}
\end{table*}
\begin{table*}
\setstretch{1.2}
\captionof{table}{Photometric data used for the SED-fit of \HD.}
\label{tab:data:HD:phot}
\vspace*{-5pt}
\begin{center}
\begin{tabular}{l l r r r c c}
\toprule
\toprule
    System & Passband & Magnitude & Uncertainty & Type & Reference\\ 
\midrule
     2MASS &         H & 10.816  &   0.028 &  magnitude  & \cite[2MASS: II/246/out]{Cutri2003_2MASS}\\ 
     2MASS &         J & 10.641  &   0.023 &  magnitude  & \cite[2MASS: II/246/out]{Cutri2003_2MASS}\\ 
     2MASS &         K & 10.907  &   0.025 &  magnitude  & \cite[2MASS: II/246/out]{Cutri2003_2MASS}\\ 
   Johnson &       B$-$V & -0.234  &       &      color  & \cite[II/168/ubvmeans]{Mermilliod2006_UVB}\\ 
   Johnson &       U$-$B & -1.17   &       &      color  & \cite[II/168/ubvmeans]{Mermilliod2006_UVB}\\ 
   Johnson &         V & 10.05   &       &  magnitude  & \cite[II/168/ubvmeans]{Mermilliod2006_UVB}\\ 
Stroemgren &   H$_\beta$   &  2.57   &       &      color  & \cite[J/A+A/580/A23/catalog]{Paunzen2015_Stromgen}\\ 
Stroemgren &       b$-$y & -0.114  &   0.003 &      color  & \cite[J/A+A/580/A23/catalog]{Paunzen2015_Stromgen}\\ 
Stroemgren &        c1 & -0.214  &   0.013 &      color  & \cite[J/A+A/580/A23/catalog]{Paunzen2015_Stromgen}\\ 
Stroemgren &        m1 &  0.048  &   0.002 &      color  & \cite[J/A+A/580/A23/catalog]{Paunzen2015_Stromgen}\\ 
Stroemgren &         y & 10.035  &   0.009 &  magnitude  & \cite[J/A+A/580/A23/catalog]{Paunzen2015_Stromgen}\\ 
      WISE &        W1 & 10.954  &   0.023 &  magnitude  & \cite[AllWISE: II/328/allwise]{Cutri2012_WISE}\\ 
      WISE &        W2 & 11.045  &   0.021 &  magnitude  & \cite[AllWISE: II/328/allwise]{Cutri2012_WISE}\\ 
   IUE box & 1300-1800 &  6.306  &    0.02 &  magnitude  & \cite[VI/110/inescat, SWP08275LL]{Wamsteker2000_INES}\\ 
   IUE box & 2000-2500 &  7.164  &    0.02 &  magnitude  & \cite[VI/110/inescat, LWR07210LL]{Wamsteker2000_INES}\\ 
   IUE box & 2500-3000 &  7.547  &    0.02 &  magnitude  & \cite[VI/110/inescat, LWR07210LL]{Wamsteker2000_INES}\\ 
\textit{Gaia}   G   &           &  9.9636 & 0.0011  &  Magnitude  & \cite[I/345/gaia2]{Gaia2018_VizieR}\\ 
\textit{Gaia}   GBP &           &  9.8227 & 0.0038  &  Magnitude  & \cite[I/345/gaia2]{Gaia2018_VizieR}\\ 
\textit{Gaia}   GRP &           & 10.2446 & 0.0015  &  Magnitude  & \cite[I/345/gaia2]{Gaia2018_VizieR}\\ 
   Johnson &       B$-$V & -0.258  &       &    color    & \cite{Menzies1990_Johnson}\\ 
   Johnson &       U$-$B & -1.165  &       &    color    & \cite{Menzies1990_Johnson}\\ 
   Johnson &         V & 10.01   &       & magnitude   & \cite{Menzies1990_Johnson}\\ 
   Johnson &      B$-$V & -0.269  &       &    color    & \cite{Kilkenny1998}\\ 
   Johnson &       U$-$B & -1.184  &       &    color    & \cite{Kilkenny1998}\\ 
   Johnson &       V$-$R & -0.115  &       &    color    & \cite{Kilkenny1998}\\ 
   Johnson &       V$-$I & -0.276  &       &    color    & \cite{Kilkenny1998}\\ 
   Johnson &         V & 10.039  &       & magnitude   & \cite{Kilkenny1998}\\ 
\bottomrule 
\end{tabular}
\end{center}
\end{table*}
%
%
\subsection{Comparison with literature}
\label{sect:abu:literature}
\begin{figure*}
\begin{center}
\includegraphics[width=1\textwidth]{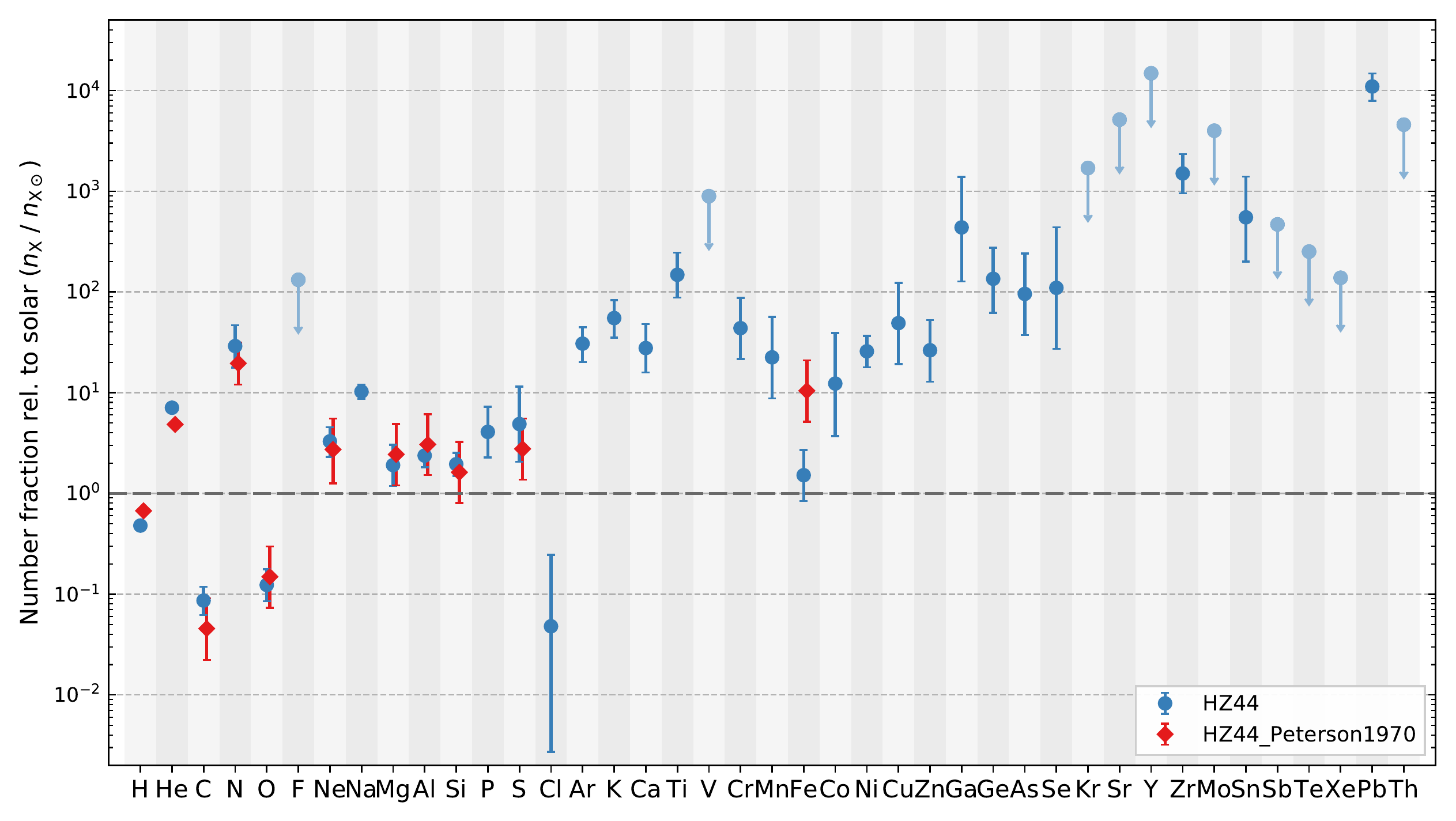}
\includegraphics[width=1\textwidth]{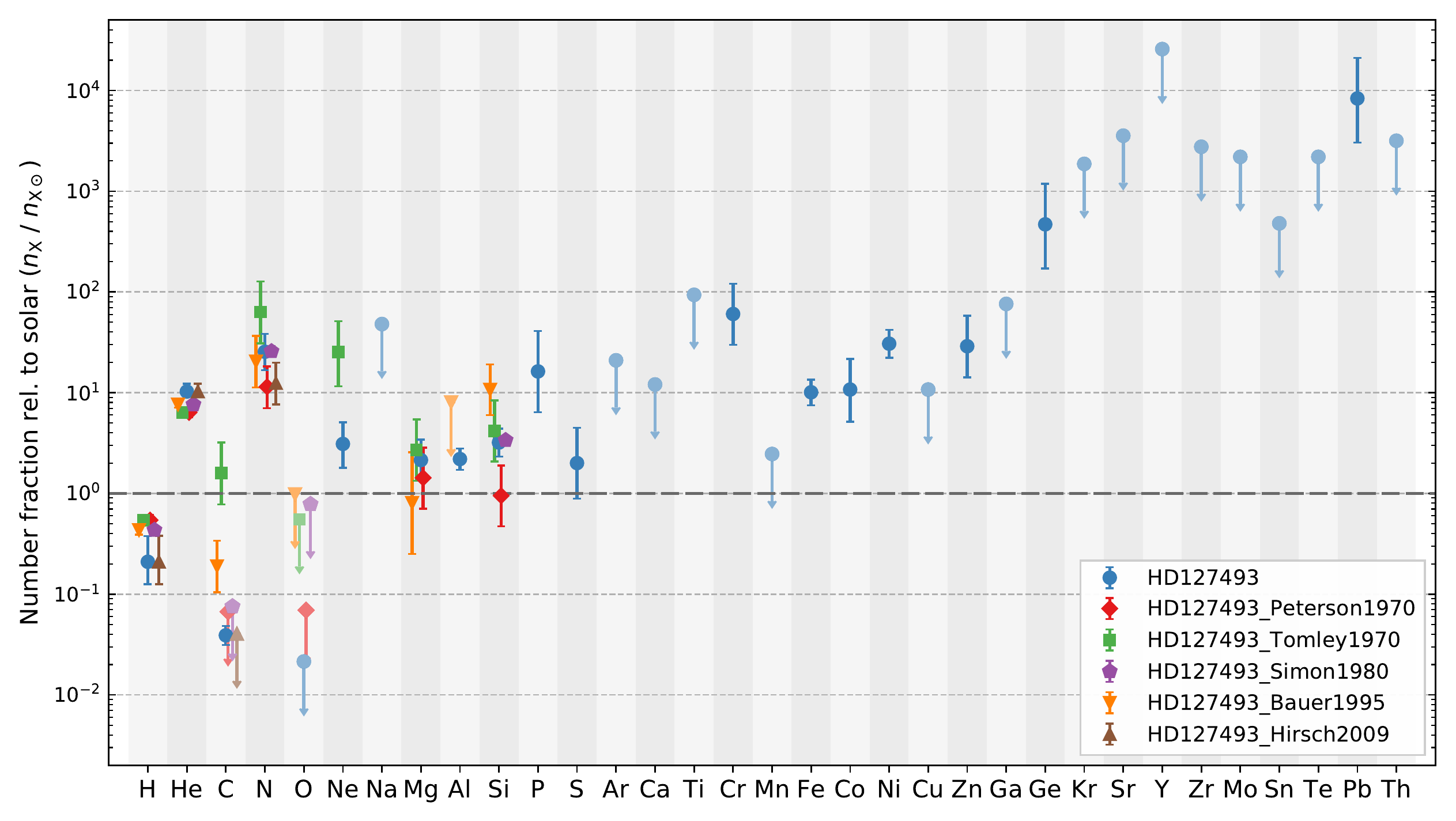}
\begin{minipage}{0.95\textwidth}
\vspace{5pt}
\captionof{figure}{Same as Fig.~\ref{fig:abupattern} but for the comparison of abundances derived in this paper with literature values.}
\label{f:compare_literature}
\end{minipage}
\end{center}
\end{figure*}

Figure \ref{f:compare_literature} shows the comparison of abundances determined in this paper with literature values for \HZ\ and \HD. 
\smallskip\\
The only previous metal analysis of \HZ\ was performed by \cite{Peterson1970} using the curve-of-growth method. 
Their results for C, N, O, Ne, Mg, Al, Si, and S are consistent with the values presented in this paper considering 1-$\sigma$ uncertainties. Only their H abundance (based on early ATLAS model atmospheres) and Fe abundance (based on three weak optical Fe\,\textsc{iii} lines with at the time uncertain oscillator strengths) are overestimated compared to ours.
\smallskip\\
\cite{Peterson1970} also performed a curve-of-growth analysis of optical spectra for \HD\ (including C, N, Mg, and Si) which agrees well with the abundances derived here.
A similar analysis was performed by \cite{Tomley1970}; his abundance results for C and Ne are higher by about 1\,dex while the abundances of N, Mg, and Si match within the respective uncertainties.
C and Si abundance determinations from early NLTE models by \cite{bauer95} are higher by $\sim$0.5\,dex whereas their N and Mg abundances match well.
The C and N abundances derived by \cite{hirsch09} are consistent with our results.

%

%
\subsection{Abundances and stellar spectra}
\label{sect:fullspectra}

\begin{table*}
\captionof{table}{
Abundances of \HZ\ and \HD\ as derived from visual/UVA and FUV data. }
\label{tab:abundances_units}
\vspace{-9pt}
\setstretch{1.22}
\begin{center}
\begin{tabular}{lrrrrrrrr}
\toprule
\toprule
 \hspace{-2pt} & \multicolumn{2}{c}{Abundance ($\log n_\mathrm{X}/n_\mathrm{H}$)} & \multicolumn{2}{c}{Mass fraction ($\beta _\mathrm{X}$)}& \multicolumn{2}{c}{Number fraction ($\log n _\mathrm{X}$)} & \multicolumn{2}{c}{Abundance ($\log n _\mathrm{X} / n _\mathrm{X,\odot}$)}\\[0pt]
Element        \hspace{-10pt} & \small{\HZ} & \small{\HD}  & \small{\HZ} & \small{\HD} & \small{\HZ} & \small{\HD} & \small{\HZ} & \small{\HD} \\
\midrule
H  &      0.00$^{+0.00}_{-0.00}$ &      0.00$^{+0.00}_{-0.00}$ &     -0.79$^{+0.03}_{-0.03}$ &     -1.26$^{+0.25}_{-0.21}$ &     -0.36$^{+0.03}_{-0.03}$ &     -0.71$^{+0.26}_{-0.22}$ &                            -0.32$^{+0.03}_{-0.03}$ &                            -0.68$^{+0.26}_{-0.22}$ \\
He &      0.10$^{+0.05}_{-0.05}$ &      0.62$^{+0.30}_{-0.30}$ &     -0.09$^{+0.01}_{-0.01}$ &     -0.03$^{+0.02}_{-0.01}$ &     -0.26$^{+0.02}_{-0.02}$ &     -0.09$^{+0.08}_{-0.04}$ &                             0.85$^{+0.02}_{-0.02}$ &                             1.01$^{+0.08}_{-0.05}$ \\
C  &     -4.31$^{+0.13}_{-0.13}$ &     -4.30$^{+0.08}_{-0.08}$ &     -4.02$^{+0.13}_{-0.13}$ &     -4.48$^{+0.11}_{-0.09}$ &     -4.67$^{+0.13}_{-0.13}$ &     -5.01$^{+0.08}_{-0.08}$ &                            -1.06$^{+0.14}_{-0.14}$ &                            -1.41$^{+0.09}_{-0.10}$ \\
N  &     -2.39$^{+0.20}_{-0.20}$ &     -2.08$^{+0.17}_{-0.17}$ &     -2.03$^{+0.20}_{-0.20}$ &     -2.19$^{+0.18}_{-0.18}$ &     -2.74$^{+0.20}_{-0.20}$ &     -2.80$^{+0.17}_{-0.17}$ &                             1.46$^{+0.21}_{-0.21}$ &                             1.41$^{+0.18}_{-0.18}$ \\
O  &     -3.90$^{+0.15}_{-0.15}$ &  <-4.30$^{+0.20}_{}$ &     -3.49$^{+0.15}_{-0.15}$ &  <-4.35$^{+0.21}_{}$ &     -4.26$^{+0.15}_{-0.15}$ &  <-5.01$^{+0.20}_{}$ &                            -0.91$^{+0.16}_{-0.16}$ &                         <-1.67$^{+0.20}_{}$ \\
F  &  <-5.00$^{+0.40}_{}$ &                             &  <-4.51$^{+0.40}_{}$ &                             &  <-5.36$^{+0.40}_{}$ &                             &                          <2.12$^{+0.45}_{}$ &                                                    \\
Ne &     -3.23$^{+0.11}_{-0.11}$ &     -2.90$^{+0.20}_{-0.20}$ &     -2.72$^{+0.11}_{-0.11}$ &     -2.85$^{+0.21}_{-0.21}$ &     -3.59$^{+0.11}_{-0.11}$ &     -3.61$^{+0.20}_{-0.20}$ &                             0.52$^{+0.14}_{-0.15}$ &                             0.49$^{+0.21}_{-0.24}$ \\
Na &     -4.43$^{+0.06}_{-0.06}$ &  <-3.40$^{+0.30}_{}$ &     -3.86$^{+0.06}_{-0.06}$ &  <-3.29$^{+0.30}_{}$ &     -4.79$^{+0.06}_{-0.06}$ &  <-4.11$^{+0.30}_{}$ &                             1.01$^{+0.07}_{-0.07}$ &                          <1.68$^{+0.30}_{}$ \\
Mg &     -3.80$^{+0.20}_{-0.20}$ &     -3.39$^{+0.20}_{-0.20}$ &     -3.20$^{+0.20}_{-0.20}$ &     -3.26$^{+0.21}_{-0.21}$ &     -4.16$^{+0.20}_{-0.20}$ &     -4.10$^{+0.20}_{-0.20}$ &                             0.28$^{+0.20}_{-0.21}$ &                             0.33$^{+0.20}_{-0.21}$ \\
Al &     -4.86$^{+0.11}_{-0.11}$ &     -4.53$^{+0.10}_{-0.10}$ &     -4.21$^{+0.11}_{-0.11}$ &     -4.35$^{+0.12}_{-0.11}$ &     -5.21$^{+0.11}_{-0.11}$ &     -5.24$^{+0.10}_{-0.10}$ &                             0.38$^{+0.11}_{-0.11}$ &                             0.34$^{+0.10}_{-0.11}$ \\
Si &     -3.88$^{+0.11}_{-0.11}$ &     -3.31$^{+0.14}_{-0.14}$ &     -3.22$^{+0.11}_{-0.11}$ &     -3.11$^{+0.15}_{-0.14}$ &     -4.24$^{+0.11}_{-0.11}$ &     -4.02$^{+0.14}_{-0.13}$ &                             0.29$^{+0.11}_{-0.12}$ &                             0.51$^{+0.14}_{-0.14}$ \\
P  &     -5.66$^{+0.25}_{-0.25}$ &     -4.70$^{+0.40}_{-0.40}$ &     -4.96$^{+0.25}_{-0.25}$ &     -4.46$^{+0.40}_{-0.41}$ &     -6.02$^{+0.25}_{-0.25}$ &     -5.41$^{+0.40}_{-0.40}$ &                             0.61$^{+0.25}_{-0.25}$ &                             1.21$^{+0.40}_{-0.40}$ \\
S  &     -3.87$^{+0.37}_{-0.37}$ &     -3.90$^{+0.35}_{-0.35}$ &     -3.16$^{+0.37}_{-0.37}$ &     -3.65$^{+0.35}_{-0.36}$ &     -4.23$^{+0.37}_{-0.37}$ &     -4.61$^{+0.35}_{-0.35}$ &                             0.69$^{+0.37}_{-0.38}$ &                             0.30$^{+0.35}_{-0.35}$ \\
Cl &     -7.50$^{+0.70}_{-0.70}$ &                             &     -6.74$^{+0.70}_{-0.70}$ &                             &     -7.86$^{+0.70}_{-0.70}$ &                             &                            -1.32$^{+0.71}_{-1.25}$ &                                                    \\
Ar &     -3.79$^{+0.11}_{-0.11}$ &  <-3.60$^{+0.20}_{}$ &     -2.98$^{+0.11}_{-0.11}$ &  <-3.25$^{+0.21}_{}$ &     -4.15$^{+0.11}_{-0.11}$ &  <-4.31$^{+0.20}_{}$ &                             1.49$^{+0.16}_{-0.18}$ &                          <1.32$^{+0.23}_{}$ \\
K  &     -4.91$^{+0.16}_{-0.16}$ &                             &     -4.11$^{+0.16}_{-0.16}$ &                             &     -5.27$^{+0.16}_{-0.16}$ &                             &                             1.74$^{+0.18}_{-0.19}$ &                                                    \\
Ca &     -3.90$^{+0.24}_{-0.24}$ &  <-3.90$^{+0.20}_{}$ &     -3.08$^{+0.24}_{-0.24}$ &  <-3.55$^{+0.21}_{}$ &     -4.25$^{+0.24}_{-0.24}$ &  <-4.61$^{+0.20}_{}$ &                             1.44$^{+0.24}_{-0.24}$ &                          <1.08$^{+0.20}_{}$ \\
Ti &     -4.56$^{+0.22}_{-0.22}$ &  <-4.40$^{+0.25}_{}$ &     -3.67$^{+0.22}_{-0.22}$ &  <-3.98$^{+0.26}_{}$ &     -4.92$^{+0.22}_{-0.22}$ &  <-5.11$^{+0.25}_{}$ &                             2.17$^{+0.22}_{-0.23}$ &                          <1.97$^{+0.25}_{}$ \\
V  &  <-4.80$^{+0.40}_{}$ &                             &  <-3.88$^{+0.40}_{}$ &                             &  <-5.16$^{+0.40}_{}$ &                             &                          <2.95$^{+0.40}_{}$ &                                                    \\
Cr &     -4.40$^{+0.30}_{-0.30}$ &     -3.90$^{+0.30}_{-0.30}$ &     -3.47$^{+0.30}_{-0.30}$ &     -3.44$^{+0.30}_{-0.31}$ &     -4.76$^{+0.30}_{-0.30}$ &     -4.61$^{+0.30}_{-0.30}$ &                             1.64$^{+0.30}_{-0.31}$ &                             1.78$^{+0.30}_{-0.31}$ \\
Mn &     -4.90$^{+0.40}_{-0.40}$ &  <-5.50$^{+0.30}_{}$ &     -3.95$^{+0.40}_{-0.40}$ &  <-5.02$^{+0.30}_{}$ &     -5.26$^{+0.40}_{-0.40}$ &  <-6.21$^{+0.30}_{}$ &                             1.35$^{+0.40}_{-0.41}$ &                          <0.39$^{+0.30}_{}$ \\
Fe &     -4.00$^{+0.25}_{-0.25}$ &     -2.82$^{+0.12}_{-0.12}$ &     -3.04$^{+0.25}_{-0.25}$ &     -2.33$^{+0.14}_{-0.13}$ &     -4.36$^{+0.25}_{-0.25}$ &     -3.53$^{+0.12}_{-0.12}$ &                             0.18$^{+0.25}_{-0.26}$ &                             1.00$^{+0.12}_{-0.13}$ \\
Co &     -5.60$^{+0.50}_{-0.50}$ &     -5.30$^{+0.30}_{-0.30}$ &     -4.62$^{+0.50}_{-0.50}$ &     -4.78$^{+0.30}_{-0.31}$ &     -5.96$^{+0.50}_{-0.50}$ &     -6.01$^{+0.30}_{-0.30}$ &                             1.09$^{+0.50}_{-0.52}$ &                             1.03$^{+0.30}_{-0.32}$ \\
Ni &     -4.05$^{+0.15}_{-0.15}$ &     -3.61$^{+0.13}_{-0.13}$ &     -3.07$^{+0.15}_{-0.15}$ &     -3.10$^{+0.15}_{-0.14}$ &     -4.41$^{+0.15}_{-0.15}$ &     -4.33$^{+0.13}_{-0.13}$ &                             1.41$^{+0.15}_{-0.16}$ &                             1.49$^{+0.14}_{-0.14}$ \\
Cu &     -5.80$^{+0.40}_{-0.40}$ &  <-6.10$^{+0.40}_{}$ &     -4.79$^{+0.40}_{-0.40}$ &  <-5.55$^{+0.40}_{}$ &     -6.16$^{+0.40}_{-0.40}$ &  <-6.81$^{+0.40}_{}$ &                             1.69$^{+0.40}_{-0.41}$ &                          <1.03$^{+0.40}_{}$ \\
Zn &     -5.70$^{+0.30}_{-0.30}$ &     -5.30$^{+0.30}_{-0.30}$ &     -4.67$^{+0.30}_{-0.30}$ &     -4.74$^{+0.30}_{-0.31}$ &     -6.06$^{+0.30}_{-0.30}$ &     -6.01$^{+0.30}_{-0.30}$ &                             1.42$^{+0.30}_{-0.31}$ &                             1.46$^{+0.30}_{-0.31}$ \\
Ga &     -6.00$^{+0.50}_{-0.50}$ &  <-6.40$^{+0.40}_{}$ &     -4.95$^{+0.50}_{-0.50}$ &  <-5.81$^{+0.40}_{}$ &     -6.36$^{+0.50}_{-0.50}$ &  <-7.11$^{+0.40}_{}$ &                             2.64$^{+0.50}_{-0.54}$ &                          <1.88$^{+0.40}_{}$ \\
Ge &     -5.90$^{+0.30}_{-0.30}$ &     -5.00$^{+0.40}_{-0.40}$ &     -4.83$^{+0.30}_{-0.30}$ &     -4.39$^{+0.40}_{-0.41}$ &     -6.26$^{+0.30}_{-0.30}$ &     -5.71$^{+0.40}_{-0.40}$ &                             2.13$^{+0.31}_{-0.34}$ &                             2.67$^{+0.40}_{-0.44}$ \\
As &     -7.40$^{+0.40}_{-0.40}$ &                             &     -6.31$^{+0.40}_{-0.40}$ &                             &     -7.76$^{+0.40}_{-0.40}$ &                             &                             1.98$^{+0.40}_{-0.41}$ &                                                    \\
Se &     -6.30$^{+0.60}_{-0.60}$ &                             &     -5.19$^{+0.60}_{-0.60}$ &                             &     -6.66$^{+0.60}_{-0.60}$ &                             &                             2.04$^{+0.60}_{-0.61}$ &                                                    \\
Kr &  <-5.20$^{+0.60}_{}$ &  <-4.80$^{+0.40}_{}$ &  <-4.07$^{+0.60}_{}$ &  <-4.13$^{+0.40}_{}$ &  <-5.56$^{+0.60}_{}$ &  <-5.51$^{+0.40}_{}$ &                          <3.23$^{+0.60}_{}$ &                          <3.27$^{+0.40}_{}$ \\
Sr &  <-5.10$^{+0.60}_{}$ &  <-4.90$^{+0.30}_{}$ &  <-3.95$^{+0.60}_{}$ &  <-4.21$^{+0.30}_{}$ &  <-5.46$^{+0.60}_{}$ &  <-5.61$^{+0.30}_{}$ &                          <3.71$^{+0.60}_{}$ &                          <3.55$^{+0.30}_{}$ \\
Y  &  <-5.30$^{+0.20}_{}$ &  <-4.70$^{+0.30}_{}$ &  <-4.14$^{+0.20}_{}$ &  <-4.01$^{+0.30}_{}$ &  <-5.66$^{+0.20}_{}$ &  <-5.41$^{+0.30}_{}$ &                          <4.17$^{+0.20}_{}$ &                          <4.41$^{+0.30}_{}$ \\
Zr &     -5.92$^{+0.19}_{-0.19}$ &  <-5.30$^{+0.20}_{}$ &     -4.75$^{+0.19}_{-0.19}$ &  <-4.60$^{+0.21}_{}$ &     -6.28$^{+0.19}_{-0.19}$ &  <-6.01$^{+0.20}_{}$ &                             3.18$^{+0.19}_{-0.20}$ &                          <3.44$^{+0.20}_{}$ \\
Mo &  <-6.20$^{+0.40}_{}$ &  <-6.10$^{+0.40}_{}$ &  <-5.01$^{+0.40}_{}$ &  <-5.37$^{+0.40}_{}$ &  <-6.56$^{+0.40}_{}$ &  <-6.81$^{+0.40}_{}$ &                          <3.60$^{+0.40}_{}$ &                          <3.34$^{+0.40}_{}$ \\
Sn &     -6.90$^{+0.40}_{-0.40}$ &  <-6.60$^{+0.40}_{}$ &     -5.61$^{+0.40}_{-0.40}$ &  <-5.78$^{+0.40}_{}$ &     -7.26$^{+0.40}_{-0.40}$ &  <-7.31$^{+0.40}_{}$ &                             2.74$^{+0.40}_{-0.44}$ &                          <2.68$^{+0.40}_{}$ \\
Sb &  <-8.00$^{+0.50}_{}$ &                             &  <-6.70$^{+0.50}_{}$ &                             &  <-8.36$^{+0.50}_{}$ &                             &                          <2.67$^{+0.50}_{}$ &                                                    \\
Te &  <-7.10$^{+0.40}_{}$ &  <-5.80$^{+0.40}_{}$ &  <-5.78$^{+0.40}_{}$ &  <-4.95$^{+0.40}_{}$ &  <-7.46$^{+0.40}_{}$ &  <-6.51$^{+0.40}_{}$ &                          <2.40$^{+0.40}_{}$ &                          <3.34$^{+0.40}_{}$ \\
Xe &  <-7.30$^{+0.40}_{}$ &                             &  <-5.97$^{+0.40}_{}$ &                             &  <-7.66$^{+0.40}_{}$ &                             &                          <2.14$^{+0.40}_{}$ &                                                    \\
Pb &     -5.89$^{+0.09}_{-0.09}$ &     -5.65$^{+0.40}_{-0.40}$ &     -4.36$^{+0.09}_{-0.09}$ &     -4.59$^{+0.40}_{-0.41}$ &     -6.25$^{+0.09}_{-0.09}$ &     -6.36$^{+0.40}_{-0.40}$ &                             4.04$^{+0.13}_{-0.14}$ &                             3.92$^{+0.40}_{-0.44}$ \\
Th &  <-8.00$^{+0.30}_{}$ &  <-7.80$^{+0.30}_{}$ &  <-6.42$^{+0.30}_{}$ &  <-6.69$^{+0.30}_{}$ &  <-8.36$^{+0.30}_{}$ &  <-8.51$^{+0.30}_{}$ &                          <3.66$^{+0.31}_{}$ &                          <3.50$^{+0.31}_{}$ \\
\bottomrule
\end{tabular}
\end{center}
\vspace{-12pt}
\tablefoot{Abundances are given as logarithmic number ratio of element X relative to hydrogen $\log n_\mathrm{X}/n_\mathrm{H}$, logarithmic mass fraction $\beta _\mathrm{X}$, logarithmic number fraction $\log n _\mathrm{X}$, and logarithmic number fraction relative to solar values $\log n _\mathrm{X} / n _{\mathrm{X,}\odot}$. Uncertainties are given as standard deviation between single line fits. If an  abundance was ``fit by eye'' the uncertainties are similarly estimated. The He abundance for \HD\ is from \cite{hirsch09}.}
\end{table*}

This section presents our final abundance values (Table \ref{tab:abundances_units}) as well as a comparison between the full observed and final synthetic spectra of \HZ\ and \HD\ (Fig. \ref{fig:full_spectra1} to Fig. \ref{fig:full_spectra_last}). In the synthetic spectra, elements with upper limits only are included at their upper limit.
The synthetic spectra are convolved with a Gaussian kernel (constant for GHRS, but wavelength-depended for all echelle spectrographs) to match the resolution of the respective spectrograph.
The strongest photospheric metal lines are labeled with magenta marks, interstellar lines are labeled with green marks. At the bottom of each spectral range we also show the residual between the observation and our final model.
A proper normalization of the HIRES spectrum of \HZ\ was only possible using the final synthetic spectrum as a template. Thus the shape of broad hydrogen and helium lines in the HIRES spectra is adjusted during the normalization procedure to fit the shape of the synthetic spectrum.
However, the shape of the sharp metal lines, that are of interest in the HIRES spectra, are not affected by the normalization procedure.
The optical spectra of \HZ\ and \HD\ are shown up to 6710\,\AA\ since the number of metal lines at longer wavelengths is very limited.

\captionsetup[ContinuedFloat]{labelformat=continued}
\begin{figure*}
\centering
\includegraphics[width=24.2cm,angle =90,page=1]{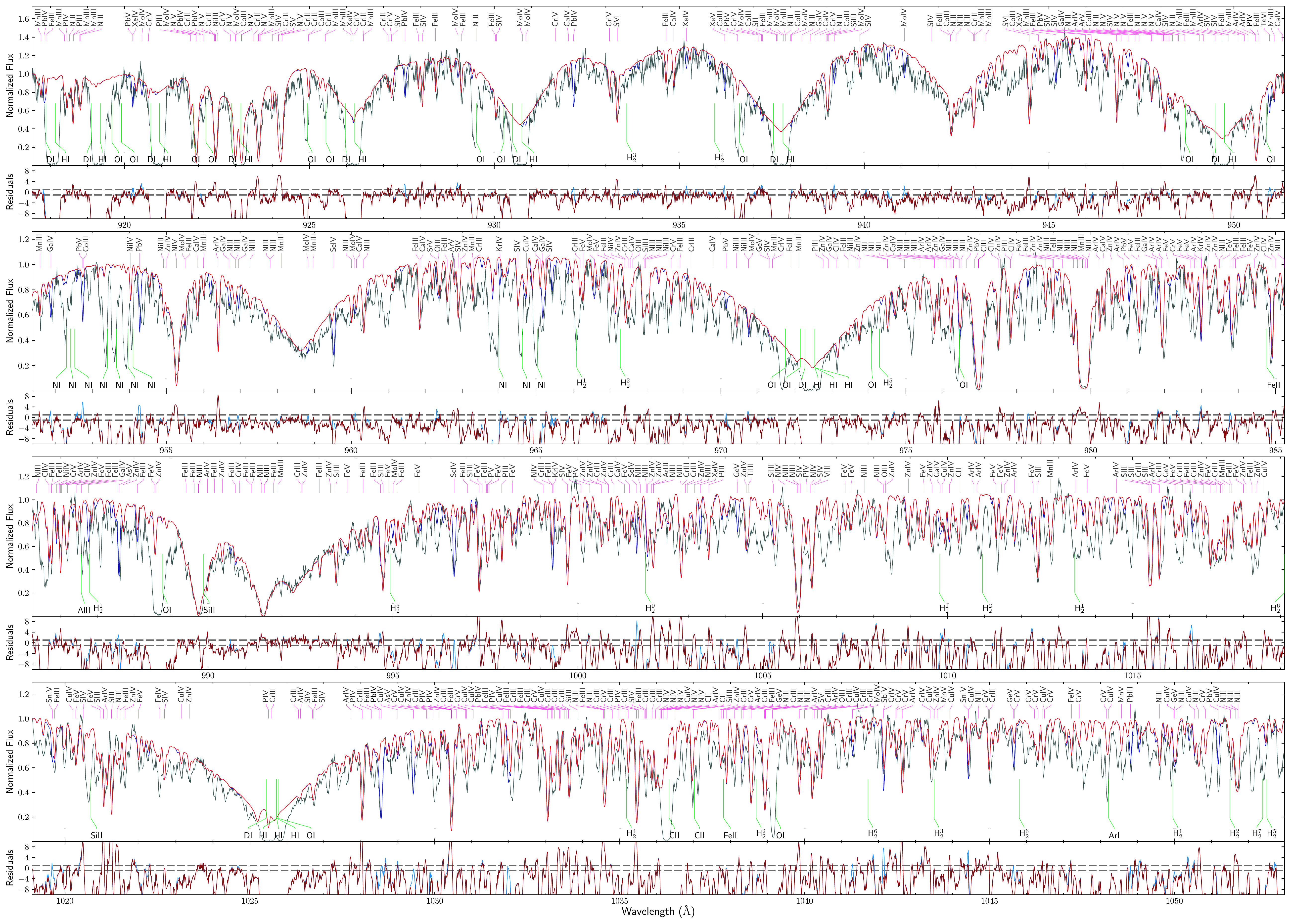}
\caption{FUSE spectrum of \HZ\ (gray) and the final model (red, with heavy metals: blue).}\label{fig:full_spectra1}
\end{figure*}
\begin{figure*}
\ContinuedFloat
\centering
\includegraphics[width=24.2cm,angle =90,page=2]{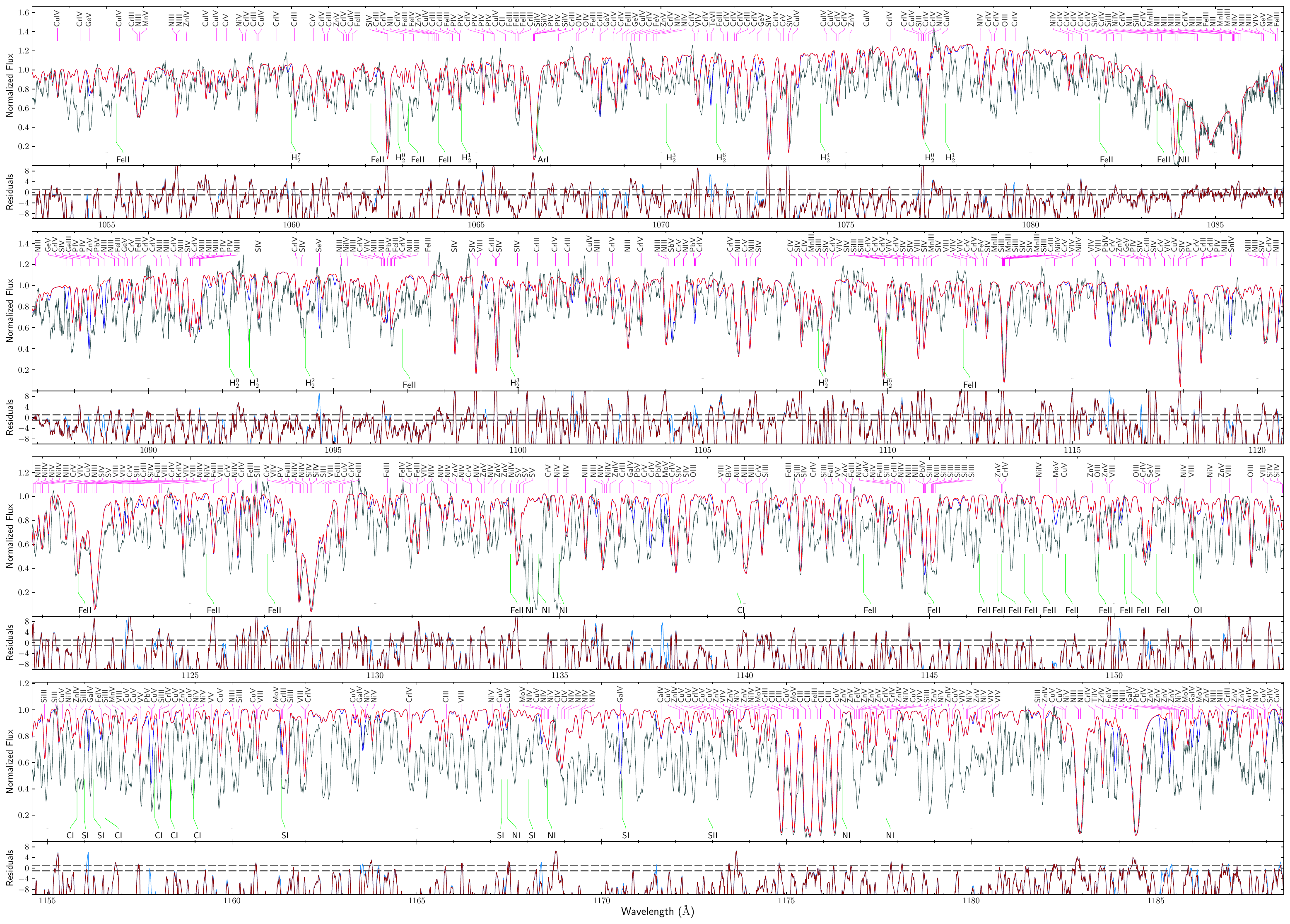}
\caption{FUSE spectrum of \HZ\ (gray) and the final model (red, with heavy metals: blue).}

\end{figure*}

\captionsetup[ContinuedFloat]{labelformat=continued}
\begin{figure*}
\centering
\includegraphics[width=24.2cm,angle =90,page=1]{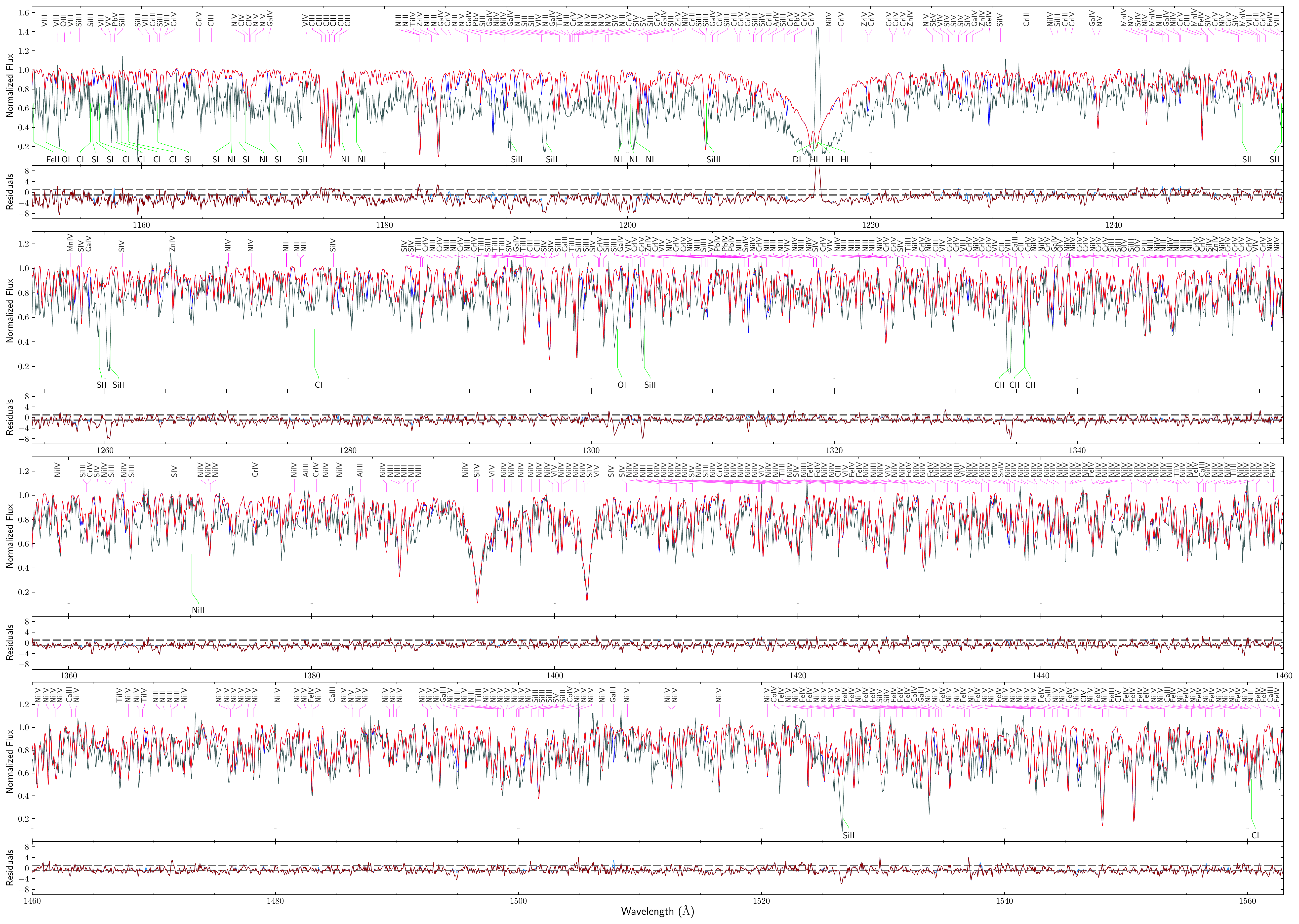}
\caption{IUE spectrum of \HZ\ (gray) and the final model (red, with heavy metals: blue).}
\end{figure*}
\begin{figure*}
\ContinuedFloat
\centering
\includegraphics[width=24.2cm,angle =90,page=2]{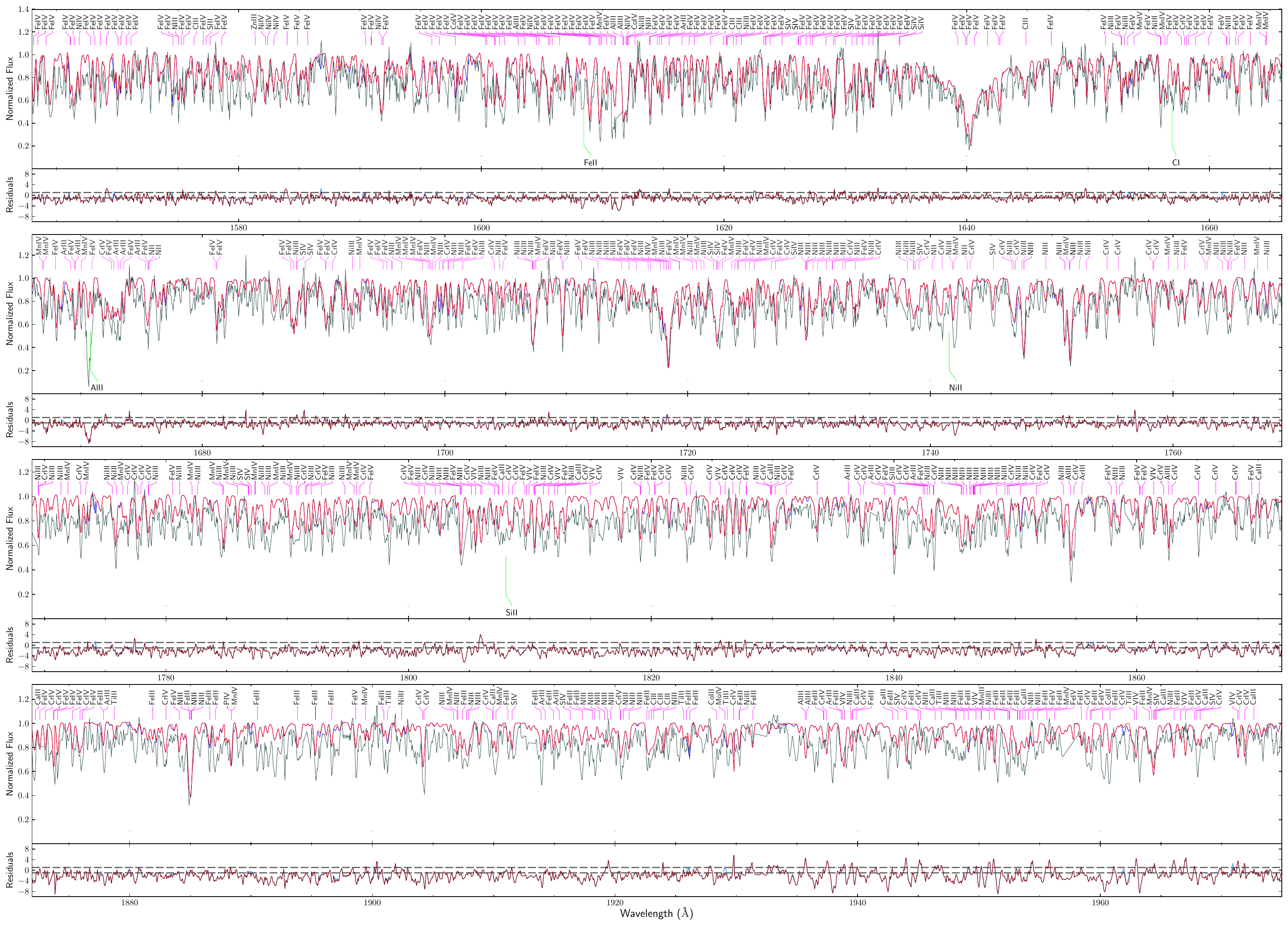}
\caption{IUE spectrum of \HZ\ (gray) and the final model (red, with heavy metals: blue).}
\end{figure*}
\captionsetup[ContinuedFloat]{labelformat=continued}
\begin{figure*}
\centering
\includegraphics[width=24.2cm,angle =90,page=1]{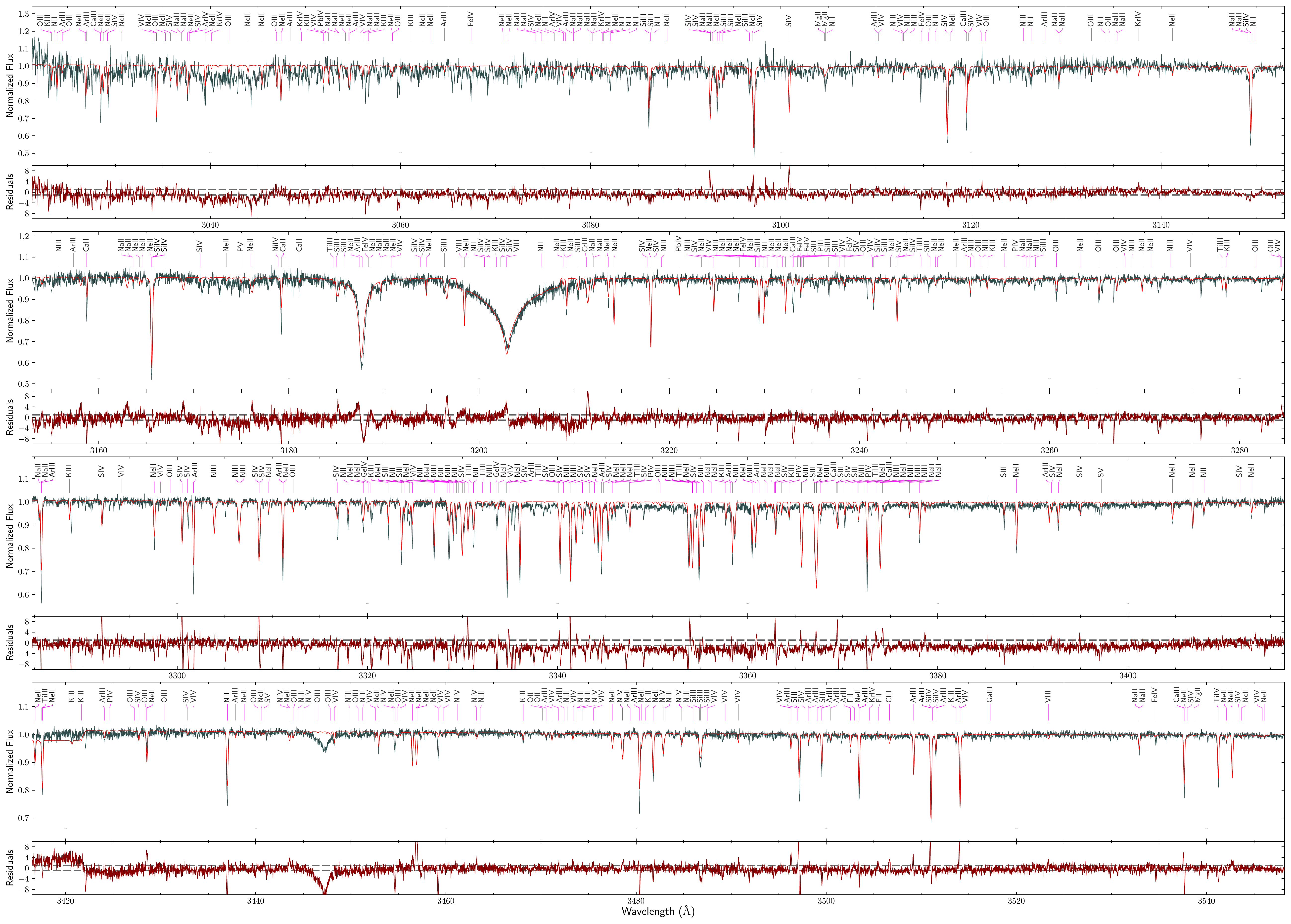}
\caption{HIRES spectrum of \HZ\ (gray) and the final model (red).}
\end{figure*}
\begin{figure*}
\ContinuedFloat
\centering
\includegraphics[width=24.2cm,angle =90,page=2]{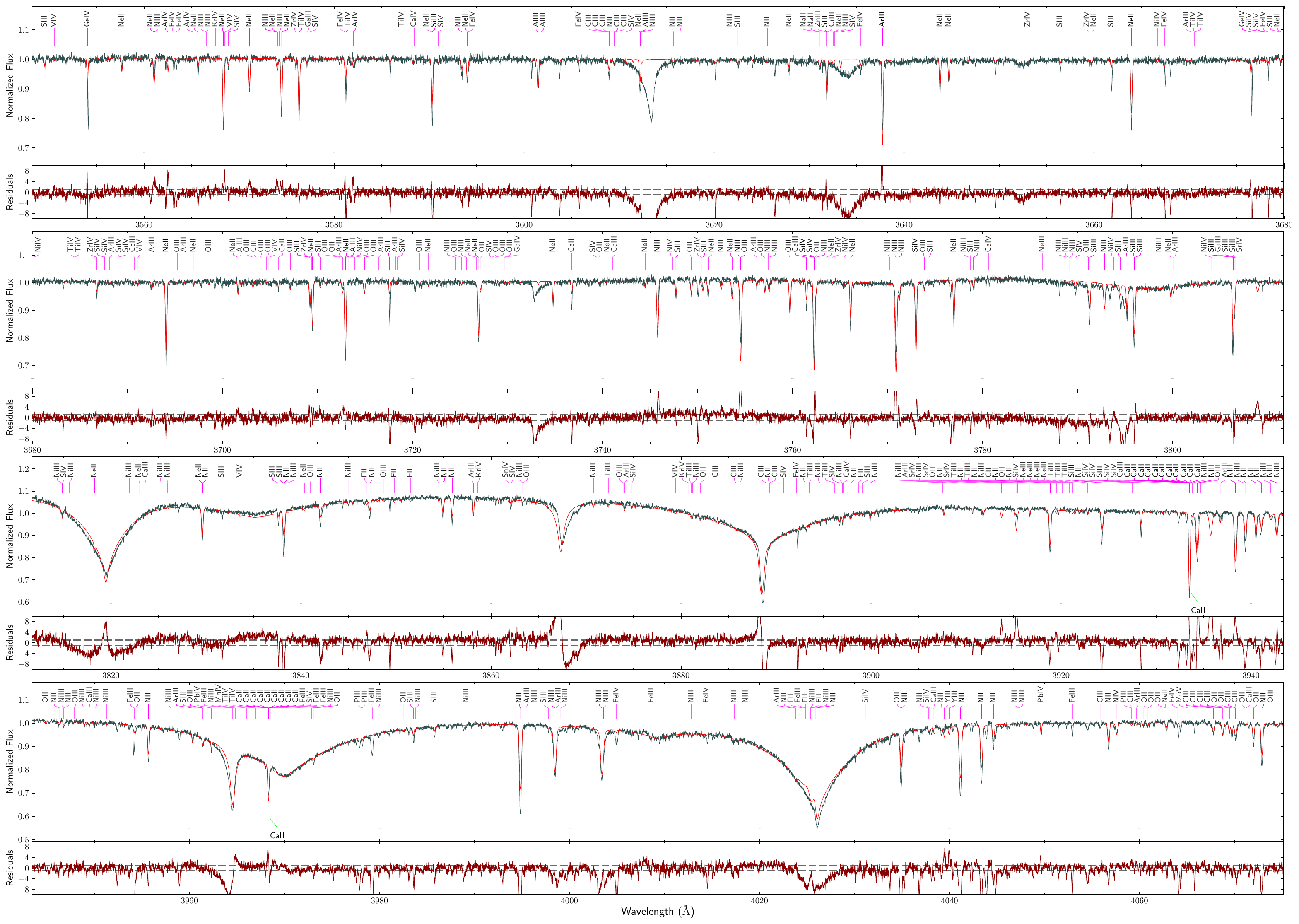}
\caption{HIRES spectrum of \HZ\ (gray) and the final model (red).}
\end{figure*}
\begin{figure*}
\ContinuedFloat
\centering
\includegraphics[width=24.2cm,angle =90,page=3]{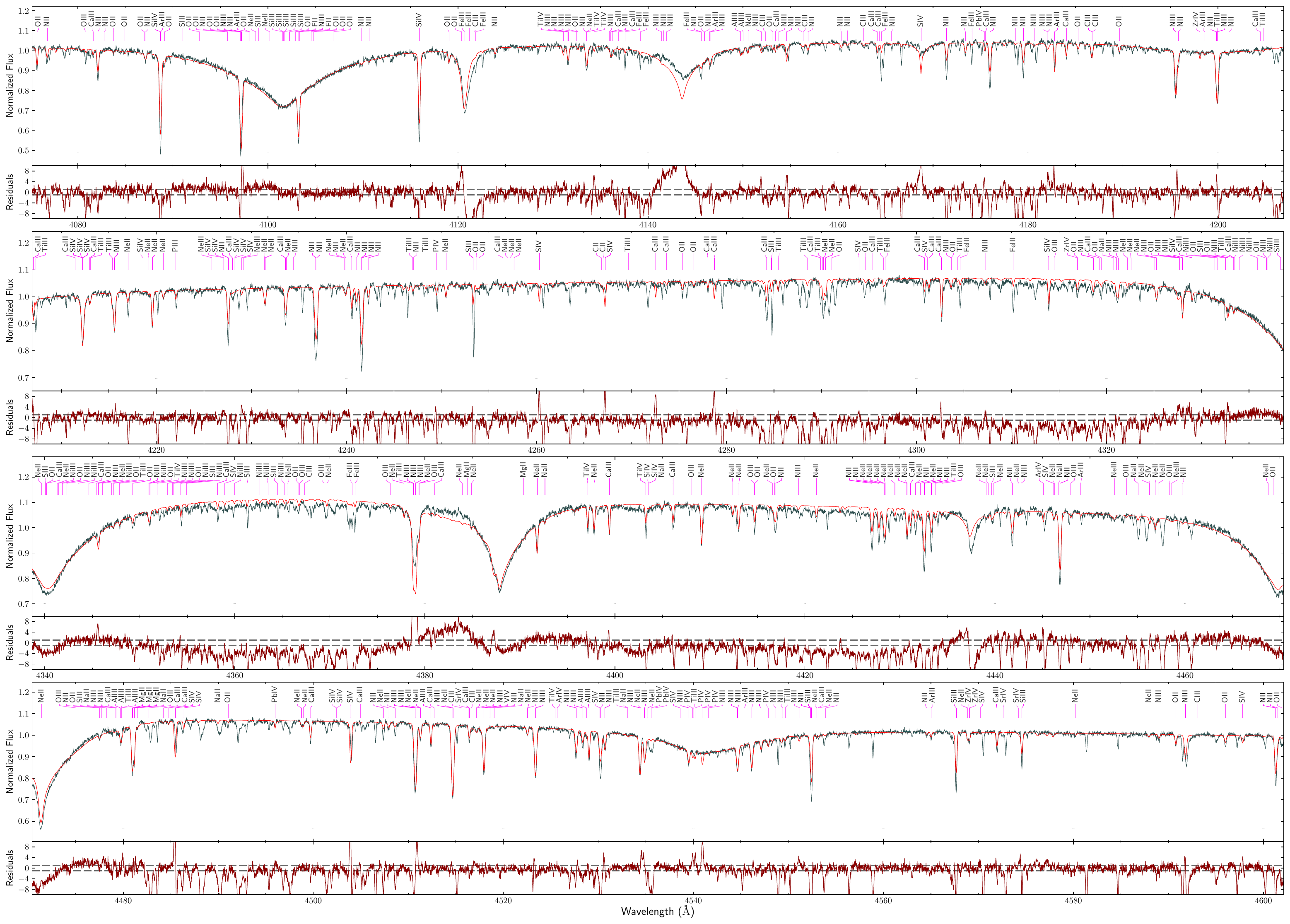}
\caption{HIRES spectrum of \HZ\ (gray) and the final model (red).}
\end{figure*}
\begin{figure*}
\ContinuedFloat
\centering
\includegraphics[width=24.2cm,angle =90,page=4]{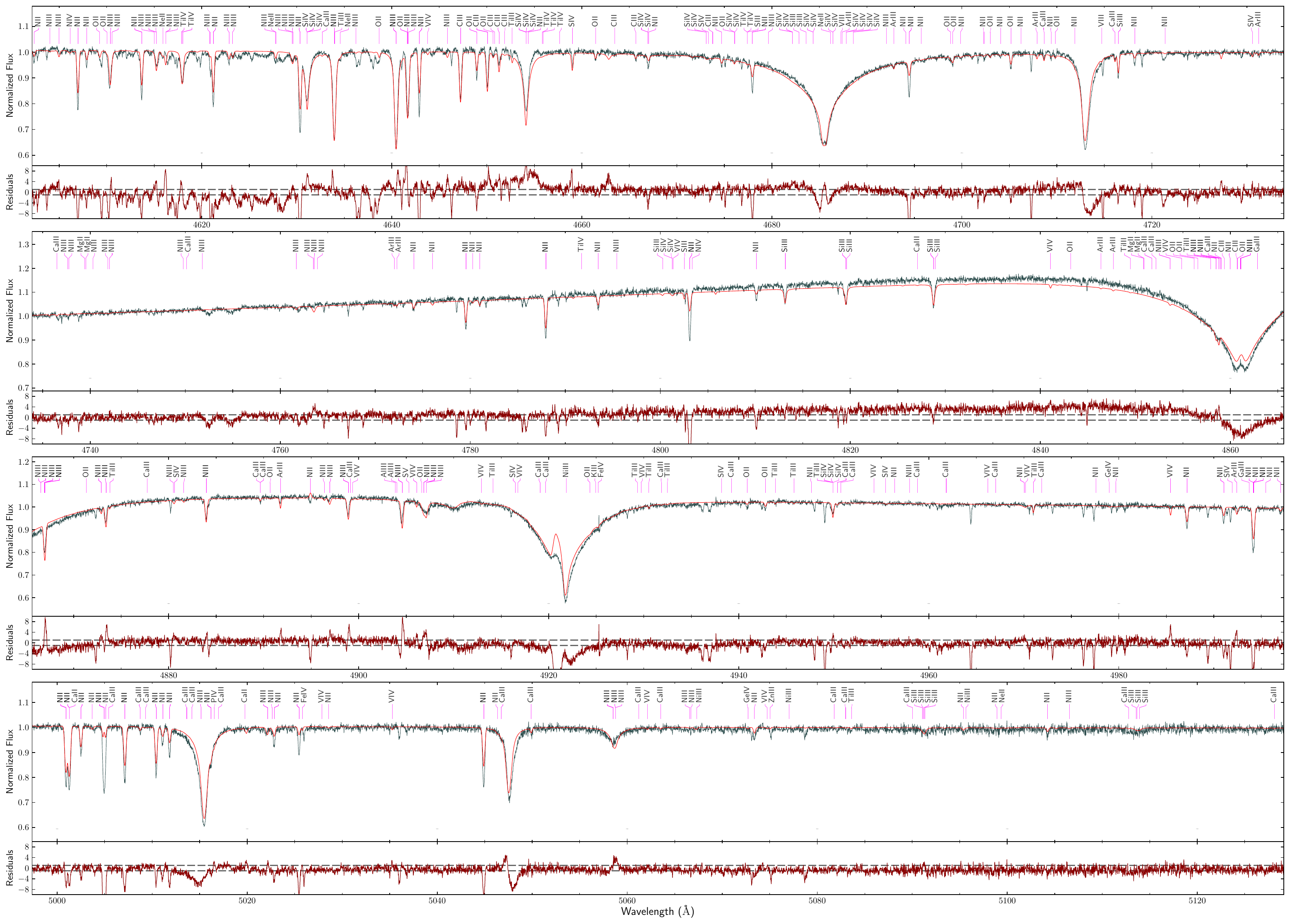}
\caption{HIRES spectrum of \HZ\ (gray) and the final model (red).}
\end{figure*}
\begin{figure*}
\ContinuedFloat
\centering
\includegraphics[width=24.2cm,angle =90,page=5]{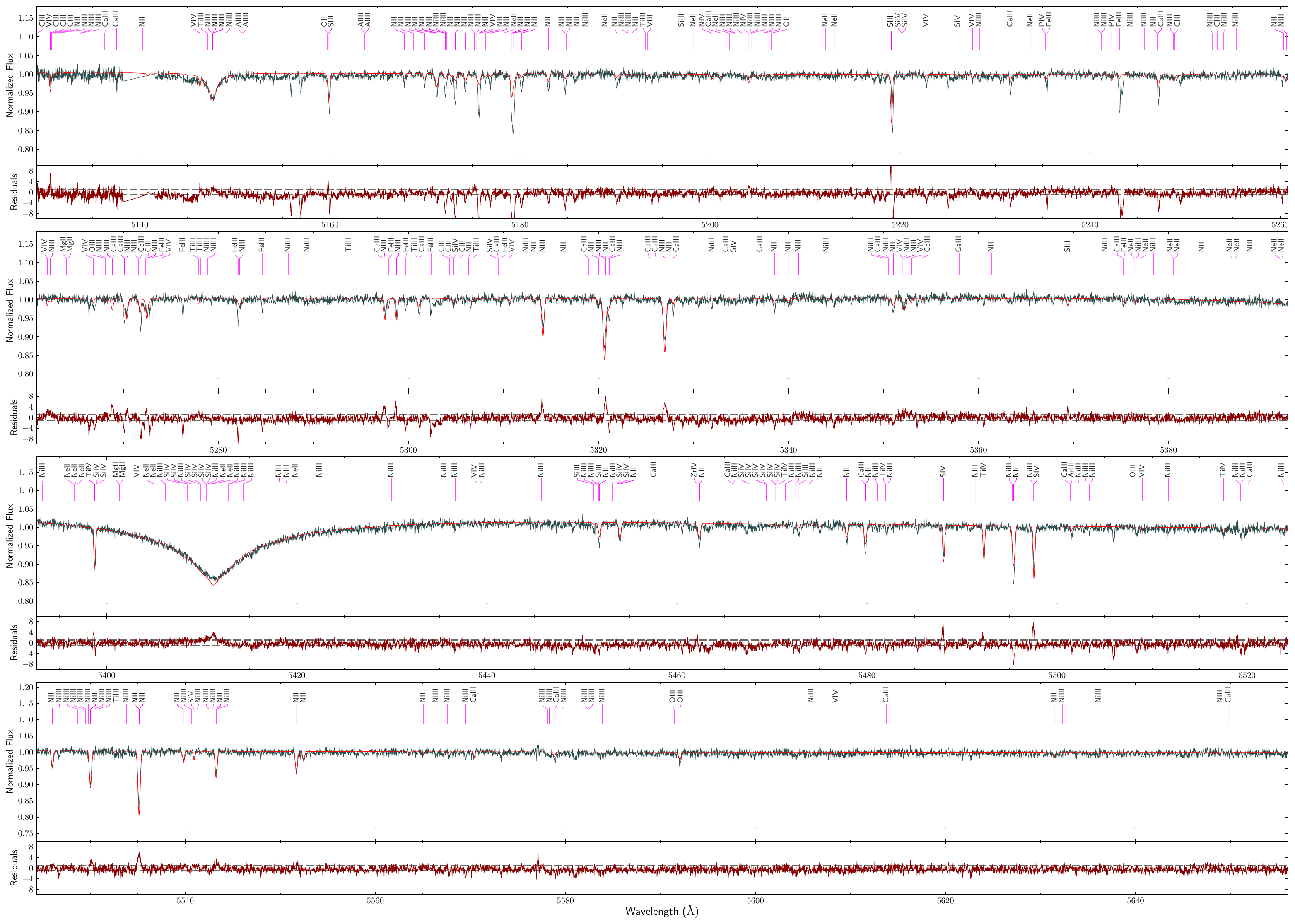}
\caption{HIRES spectrum of \HZ\ (gray) and the final model (red).}
\end{figure*}
\begin{figure*}
\ContinuedFloat
\centering
\includegraphics[width=24.2cm,angle =90,page=6]{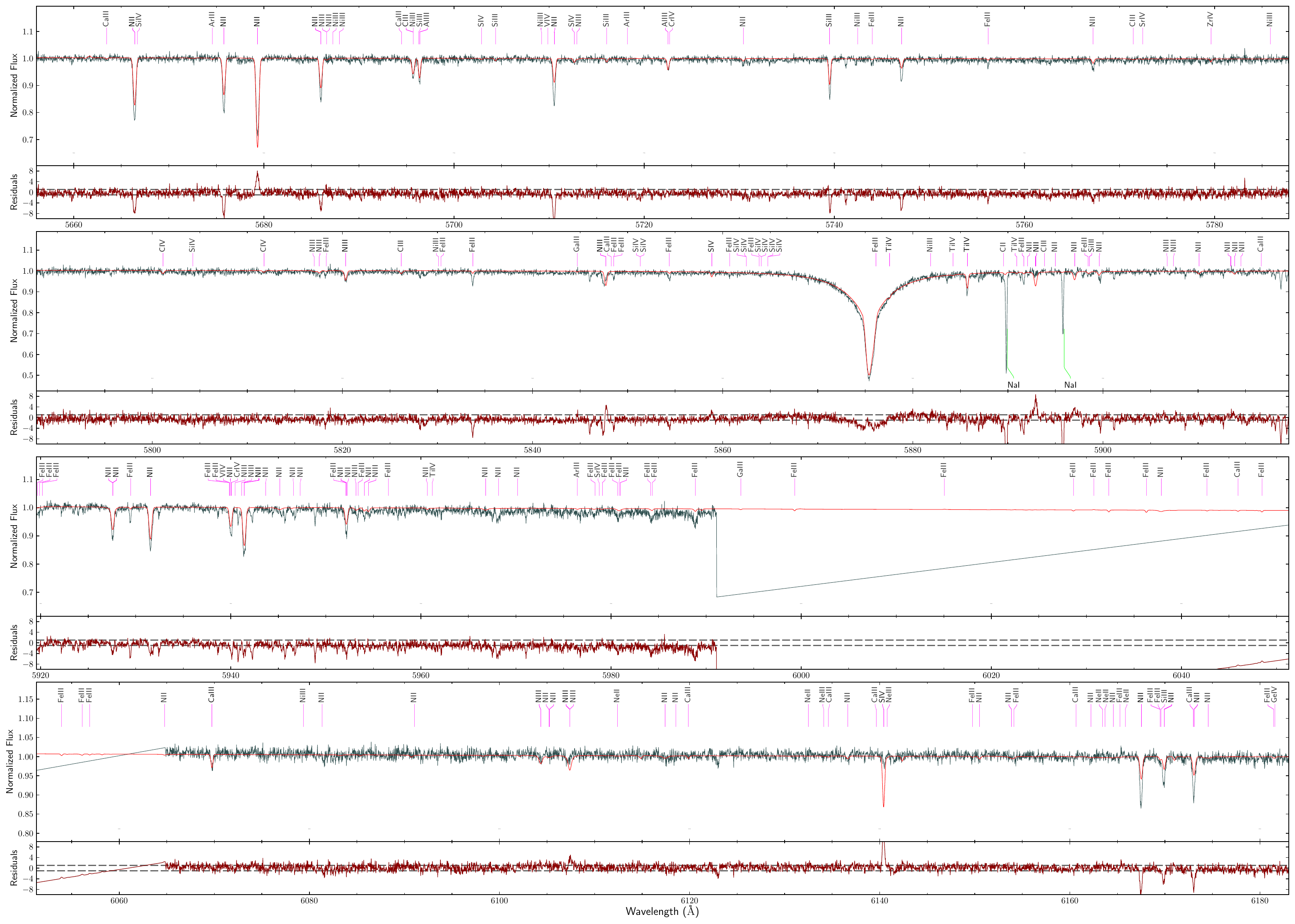}
\caption{HIRES spectrum of \HZ\ (gray) and the final model (red).}
\end{figure*}
\begin{figure*}
\ContinuedFloat
\centering
\includegraphics[width=24.2cm,angle =90,page=7]{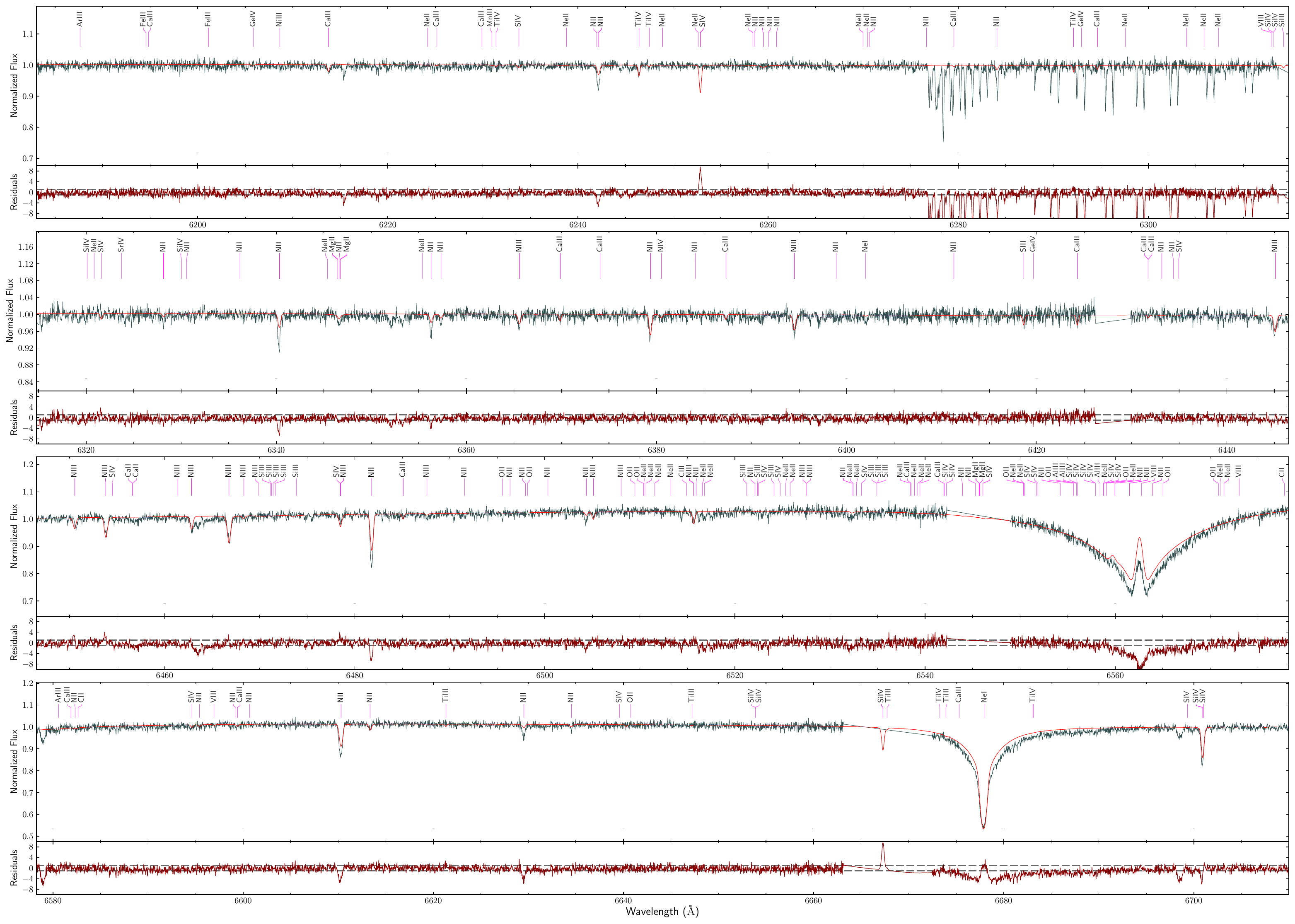}
\caption{HIRES spectrum of \HZ\ (gray) and the final model (red).}
\end{figure*}
\captionsetup[ContinuedFloat]{labelformat=continued}
\begin{figure*}
\centering
\includegraphics[width=24cm,angle =90,page=1]{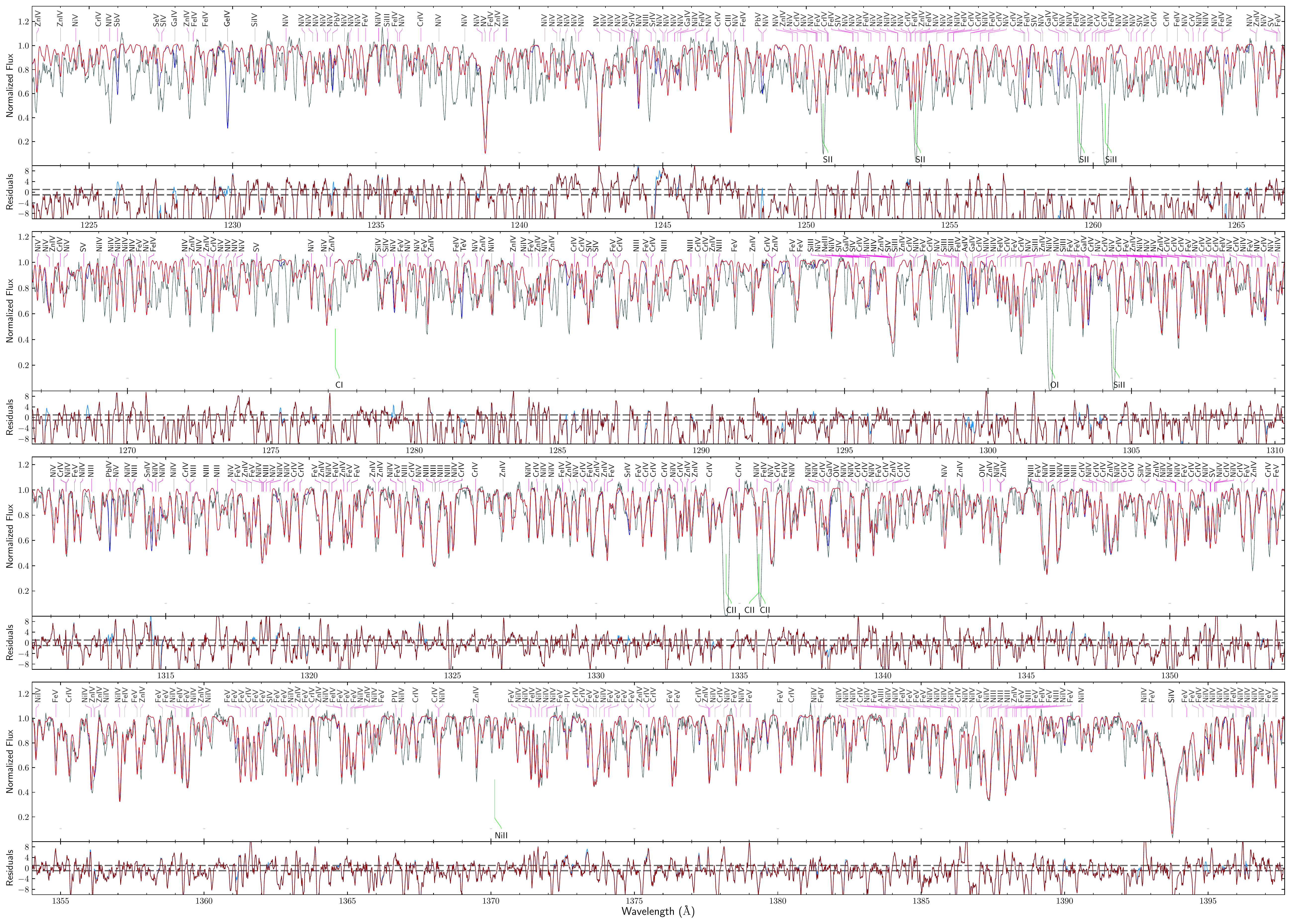}
\caption{GHRS spectrum of \HD\ (gray) and the final model (red, with heavy metals: blue).}
\end{figure*}
\begin{figure*}
\ContinuedFloat
\centering
\includegraphics[width=24.2cm,angle =90,page=2]{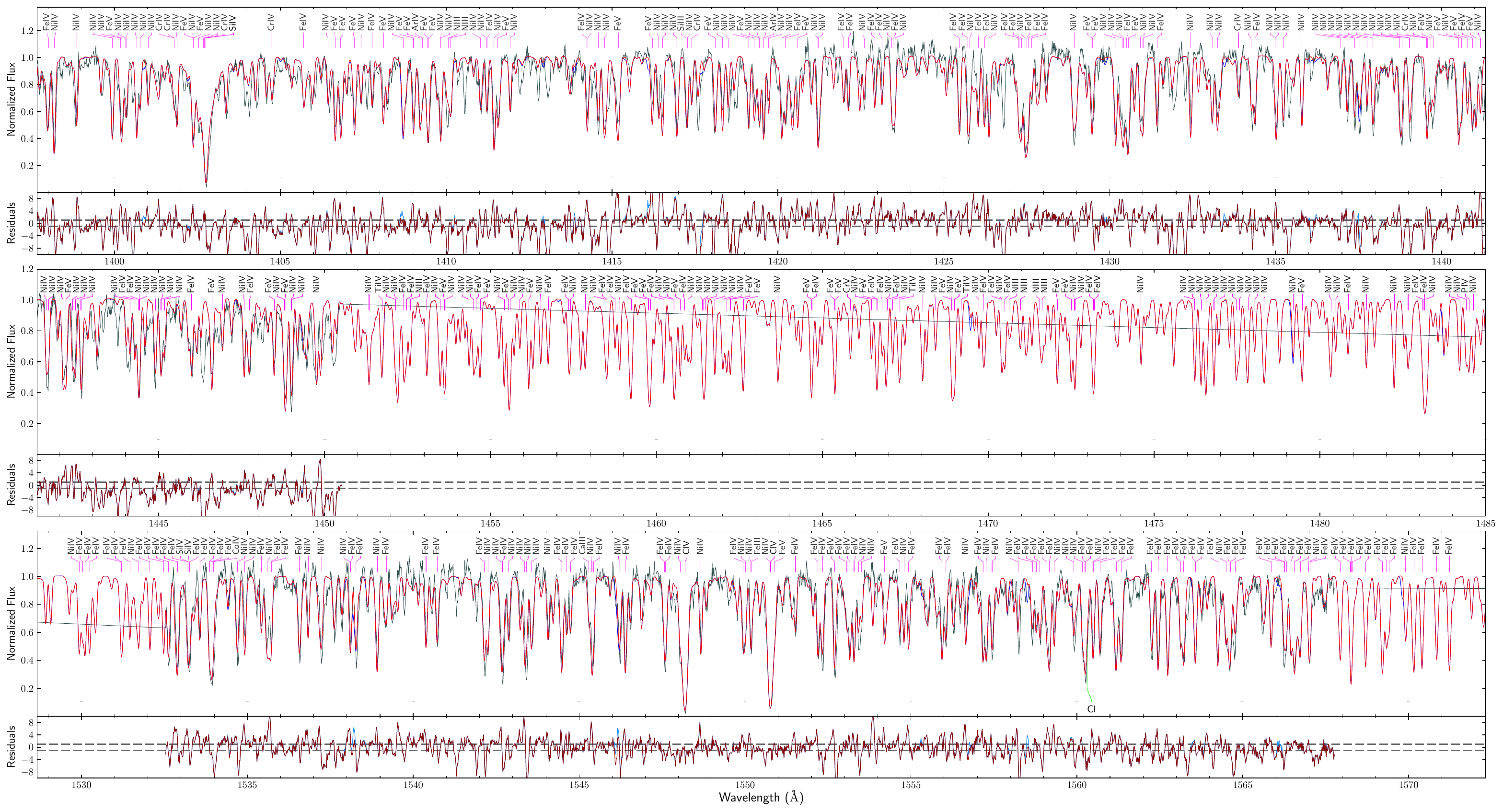}
\caption{GHRS spectrum of \HD\ (gray) and the final model (red, with heavy metals: blue).}
\end{figure*}
\begin{figure*}
\ContinuedFloat
\centering
\includegraphics[width=24.2cm,angle =90,page=3]{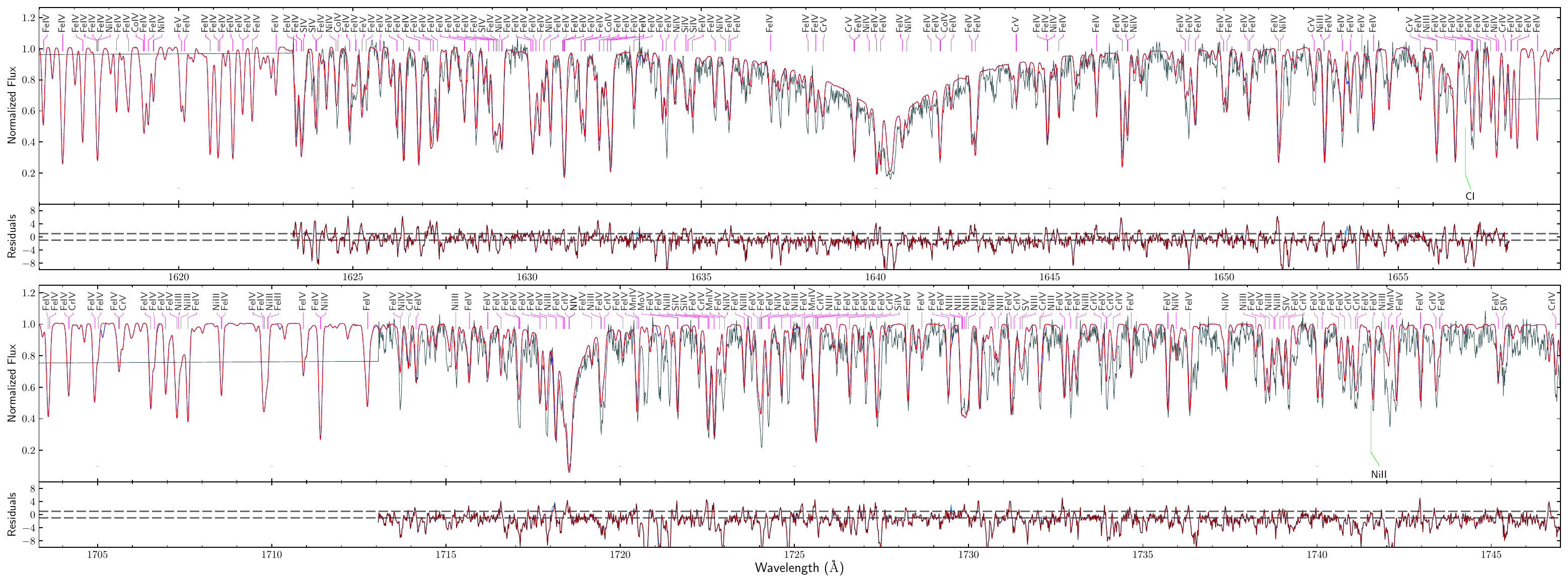}
\caption{GHRS spectrum of \HD\ (gray) and the final model (red, with heavy metals: blue).}
\end{figure*}
\captionsetup[ContinuedFloat]{labelformat=continued}
\begin{figure*}
\centering
\includegraphics[width=24.2cm,angle =90,page=1]{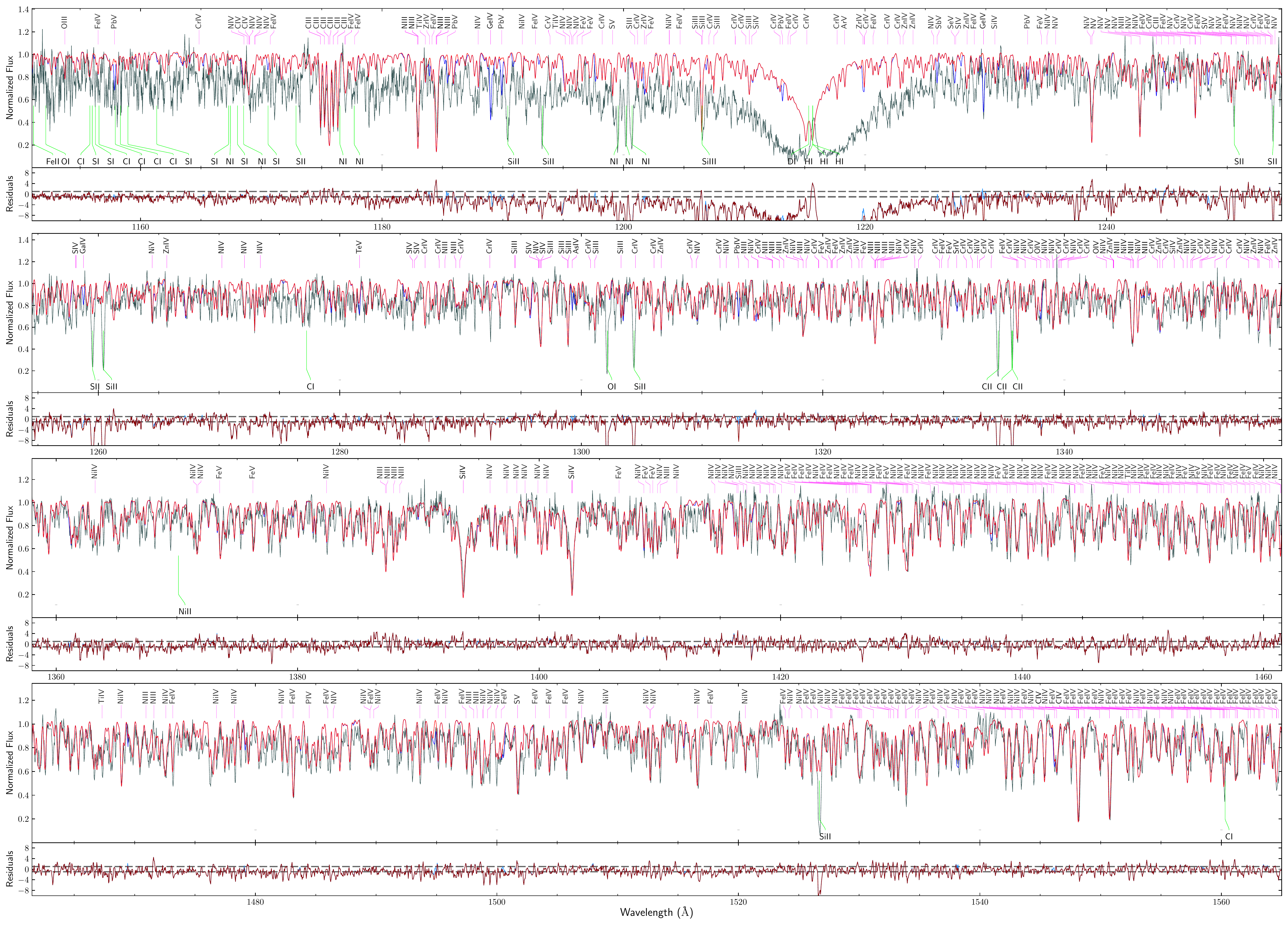}
\caption{IUE spectrum of \HD\ (gray) and the final model (red, with heavy metals: blue).}
\end{figure*}
\begin{figure*}
\ContinuedFloat
\centering
\includegraphics[width=24.2cm,angle =90,page=2]{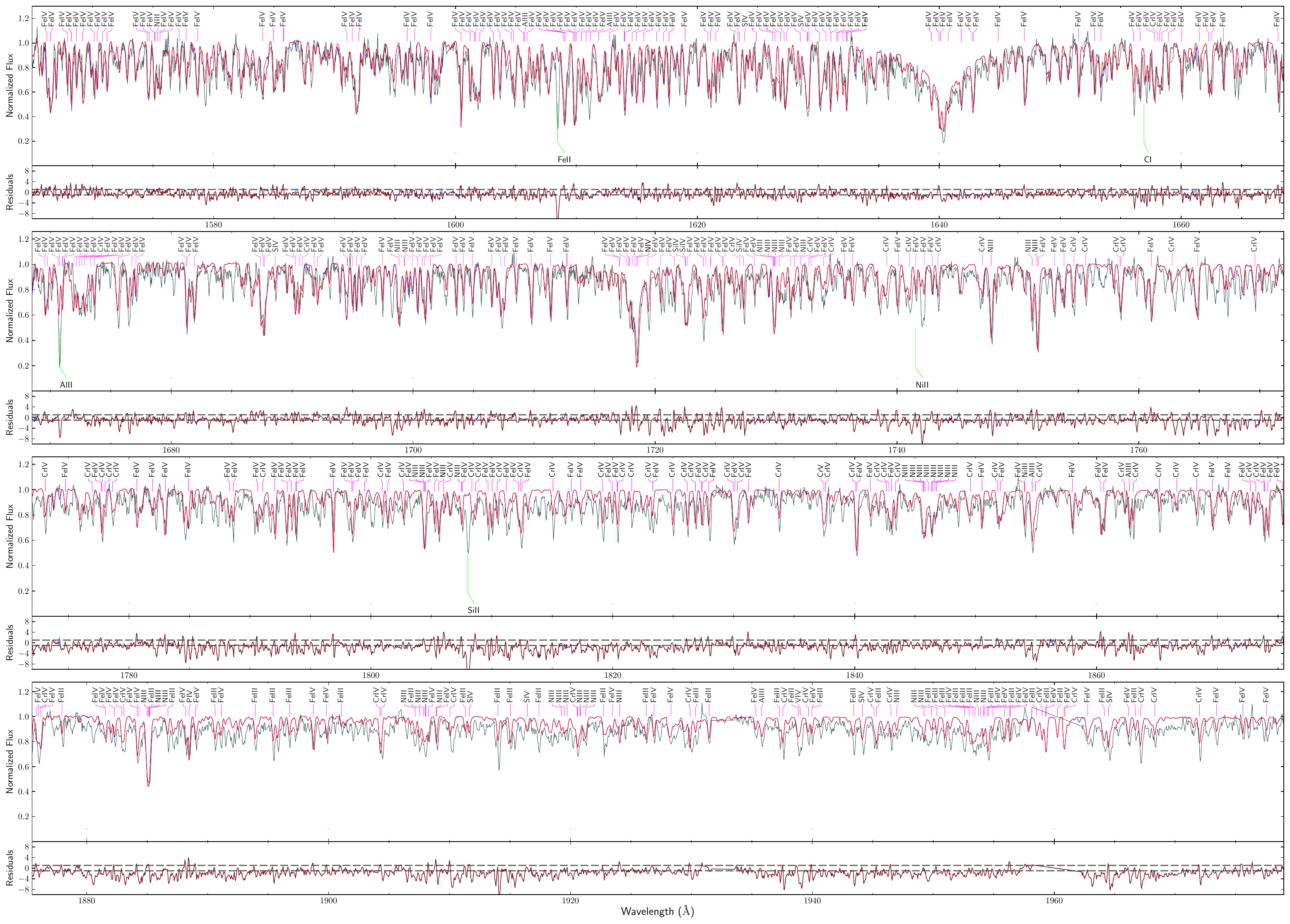}
\caption{IUE spectrum of \HD\ (gray) and the final model (red, with heavy metals: blue).}
\end{figure*}
\captionsetup[ContinuedFloat]{labelformat=continued}
\begin{figure*}
\ContinuedFloat
\centering
\includegraphics[width=24.2cm,angle =90,page=1]{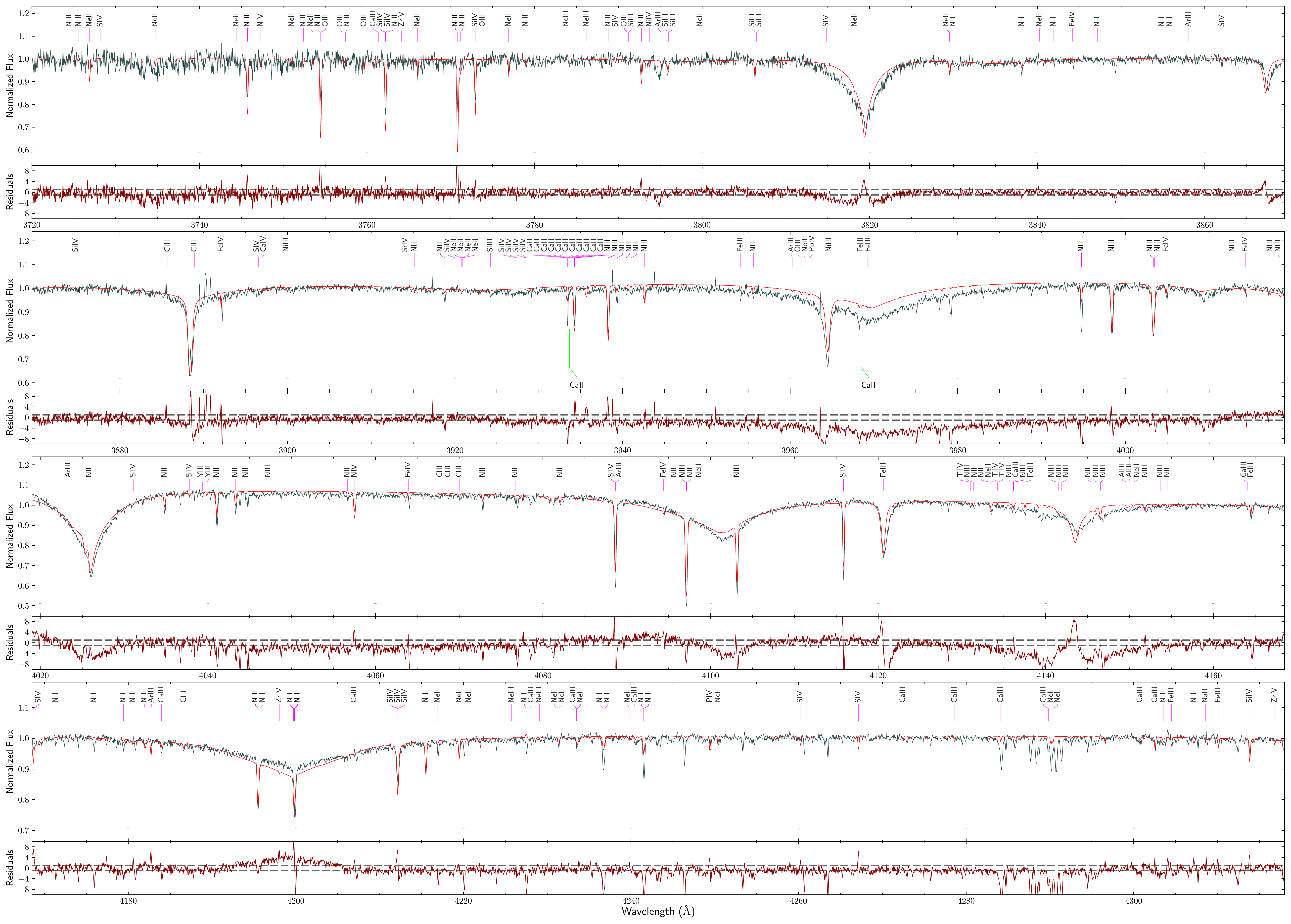}
\caption{FEROS spectrum of \HD\ (gray) and the final model (red).}\label{fig:full_spectra_last}
\end{figure*}
\begin{figure*}
\ContinuedFloat
\centering
\includegraphics[width=24.2cm,angle =90,page=2]{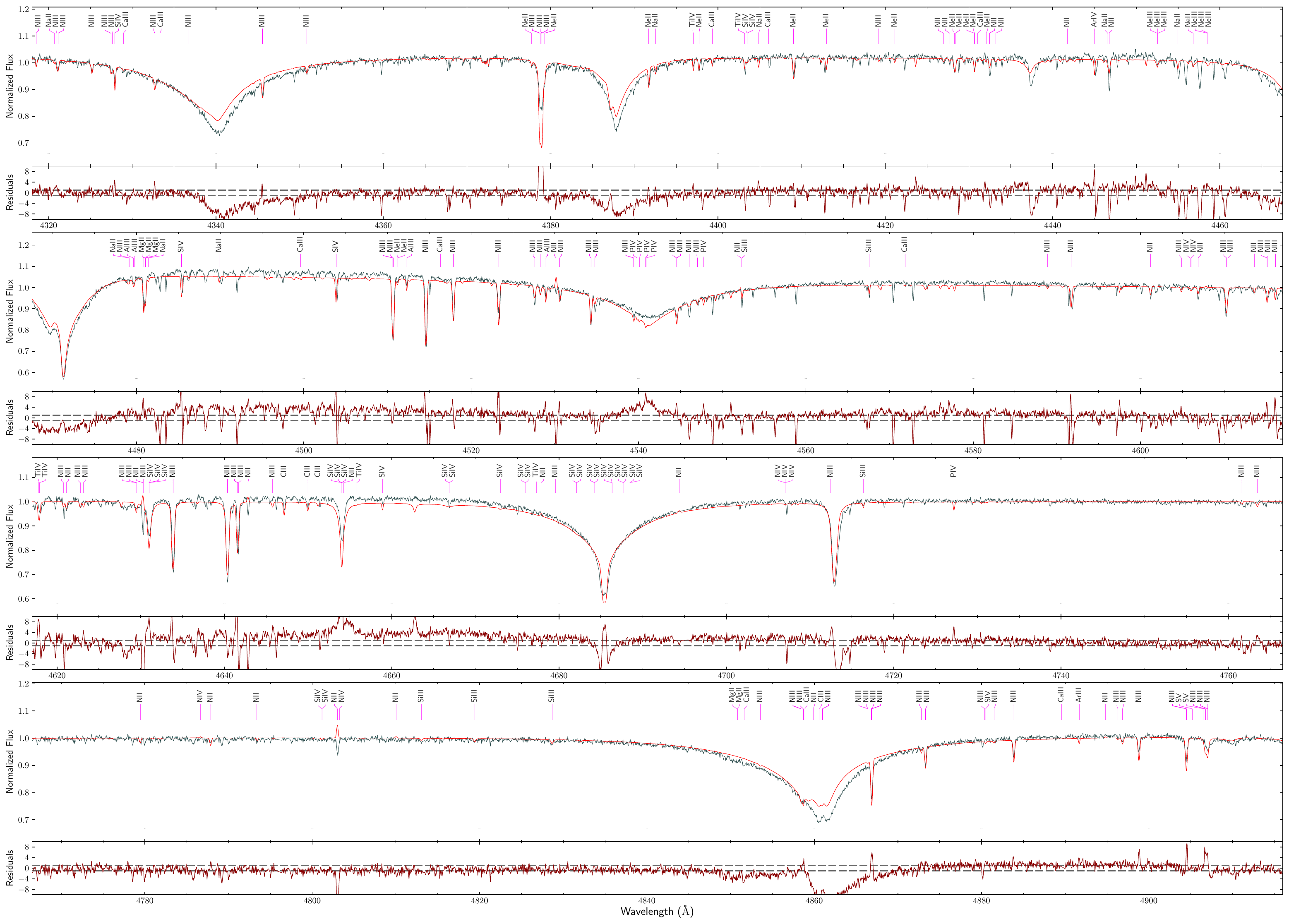}
\caption{FEROS spectrum of \HD\ (gray) and the final model (red).}
\end{figure*}
\begin{figure*}
\ContinuedFloat
\centering
\includegraphics[width=24.2cm,angle =90,page=3]{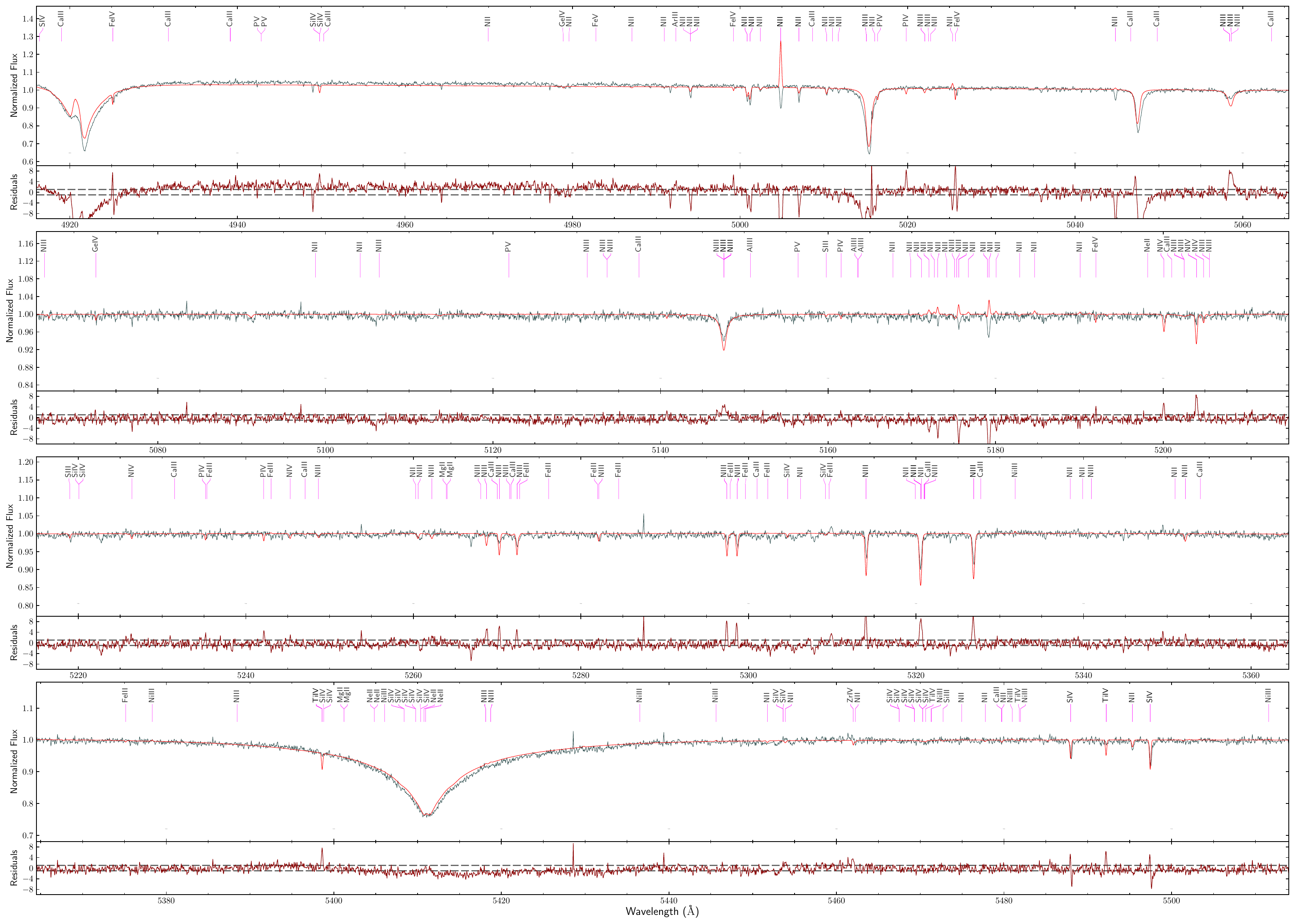}
\caption{FEROS spectrum of \HD\ (gray) and the final model (red).}
\end{figure*}
\begin{figure*}
\ContinuedFloat
\centering
\includegraphics[width=24.2cm,angle =90,page=4]{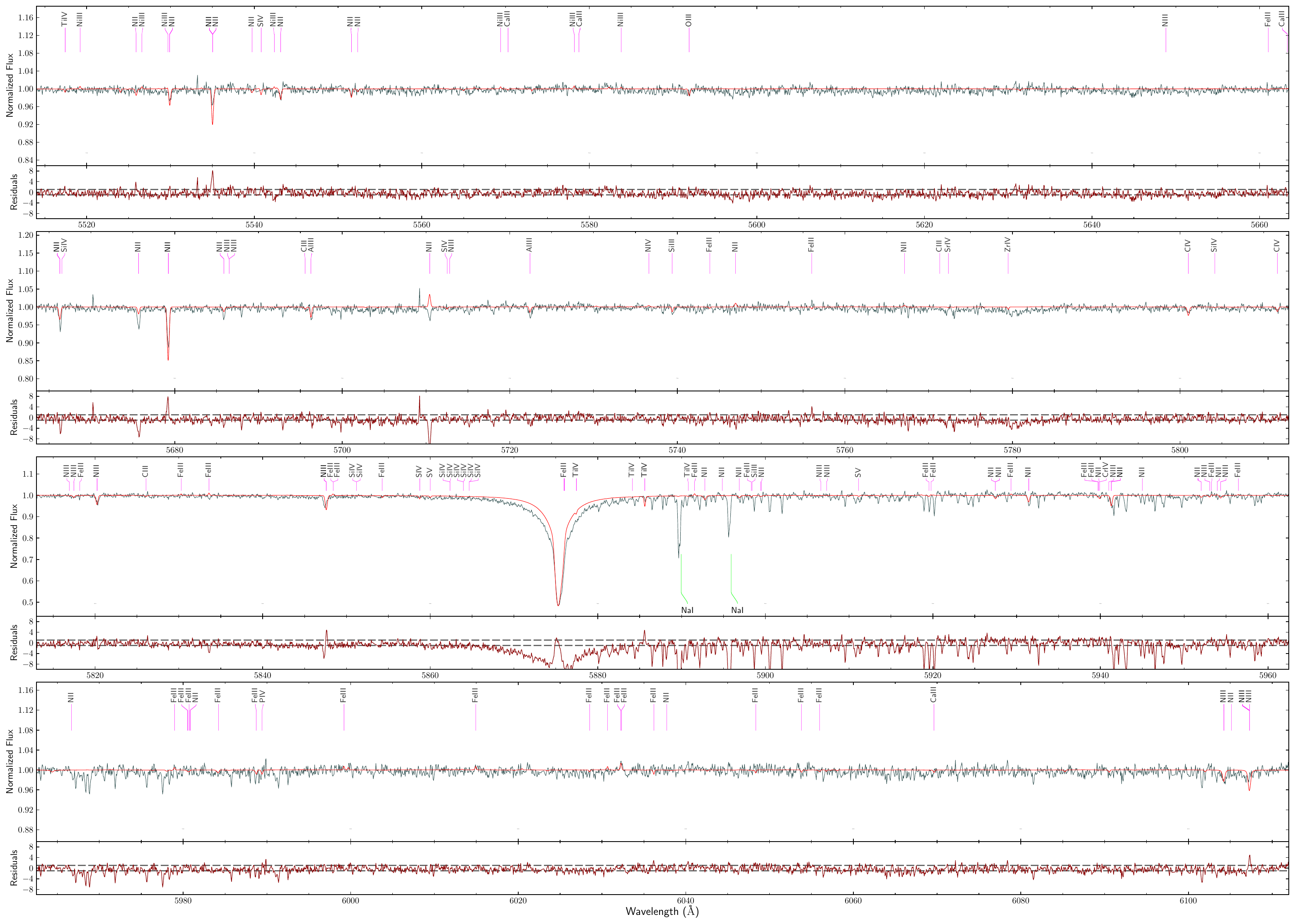}
\caption{FEROS spectrum of \HD\ (gray) and the final model (red).}
\end{figure*}
\begin{figure*}
\ContinuedFloat
\centering
\includegraphics[width=24.2cm,angle =90,page=5]{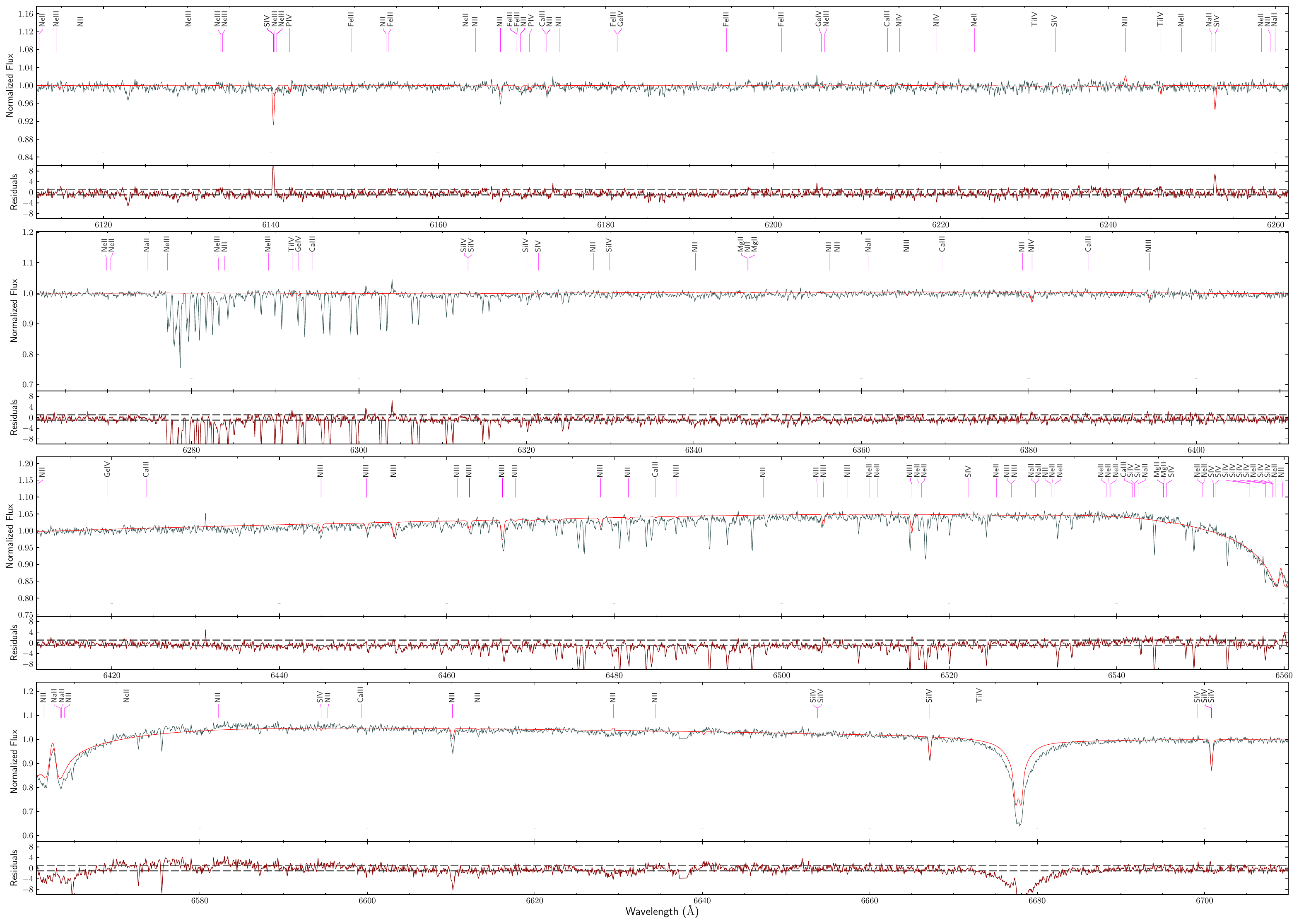}
\caption{FEROS spectrum of \HD\ (gray) and the final model (red).}
\end{figure*}

\end{appendix}
%
\end{document}